\documentclass[superscriptaddress,
showpacs,preprintnumbers,nofootinbib,eqsecnum,amsmath,amssymb,aps,11pt, groupedaddress, floatfix]{revtex4-2}
\usepackage[margin=1.0in]{geometry}
\usepackage{tikz,pgfplots}
\usepackage{amsmath,amssymb,bm}
\usepackage{graphicx}
\usepackage{epstopdf}
\usepackage[caption=false]{subfig}
\usepackage{braket}
\usepackage{dcolumn}
\usepackage{cancel}
\input{colordvi.tex}
\usepackage{tabularx}
\usepackage{hyperref}
\usepackage{cleveref}
\usepackage{mathtools}
\usepackage[utf8]{inputenc}
\usepackage{xcolor}
\usepackage{soul}
\usepackage{pifont}
\usepackage{ragged2e}
\definecolor{nicered}{rgb}{0,0.4,0.8}
\definecolor{nicegreen}{rgb}{.1,.5,.1}
\definecolor{darkblue}{rgb}{0,0,.8}
\hypersetup{colorlinks, citecolor=nicered ,linkcolor=darkblue, urlcolor=nicered}

\usepackage{multirow, makecell, cellspace, bigstrut}
\usepackage{pict2e}
\usepackage{booktabs}
\usepackage{diagbox}

\newcolumntype{C}{>{\centering\arraybackslash}X}

\usepackage{subcaption}
\usepackage{slashed}

\newlength{\hkpanelw}

\usepackage{soul}

\begin{document}


\title{\large \bf Dark Matter Interpretations of the Galactic Center Gamma-Ray Excess with 18 Years of Fermi-LAT Data and the FL16Y Source Catalog}

\author{Sang Hwan Kim$^{1,2}$}
\email{sanghwankim97@yonsei.ac.kr}

\author{Hae Barg Kang$^1$} 
\email{hamsark@jbnu.ac.kr}

\author{Seodong Shin$^{1,3,4}$}
\email{sshin@jbnu.ac.kr}

\affiliation{
$^1$Laboratory for Symmetry and Structure of the Universe, Department of Physics, Jeonbuk National University, Jeonju, Jeonbuk 54896, Korea \\
$^2$Department of Physics and IPAP, Yonsei University, Seoul 03722, Korea \\
$^3$School of Physics, Korea Institute for Advanced Study, Seoul 02455, Korea \\
$^4$Center for Theoretical Physics of the Universe, Institute for Basic Science, Daejeon 34126, Korea}

\begin{abstract}
We investigate dark matter (DM) interpretations of the Galactic
Center Excess (GCE) using 17.7 years of Fermi-LAT data and the FL16Y source catalog. 
We examine the dependence on Galactic
diffuse emission using 80 models and include correlated systematic uncertainties estimated from control regions along the Galactic plane. 
We derive updated combined 95\% CL upper limits on the annihilation cross section from 30 dwarf spheroidal galaxies (dSphs), including uncertainties in their stellar-kinematic $J$ factors, and compare the GCE-preferred regions with these limits and AMS-02 antiproton constraints.
For self-conjugate DM, we consider direct two-body annihilation, effective four-body annihilation into $4b$, $4\tau$, and $2b2\tau$,
and cascade annihilation through on-shell scalar mediators.
Using Model~X, which gives the best fit to the Galactic center
data among the 80 diffuse emission models considered in our
template analysis including a DM component, we find that the
dSph limits exclude the 68\% CL GCE-preferred regions for the
$b\bar b$, $c\bar c$, light-quark, and gluon channels, as well
as the effective $4b$ and $2b2\tau$ channels.
Portions of the $\tau^+\tau^-$ and effective $4\tau$ regions remain compatible with these limits, although their best-fit points lie above them. Heavy two-body final states are less constrained but provide poorer fits to the GCE spectrum.
Including the mediator mass dependence changes both the
GCE-preferred regions and the complementary constraints.
In particular, part of the 68\% CL preferred region for the $4b$ cascade remains below the dSph limits for $r=m_{\rm mediator}/m_{\rm DM} \lesssim 0.4$.
Among the four-body benchmarks, the $4\tau$ channel retains a comparatively large preferred region compatible with both constraints. With the adopted BIG propagation model, the antiproton bounds are generally weaker but constrain part of the $4b$ cascade region. 
Finally, we examine a benchmark of the secluded singlet fermionic dark matter model that realizes a predominantly four-bottom cascade.
\end{abstract}

\maketitle

\tableofcontents

\newpage

\section{Introduction}

Indirect detection searches for signatures of dark matter (DM)
annihilation or decay in the present Universe through observations
of the resulting Standard Model (SM) particles.
This approach probes non-gravitational interactions between DM and
SM particles. Since DM constitutes most of the matter in the
Universe and forms gravitationally bound halos, regions of high
DM density are promising targets, particularly for annihilation
signals.
Depending on the interactions involved, messengers such as photons, neutrinos, positrons, and antiprotons can provide information on the nature of DM, motivating a variety of observational
programs.~\footnote{In a broader sense, indirect probes of DM also
include its effects on stellar evolution and supernovae, as well
as cosmological signatures associated with the dark sector,
including gravitational waves.}
Indirect detection can also probe the annihilation processes
responsible for the relic abundance in thermal freeze-out
scenarios, including those involving weakly interacting massive
particles (WIMPs). The connection between the present-day
annihilation rate and the relic abundance depends on the
annihilation mechanism and its velocity dependence.

However, the expected DM signals are generally faint compared with astrophysical backgrounds.
Their identification therefore requires a reliable understanding
of both the backgrounds and instrumental systematic uncertainties.
Despite these difficulties, indirect searches have placed strong
constraints on particle DM candidates~\cite{Cooley:2022ufh}.
Several observations have also reported excesses over the
predicted backgrounds that could admit DM interpretations.
However, unresolved sources and uncertainties in background
modeling remain possible explanations, and none of these excesses has been conclusively identified as a DM signal.
Resolving these ambiguities remains an important objective of
indirect detection studies.
The Snowmass 2021 white paper provides a detailed discussion
of these possible signals~\cite{Leane:2022bfm}.

One of the most widely discussed examples is the Galactic Center
Excess (GCE), an excess of $\gamma$-ray emission toward the Galactic
Center (GC), with a prominent component at energies of a few GeV,
relative to modeled astrophysical backgrounds.
It was first reported in an analysis of Fermi Large Area Telescope
(Fermi-LAT) data in 2009~\cite{Goodenough:2009gk}.
A subsequent analysis by the Fermi-LAT Collaboration also found
additional emission toward the GC, while emphasizing its
dependence on the interstellar-emission
model~\cite{Fermi-LAT:2015sau}.
The GC is a promising target for DM annihilation searches because the DM density is expected to increase toward the Galactic center.
Annihilation into SM particles such as $b\bar b$ can produce photons through hadronization and subsequent hadron decays, with additional contributions from final state radiation, as discussed
in Ref.~\cite{Goodenough:2009gk}. 
Early analyses also found an approximately spherically symmetric
morphology consistent with an annihilation template based on a
generalized Navarro--Frenk--White (NFW) density profile~\cite{Daylan:2014rsa}.

Astrophysical explanations have also been proposed, including a
population of unresolved millisecond pulsars (MSPs)~\cite{Abazajian:2010zy} and past cosmic-ray (CR)
outbursts~\cite{Carlson:2014cwa}.
Studies of the diffuse emission and GCE morphology have supported
astrophysical interpretations under particular modeling
assumptions~\cite{Bartels:2017vsx,Ploeg:2020jeh,Manconi:2024tgh}.
The MSP interpretation can be tested by comparing the required
pulsar luminosity function with those inferred from globular
clusters and the Galactic
disk~\cite{Hooper:2013nhl,Zhong:2019ycb,Dinsmore:2021nip}.
The abundance of low-mass $X$-ray binaries, which can evolve into
MSPs through accretion-driven recycling, provides another
test~\cite{Zhong:2019ycb,Abazajian:2012pn}.
Spatial variations in the GCE spectrum can also constrain CR
outburst scenarios~\cite{Carlson:2014cwa,Calore:2014xka}.
Conversely, an analysis using diffuse-emission models constrained
by CR observations found that the high-energy GCE spectrum
disfavors known MSP populations as its sole origin and can be
described by DM annihilation~\cite{Cholis:2021rpp}.
A recent morphological analysis also found that triaxial DM halo
templates provide a better fit than the stellar templates
considered in that study~\cite{Hu:2026apn}.
The origin of the excess therefore remains unresolved.
Further studies of Galactic diffuse emission, the GCE morphology,
extended and point-source templates, and the pulsar luminosity function are needed.

In addition to studies of unresolved sources, background systematics, and morphology, DM interpretations of the GCE can be tested with independent astrophysical observations.
These include gamma rays from dwarf spheroidal galaxies (dSphs)
and measurements of antiproton and positron fluxes or flux ratios
near the Earth.~\footnote{Some constraints can be weakened in dark sector scenarios with additional particles, for example through long-lived mediators~\cite{Kim:2017qaw} or predominantly leptonic decay channels~\cite{Arkani-Hamed:2008hhe}, depending on the observable under consideration.}
Following this approach, we re-examine the DM interpretation of the GCE through an event-level analysis of 17.7 years of
Fermi-LAT data.
We use the Galactic diffuse emission templates developed in
Ref.~\cite{Cholis:2021rpp}, which incorporate cosmic-ray (CR) measurements from the Alpha Magnetic Spectrometer (AMS-02).
We also adopt the FL16Y source list~\cite{Ballet:2026kpq}, an interim release based on 16 years of LAT data that precedes
the planned 5FGL catalog.
Relative to the 14-year 4FGL-DR4 catalog, FL16Y includes additional
sources and revised source localizations.
These changes affect the point source masks, particularly at
low energies, where the LAT point spread function is broader.
We also use AMS-02 antiproton measurements, reconstructing the experimental covariance matrix from the published uncertainty information.

We specifically investigate annihilation into four-body final
states, such as $b\bar b b\bar b$, denoted by $4b$ below.
Such final states commonly arise through cascade decays of dark
sector particles in secluded DM scenarios, where the relic
abundance can be determined primarily by annihilation within the
dark sector~\cite{Pospelov:2007mp,Kim:2009ke,Kim:2016csm,
Elor:2015tva,DEramo:2026tvx}.
At a fixed DM mass, the available energy is shared among four
primary SM particles rather than two, generally producing a
softer photon spectrum than direct annihilation into $b\bar b$.
Consequently, four-body channels can favor larger DM masses when fitted to the GCE, as found in earlier studies~\cite{Cline:2015qha,Dutta:2015ysa}.
The detailed photon spectrum depends on the annihilation and
decay kinematics.
For example, in the cascade
$\chi \bar \chi \to \phi \phi \to b \bar b b \bar b$,
the mediator ($\phi$) boost depends on the mass ratio:
$\gamma_\phi = m_\chi/m_\phi$ for nonrelativistic annihilation into
two identical on-shell mediators.
Varying this ratio changes the photon spectrum, as demonstrated for several benchmarks in Ref.~\cite{Profumo:2017obk}.

We investigate these mediator effects systematically using the
updated GCE spectrum.
As a reference framework, we consider the singlet
fermionic dark matter (SFDM)
model~\cite{Kim:2008pp,Kim:2009ke,Kim:2016csm},
a representative Higgs portal scenario.
In this model, four-body final states can arise from DM annihilation into pairs of scalar mass eigenstates $h_i h_j$
with $i,j=1,2$, when kinematically accessible.
Here $h_1$ and $h_2$ are the SM-like and singlet-like Higgs bosons,
respectively, whose subsequent decays can produce two pairs of
SM particles.

We compare the GCE-preferred parameter regions with constraints
from Fermi-LAT observations of dSphs and AMS-02 antiproton
measurements.~\footnote{See also Ref.~\cite{DiMauro:2021qcf}
for a complementary analysis of two-body annihilation channels.}
The high mass-to-light ratios of dSphs indicate that these systems
are DM dominated. 
Their low gas content and limited recent star
formation also imply weak intrinsic $\gamma$-ray backgrounds,
making them promising targets for DM annihilation searches.
We analyze 17.7 years of Fermi-LAT data for 30 dSphs with
stellar-kinematic $J$-factor estimates and include the
corresponding uncertainties in the combined limits.
In addition to updating the constraints on two-body channels,
including $b\bar b$ and $\tau^+\tau^-$, we derive limits for
the $4b$, $2b2\tau$, and $4\tau$ channels using effective
four-body spectra and on-shell cascades over a range of
mediator-to-DM mass ratios.
The analysis uses the open-source Python package
\texttt{Fermipy}~\cite{Wood:2017yyb}.
We use Refs.~\cite{McDaniel:2023bju,Circiello:2026inp}
as methodological references, while adopting our own choices
for the source catalogs, residual-source procedure, and
annihilation spectra.
In particular, we use \texttt{CosmiXs}~\cite{Arina:2023eic}
for the two-body spectra and custom spectra for the four-body
channels.

Dark matter annihilation into colored particles can also be tested through the antiproton flux and antiproton-to-proton flux ratio,
$\bar p/p$, measured by experiments such as
AMS-02~\cite{AMS:2021nhj}.~\footnote{Purely leptonic channels,
including $e^+e^-$, $\mu^+\mu^-$, and $\tau^+\tau^-$, can also
produce antiprotons through electroweak radiation followed by
hadronic decays of the emitted weak bosons, where kinematically
relevant~\cite{Garny:2011cj}.}
Several analyses have reported an excess in the AMS-02 antiproton
data over the predicted secondary flux from CR collisions with
interstellar gas.
This excess has been fitted by DM annihilation, for example into
$b\bar b$, with parameters compatible with some GCE
interpretations~\cite{Cuoco:2016eej,Cui:2016ppb}.
However, its significance depends on uncertainties in secondary antiproton production, CR propagation, primary CR spectra and source populations, solar modulation, and the correlations among
experimental systematic uncertainties.
We therefore use the AMS-02 measurements to derive constraints
and leave a dedicated investigation of a possible antiproton
excess for future work.

We use the four-year AMS-02 antiproton dataset~\cite{AMS:2016oqu}, for which previous studies have constructed estimates of the experimental covariance
matrix~\cite{Boudaud:2019efq}.
The antiproton fluxes are calculated with the semi-analytical code \texttt{USINE}~\cite{Maurin:2018rmm} using the
\texttt{BIG} propagation benchmark~\cite{Genolini:2019ewc}.
Our likelihood includes a reconstructed experimental covariance
matrix with correlations between energy bins and an external constraint on the halo half-height $L$, as described in
Sec.~\ref{sec:AMS-02}.
We do not include an additional model covariance matrix for
uncertainties in CR propagation, primary CR fluxes, solar
modulation, or antiproton production cross sections.
Including these uncertainties could weaken the limits, although
their quantitative effect requires a dedicated reanalysis.
In our analysis setup, the antiproton constraints are generally
weaker than the dSph constraints in the parameter regions relevant
to the GCE and have little effect on the low-mass leptonic
benchmarks.
For example, at $m_\chi = 40~\mathrm{GeV}$ in the $b\bar b$ channel,
our 95\% CL upper limits from antiprotons and dSphs are
approximately $1.5\times10^{-26}$ and
$9.2\times10^{-27}~\mathrm{cm^3\,s^{-1}}$, respectively.

Finally, we apply these results to an SFDM benchmark.
The scalar mediators can decay into several SM final states, with branching fractions determined by the model parameters.
We therefore calculate the inclusive photon spectrum by combining all relevant decay channels.~\footnote{In this model
application, the branching fractions follow from the model
parameters. This differs from phenomenological analyses that
treat the branching fractions as independent fit parameters,
such as Ref.~\cite{DiMauro:2021qcf}.}
We select a benchmark using the pure-$4b$ GCE interpretation
and its complementary constraints, and evaluate its inclusive
photon spectrum against the updated GCE data.
Comparisons with the pure-$4b$ dSph and antiproton limits are
used as approximate tests; a full likelihood analysis using
the mixed spectrum predicted by the model is left for future
work.

This paper is organized as follows.
In Sec.~\ref{sec:FermiLATandAMS02}, we briefly summarize the key features of the Fermi-LAT and AMS-02 experiments.
Section~\ref{sec:GCE} describes the GCE analysis, including diffuse emission modeling, point source masking, data preparation,
template fitting, and systematic uncertainties, and presents
the extracted GCE spectrum.
The combined gamma-ray constraints from 30 dSphs are derived
in Sec.~\ref{sec:dsph-gamma-analysis}, and the AMS-02 antiproton
analysis is described in Sec.~\ref{sec:AMS-02}.
In Sec.~\ref{sec:DMinterp}, we compare the GCE interpretations
based on two-body annihilation, effective four-body annihilation,
and cascades through on-shell mediators with the dSph and
antiproton constraints.
Section~\ref{sec:SFDM} applies the results to the SFDM model
and presents an illustrative benchmark.
We summarize our results and discuss future directions in
Sec.~\ref{sec:conclusions}.
Appendix~\ref{app:validation} presents a validation of our
dSph analysis against the results of a previous study.
Appendix~\ref{app:catalog_comparison} compares the point source
catalogs and their effects on masking in the GCE analysis,
while Appendix~\ref{app:frontback} examines the dependence
on event selection.
Finally, Appendix~\ref{app:dSphs_weak} discusses possible reasons for the weakening of the dSph limits at high DM masses when the observation interval is extended.

\section{Fermi-LAT and AMS-02}
\label{sec:FermiLATandAMS02}

In this section, we briefly summarize the key characteristics of Fermi-LAT and AMS-02 experiments 
we considered in the analysis.
Both of the experiments have been under operation for more than 15 years and provided crucial milestones in DM indirect detection.

\subsection{Fermi-LAT}

The Fermi Gamma-ray Space Telescope is an orbital observatory for studying cosmic $\gamma$-ray emission. 
Its primary instrument, the Large Area Telescope (LAT), detects photons through electron–positron pair conversion and was designed to operate from approximately $20\,\mathrm{MeV}$ to energies above $300\,\mathrm{GeV}$~\cite{Fermi-LAT:2009ihh, Fermi-LAT:2012fsm}. 
Its instantaneous field of view is approximately $2.4\,\mathrm{sr}$, corresponding to roughly one-fifth of the sky. 
The detector incorporates sixteen towers arranged in a $4\times4$ grid, with each tower combining a silicon-strip tracker with tungsten conversion foils, a $\mathrm{CsI(Tl)}$ calorimeter, and readout electronics. 
Charged-particle backgrounds are suppressed using a segmented anticoincidence detector, while the trigger and data-acquisition system coordinates event readout~\cite{Fermi-LAT:2009ihh}.

Each LAT tracker module contains 18 $x$–$y$ tracking planes formed from single-sided silicon-strip detectors. 
Tungsten conversion foils precede the first sixteen planes: twelve thin foils in the front section and four thicker foils in the back section, with individual thicknesses of $0.03$ and $0.18$ radiation lengths, respectively. 
The final two planes contain no conversion foils. This arrangement balances photon-conversion efficiency against the angular blurring caused by multiple scattering, with front-converting events generally providing a narrower point-spread function (PSF)~\cite{Fermi-LAT:2009ihh}. 
The PSF depends on photon energy, incidence angle, and event selection~\cite{Fermi-LAT:2012fsm}.
The angular resolution of the LAT improves with increasing photon energy. For front-converting events, representative 68\% (95\%) PSF containment radii decrease from approximately $0.4^\circ$ ($1.5^\circ$) at $1\,\mathrm{GeV}$ to $0.1^\circ$ ($0.3^\circ$) at $10\,\mathrm{GeV}$~\cite{Fermi-LAT:2015sau}. Back-converting events generally have a broader PSF, with its energy dependence described in Ref.~\cite{Fermi-LAT:2012fsm}. The finer angular resolution of front-converting events helps distinguish emission from compact sources from nearby sources and structured diffuse backgrounds.

The calorimeter measures the energy deposited by electromagnetic showers initiated by the conversion electrons and positrons. 
Each module contains 96 $\mathrm{CsI(Tl)}$ crystals, each measuring approximately $2.7\,\mathrm{cm}\times2.0\,\mathrm{cm}\times32.6\,\mathrm{cm}$. 
The crystals are distributed across eight layers, with perpendicular orientations in adjacent layers, providing a total depth of 8.6 radiation lengths at normal incidence~\cite{Fermi-LAT:2009ihh}. 
The measured shower profile supports reconstruction of the incident photon energy, rejection of background events, and corrections for energy leakage from the calorimeter~\cite{Fermi-LAT:2009ihh,Fermi-LAT:2012fsm}.

The event selections of LAT should be chosen according to the scientific objectives and used with the corresponding instrument response functions (IRFs). The IRFs describe the effective area, point-spread function, and energy dispersion, allowing models of the incident $\gamma$-ray flux to be translated into predicted event distributions. Pass~8~\cite{Fermi-LAT:2013jgq} provides several event classes defined by different background-rejection and event-quality selections. Within each class, events can be partitioned according to conversion location (\texttt{FRONT/BACK}), directional reconstruction quality (\texttt{PSF0--PSF3}), or energy reconstruction quality (\texttt{EDISP0--EDISP3}).~\footnote{Details are provided in the Fermi Science Support Center documentation: \url{https://fermi.gsfc.nasa.gov/ssc/data/analysis/lat_irfs/irf_overview.html}.} We adopt the \texttt{CLEAN} event class for the GC analysis and the \texttt{SOURCE} event class for the dSph analysis. The detailed event selections and corresponding IRFs are specified in the following sections.

The official Fermi-LAT document suggests that the PSF of $\texttt{FRONT}$ is intrinsically better than that of $\texttt{BACK}$~\footnote{\url{https://fermi.gsfc.nasa.gov/ssc/data/analysis/documentation/Pass8_usage.html}, }, we only use the data from $\texttt{FRONT}$ to reduce further contamination from the limited resolution of the instrument. 
The high photon statistics near the GCE peak allow us to use only \texttt{FRONT} events.
The choice between the two conversion types trades angular resolution against effective area: the thicker \texttt{BACK} foils increase the conversion probability, and hence the effective area, at the cost of an angular resolution that is worse than for \texttt{FRONT}-converting events, by less than a factor of two at $1$~GeV~\cite{Fermi-LAT:2009ihh}. Following Ref.~\cite{Cholis:2021rpp}, we use only \texttt{FRONT}-converting events in the GC analysis, which benefits from the better angular resolution in both the point-source masking and the template fit; a comparison with \texttt{FRONT+BACK} events is presented in Appendix~\ref{app:frontback}. The dSph analysis instead uses all four \texttt{PSF} event types as separate likelihood components (Sec.~\ref{sec:dsph-gamma-analysis}).

\subsection{AMS-02}

The anti-matter search telescope AMS-02 was installed on International Space Station on May 19, 2011 for long-term measurements of cosmic-ray spectra at GeV–TeV energies~\cite{AMS:2021nhj}.
Its permanent magnet has a central field of $0.14\,{\rm T}$ and surrounds six of the nine silicon tracker layers.
The magnet and tracker determine the rigidity (momentum-to-charge ratio) and charge sign of incident particles.
Additional measurements of velocity, charge magnitude, and energy help identify particle species and distinguish particles from antiparticles.
Cosmic-ray antiparticles are produced in conventional astrophysical processes, including collisions of primary cosmic rays with interstellar gas, and their fluxes are generally smaller than those of their particle counterparts.
Hence AMS-02 measurements of antiparticle spectra can be used to search for contributions from DM annihilation or decay in addition to those expected from astrophysical processes.

The transition radiation detector (TRD), located near the top of AMS-02, consists of 20 layers and distinguishes electrons and positrons from protons.
The time-of-flight (TOF) system measures particle velocities using four scintillator planes, with two above and two below the magnet.
Below the lower TOF planes, the ring-imaging Cherenkov (RICH) detector measures particle velocities and charge magnitudes.
The electromagnetic calorimeter (ECAL), located beneath the RICH, reconstructs electron and positron energies from their electromagnetic showers.
In addition to the six silicon tracker layers (L3--L8) inside the magnet bore, L1 is located above the TRD, L2 just above the magnet, and L9 between the RICH and the ECAL.
Previous constraints on the DM interpretation of the GCE derived from AMS-02 antiproton flux data were presented in Ref.~\cite{DiMauro:2021qcf}. However, since four-body annihilation channels were not considered in that study, we estimate the corresponding constraints in this work. 
The four-year AMS-02 antiproton flux observation data is from Ref. \cite{AMS:2016oqu}.

\section{Analysis of the Galactic Center Excess}
\label{sec:GCE}

Analyzing the gamma-ray emission from the GC requires careful modeling of diffuse backgrounds and point sources. 
The Galactic diffuse emission has substantial modeling uncertainties because the distributions of cosmic rays, interstellar gas, and radiation fields are not precisely known. 
The large number of point sources toward the GC further complicates the analysis, and their inferred fluxes depend on the adopted diffuse-emission model. 
To reduce the impact of uncertainties in point-source modeling, we mask regions around cataloged point sources.

Our diffuse-background model includes three components: Galactic diffuse emission, the Fermi bubbles, and isotropic emission. 
Cosmic-ray interactions with the interstellar medium (ISM) and ambient radiation fields produce Galactic diffuse emission (GDE) through $\pi^0$ decay, bremsstrahlung, and inverse Compton scattering~\cite{Fermi-LAT:2012edv, Moskalenko:2001ya, Strong:1998fr, Kamae:2006bf, Koch:1959zz, Blumenthal:1970gc, Jones:1968zza}.
The Fermi bubbles are two large gamma-ray lobes of uncertain origin extending to Galactic latitudes of approximately $b=\pm55^\circ$~\cite{Su:2010qj, Fermi-LAT:2014sfa}.
The isotropic component includes emission from unresolved extragalactic sources and diffuse extragalactic processes, as well as any approximately isotropic residual Galactic emission~\cite{Fermi-LAT:2014ryh}.
In this section, we first describe the diffuse-background models and the point-source masks based on the Fermi-LAT 16 years catalog (FL16Y)~\cite{Ballet:2026kpq}). 
We then discuss the data selection and preparation, the template-fitting procedure, and the treatment of systematic uncertainties. 
Finally, we present the inferred GCE spectrum and examine its interpretation in terms of DM annihilation.

\subsection{Modeling of Diffuse Emission Backgrounds}

\paragraph{GDE}
GDE dominates the gamma-ray emission from the GC region above $100\,{\rm MeV}$, making its modeling a major source of systematic uncertainty. 
Predicting this emission requires the cosmic-ray distributions obtained by solving the transport equation, which accounts for diffusion, convection, reacceleration, and energy losses~\cite{Strong:2007nh}. 
The predictions depend on transport parameters, such as the diffusion coefficient and Galactic-wind velocity, as well as the cosmic-ray injection spectra and source distribution. 
Further uncertainties arise from the adopted models of the interstellar gas, radiation fields, and Galactic magnetic field.

We adopt the 80 GDE models of Ref.~\cite{Cholis:2021rpp}, which explore variations in the conditions of the inner Galaxy using local cosmic-ray measurements, including those from AMS-02, as a starting point. 
The gamma-ray templates were generated using \texttt{GALPROP} v54 through its WebRun service~\cite{Strong:1998fr, Strong:2007nh, Vladimirov:2010aq}. 
We use the parameter sets and precomputed gamma-ray sky maps available at \url{https://zenodo.org/records/6423495#.Yk92zNPMKu4} to investigate the dependence of the inferred GCE spectrum on the GDE model.

The GDE has three dominant components: gamma rays from $\pi^0\rightarrow\gamma\gamma$ following collisions of cosmic-ray nuclei with interstellar gas~\cite{Kamae:2006bf}; bremsstrahlung from cosmic-ray electrons and positrons interacting with interstellar gas~\cite{Koch:1959zz, Blumenthal:1970gc}; and inverse Compton emission produced when these leptons scatter ambient photons, including starlight, infrared photons from dust, and cosmic microwave background photons~\cite{Blumenthal:1970gc, Jones:1968zza}.~\footnote{Synchrotron emission from Galactic cosmic-ray electrons occurs predominantly below the gamma-ray energies considered here. The associated energy losses are nevertheless included in the propagation calculation.}
We denote these components by $\pi^0$, bremss, and IC, respectively. 
The $\pi^0$-decay and bremsstrahlung templates have similar spatial distributions because both trace the ISM~\cite{Calore:2014xka, Cholis:2021rpp}. 
We therefore combine them into a $\pi^0+\mathrm{bremss}$ template with a common normalization, retaining the relative contributions predicted by each GDE model. 
The template-fitting procedure is described in Sec.~\ref{sec:Template fitting}.

\paragraph{Fermi bubbles}
We model the Fermi bubbles using a spatial template with uniform intensity within the adopted bubble boundaries~\cite{Su:2010qj}. 
We take the spectrum reported in Table~2 of Ref.~\cite{Fermi-LAT:2014sfa} as a reference and introduce a penalty term that constrains departures from this spectrum in the template fit. 
Applying this constraint assumes that the high-latitude spectrum is representative of the bubble emission within our region of interest. 
The penalty widths are obtained by combining the statistical and systematic uncertainties in quadrature. 

\paragraph{Isotropic emission}
We adopt a spatially uniform template for the isotropic emission, with the reference spectrum taken from the isotropic gamma-ray background (IGRB) intensities in Table~3 of Ref.~\cite{Fermi-LAT:2014ryh}. 
As with the Fermi bubbles, we introduce a penalty term that constrains departures from the reference spectrum, using the reported uncertainties to determine the penalty widths. 
The construction of these terms is described in Sec.~\ref{sec:Template fitting}.

\subsection{Point Source Masking}

We mask cataloged point sources using the Fermi-LAT 16-year source list (FL16Y), an interim release preceding the 5FGL catalog~\cite{Ballet:2026kpq}.
FL16Y contains 7220 sources detected in 16 years of LAT observations over the 50~MeV--1~TeV energy range.
It provides updated source positions and improved localization precision compared with the 14-year 4FGL-DR4 catalog~\cite{Ballet:2026kpq}.
The source list provides positions, spectral parameters, and detection significances.
Source detection is assessed using the test statistic (TS), defined as
\begin{align}
\mathrm{TS} \equiv 2\ln\left(\frac{\mathcal{L}_\theta}{\mathcal{L}_0}\right),
\end{align}
where $\mathcal{L}_\theta$ and $\mathcal{L}_0$ are the maximum likelihoods of models with and without the source being tested described by the parameter set $\theta$, respectively.
Both models include the diffuse backgrounds and the other modeled sources. 
Our source-masking radii depend on the photon energy and on the catalog detection significance, \texttt{Signif\_Avg}, expressed in Gaussian-equivalent $\sigma$ units. This quantity is not the likelihood test statistic itself: the relation $\texttt{Signif\_Avg}\simeq\sqrt{\mathrm{TS}}$ is only approximate and becomes more accurate at high significance.

To reduce contamination from cataloged point sources, we exclude pixels within an energy-dependent angular radius of each source position.
We adopt the two sets of masking radii listed in Table~3 of Ref.~\cite{Cholis:2021rpp}.
We assign the larger radius to sources with $\texttt{Signif\_Avg}$ greater than 49 and the smaller radius to the remaining point sources.~\footnote{Reference~\cite{Cholis:2021rpp} quotes the bright-source threshold as $\mathrm{TS}>49$. Using the 4FGL-DR2 catalog, we find that the cut $\texttt{Signif\_Avg} > 49$ selects 25 sources within their $60^\circ\times60^\circ$ window, matching the number reported there. We adopt this cut on the catalog significance, rather than $\texttt{Signif\_Avg}^2$.}
We use larger masks for brighter sources to reduce contamination from photons in the extended tails, or wings, of the point spread function (PSF).
This procedure reduces the sensitivity of the diffuse-emission analysis to uncertainties in point-source modeling.

Within the $60^\circ\times60^\circ$ outer region used to construct our masks, FL16Y contains 1130 catalog entries, of which 1103 are modeled as pointlike sources and 27 as extended sources \cite{Ballet:2026kpq}. Of these, 28 satisfy the bright-source criterion defined above. For comparison with the 10-year 4FGL-DR2 catalog used in Ref.~\cite{Cholis:2021rpp}, we perform nearest neighbor positional matching against the full DR2 catalog, adopting a maximum angular separation of $0.2^\circ$. This yields 838 matched and 292 unmatched FL16Y entries. Here, ``unmatched'' refers only to this positional criterion and does not necessarily imply a newly cataloged source. Separately, 111 of the 1130 FL16Y entries in this region ($9.8\%$) have no counterpart in earlier FGL catalogs recorded in the \texttt{ASSOC\_FGL} field~\cite{Ballet:2026kpq}.
  
A comparison of the source positions and detection-significance distributions in FL16Y and 4FGL-DR2 is presented in Appendix~\ref{app:catalog_comparison}.

\begin{figure}[htbp]
    \centering
    
    \subfloat[\justifying
        FRONT-converted photon counts after source and
        $|b|<2^\circ$ masking, using 4FGL-DR2 (top) and
        FL16Y (bottom).
        Columns show $0.275$--$0.357$, $1.02$--$1.32$,
        and $3.78$--$4.91$~GeV.
        Logarithmic color scales are shared between rows
        within each column.
        Percentages give the retained-pixel fraction of
        the full fitting region, including zero-count pixels.
        \label{fig:mask_comparison}
    ]{
        \includegraphics[width=0.9\linewidth]{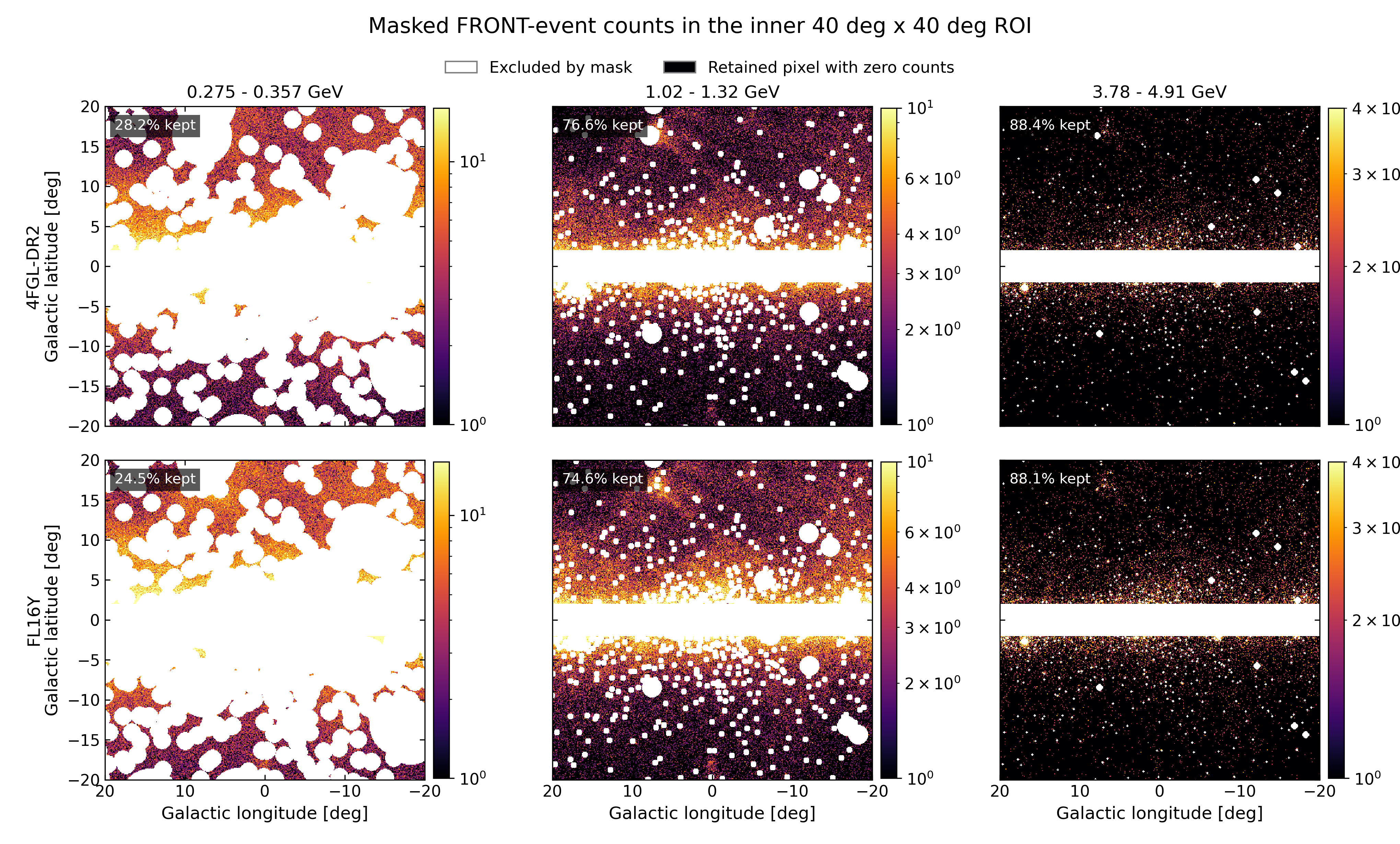}
    }
    
    \medskip 
    
    \subfloat[\justifying
        Source-mask overlap: purple indicates regions masked
        by both catalogs, while red and blue indicate regions
        masked only by DR2 and FL16Y, respectively.
        Mask radii depend on energy, with larger radii assigned
        to sources with \texttt{Signif\_Avg}$>49$.        \label{fig:mask_only_comparison}
    ]{
        \includegraphics[width=0.90\linewidth]{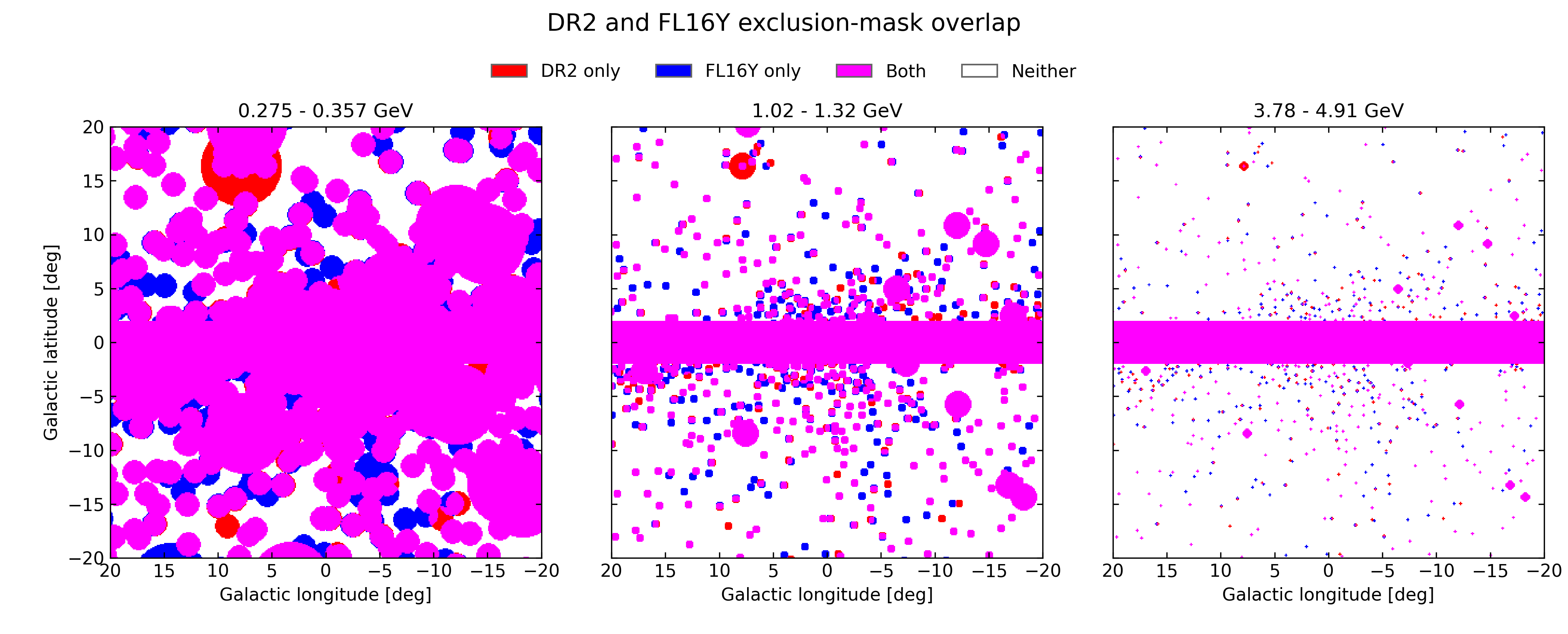}
    }
    
    \caption{\justifying
        Comparison of 4FGL-DR2 and FL16Y source masks in the
        $40^\circ\times40^\circ$ GCE fitting region.
        DR2 is the reference catalog used in Ref.~\cite{Cholis:2021rpp}, from which we adopt
        the masking prescription.
    }
    \label{fig:mask_comparison_combined}
\end{figure}

Replacing the 4FGL-DR2-based mask with the FL16Y-based mask has the largest effect at low energies.
The unmasked pixel fraction in the $40^\circ\times40^\circ$ fitting region decreases by $3.76$-$4.18$ percentage points in the three lowest-energy bins ($0.275$--$0.603$~GeV).
The difference decreases to $1$--$2$ percentage points at $1$--$2$~GeV and to $\lesssim0.1$ percentage points above $\sim10$~GeV.
Figure~\ref{fig:mask_comparison} displays the masking in three representative energy bins. Figure~\ref{fig:mask_only_comparison} shows, for the same energy bins, the pixels excluded by only one of the two masks or by both.
The larger difference at low energies reflects the broader LAT PSF and the correspondingly larger masking radii~\cite{Ballet:2026kpq}.
For a fixed data set, this additional masking can increase the statistical uncertainty of the low-energy GCE flux.

 \begin{table}[htbp]
    \centering
    \begin{tabular}{cccc}
      \hline
      Energy range [GeV] & Small-mask radius [deg] & Large-mask radius [deg] & FL16Y masked fraction [\%] \\
      \hline
      $0.275$--$0.357$   & $1.125$ & $3.750$ & $75.3$ \\
      $0.357$--$0.464$   & $0.975$ & $3.250$ & $66.6$ \\
      $0.464$--$0.603$   & $0.788$ & $2.630$ & $54.3$ \\
      $0.603$--$0.784$   & $0.600$ & $2.000$ & $39.6$ \\
      $0.784$--$1.020$   & $0.450$ & $1.500$ & $28.0$ \\
      $1.020$--$1.320$   & $0.375$ & $1.250$ & $20.3$ \\
      $1.320$--$1.720$   & $0.300$ & $1.000$ & $12.5$ \\
      $1.720$--$2.240$   & $0.225$ & $0.750$ & $10.0$ \\
      $2.240$--$2.910$   & $0.188$ & $0.625$ & $4.9$ \\
      $2.910$--$3.780$   & $0.162$ & $0.540$ & $4.7$ \\
      $3.780$--$4.910$   & $0.125$ & $0.417$ & $2.6$ \\
      $4.910$--$10.800$  & $0.100$ & $0.333$ & $2.4$ \\
      $10.800$--$23.700$ & $0.060$ & $0.200$ & $0.5$ \\
      $23.700$--$51.900$ & $0.053$ & $0.175$ & $0.5$ \\
      \hline
    \end{tabular}
    \caption{\justifying
    Circular point-source and extended source of FL16Y exclusion radii used in the
    14-bin FRONT-event GCE analysis. Small and large masks are
    assigned to sources with catalog \texttt{Signif\_Avg}$\leq49$
    and \texttt{Signif\_Avg}$>49$, respectively, following the
    implemented selection. Values are angular radii, not diameters.
    The Galactic-plane mask is applied separately. The definition of this mask size is from Ref. \cite{Cholis:2021rpp}}
    \label{tab:gce-point-source-mask-radii}
  \end{table}


\subsection{Data Preparation}

In our analysis, we consider the 17.7 years of Fermi-LAT data in the energy range of 0.275 GeV to 51.93 GeV ensuring optimal sensitivity for the GCE and minimizing contamination from instrumental backgrounds at lower energies and from reduced photon statistics at higher energies.
We follow the recommendation by the Fermi-LAT collaboration in the selected data quality criteria:
\begin{itemize}
\item We select \texttt{CLEAN}-class events (\texttt{evclass=256}) to reduce residual contamination from charged particles.
\item We retain only \texttt{FRONT}-converting events (\texttt{evtype=1}), which have better angular resolution than \texttt{BACK}-converting events.
\item We exclude events with zenith angles greater than $100^\circ$ (\texttt{zmax=100}) to reduce contamination from gamma rays produced in the Earth's atmosphere. No additional ROI-based zenith-angle cut is applied when selecting good-time intervals (\texttt{roicut=no}).
\item The production GCE event file records \texttt{TSTART=239557417} and
\texttt{TSTOP=798604698}, in mission elapsed time (MET) seconds. The first and last
good-time-interval boundaries are $239557417.49$ and $798542589.58~\mathrm{s}$,
respectively, corresponding to an elapsed span of approximately $17.7132$ Julian years.
\item We require 
\texttt{DATA\_QUAL==1}, \texttt{LAT\_CONFIG==1}, and \texttt{ABS(ROCK\_ANGLE)<52} to select good-time intervals.
\end{itemize}

In addition to the event selections above, we apply cuts on data quality and LAT operating conditions to select good-time intervals.
The ROI chosen for the raw data (not the binned maps that are used for the actual analysis) was a circular region with a radius of $30^\circ$ centered at the Galactic Center (coordinates: $l, b = 0^\circ, 0^\circ$ in Galactic coordinates). After obtaining the raw data within this region, we select the necessary region by making binned observed/expected maps. This allows us the possibility of selecting larger ROI in future research.
The events are then binned spatially in Galactic coordinates using a \texttt{CAR} projection with a pixel size of $0.1^\circ\times0.1^\circ$, and energetically from 0.275 to 51.93 GeV, with 14 energy bins. The specific definition of the energy bin is,
\begin{itemize}
    \item For the first 11 bins (0.275-4.909 GeV), $\Delta(\log_{10}(E/\text{GeV})) \approx0.114$
    \item For the last 3 bins (4.909-51.931 GeV), $\Delta(\log_{10}(E/\text{GeV})) \approx0.342$
\end{itemize}
We adopted this energy binning from Table 3 of Ref.~\cite{Cholis:2021rpp}, and the main idea of this binning is to make energy bins wider for higher photon energy, where we have relatively poor photon statistics.
We construct the data and model maps over a $60^\circ\times60^\circ$ region centered on the Galactic center first, and restrict the fitting data to the inner $40^\circ\times40^\circ$ region. This is due to the possibility of photon leakage of map because of the limited angular resolution of the LAT.
The predicted-count maps are convolved with the LAT PSF over the larger region before this restriction, allowing emission originating outside the fitting region to contribute to the predicted counts inside it.

The spectrum in each energy bin is evaluated at the geometric mean of the bin edges. We use the \texttt{P8R3\_CLEAN\_V3} instrument response functions. Because no ROI-based zenith-angle cut is applied in the selection of good-time intervals, the zenith-angle cut of $100^\circ$ is also applied in the livetime calculation (\texttt{zmax=100} in \texttt{gtltcube}).

We use the selected, masked data in the template-fitting analysis described in the next subsection.

\subsection{Template Fitting}
\label{sec:Template fitting}

We extract the GCE spectrum by fitting spatial templates to the observed $\gamma$-ray counts.
In each energy bin, the predicted counts are modeled as a linear combination of templates for the diffuse-background components and the GCE, with the instrument response taken into account.
We construct a Poisson likelihood from the observed and predicted counts in the unmasked pixels of the fitting region.

We include penalty terms that constrain the Fermi-bubble and isotropic components around their reference spectra.
The penalty widths are determined from the reported uncertainties, as described below.
These terms incorporate external constraints on the two components into the fit.

We use the affine-invariant ensemble Markov chain Monte Carlo (MCMC) sampler implemented in the \texttt{emcee} Python package~\cite{Foreman_Mackey_2013} to sample the posterior distribution of the template normalization parameters.
We obtain parameter uncertainties from the marginalized posterior distributions and report them as credible intervals.
These uncertainties are conditional on the chosen GDE model and the adopted priors and constraints.
We assess the dependence of the inferred GCE spectrum on the GDE model by repeating the fit for the 80 \texttt{GALPROP} models described above. 

In each energy bin, the expected counts in pixel $p$ are $\mu_p=\sum_k c_k\,T_{k,p}$, where $T_{k,p}$ are the exposure-weighted and PSF-convolved templates of the $\pi^0$+bremss, IC, GCE, Fermi-bubble, and isotropic components and $c_k$ are their normalizations. The fitted quantity is
\begin{align}
-2\ln\mathcal{L}=2\sum_{p}\left(\mu_p-n_p\ln\mu_p+\ln n_p!\right)+\chi^2_{\rm bub}+\chi^2_{\rm iso},
\end{align}
where $n_p$ are the observed counts and the sum runs over the unmasked pixels of the $40^\circ\times40^\circ$ region. The penalty terms are $\chi^2_X=\left[(F_X-F_X^{\rm ref})/\sigma_X\right]^2$ for $X=$ bub, iso, where $F_X$ is the mean intensity of component $X$ over the unmasked pixels implied by $c_X$, $F_X^{\rm ref}$ is its reference spectrum (Table~2 of Ref.~\cite{Fermi-LAT:2014sfa} for the bubbles and Table~3 of Ref.~\cite{Fermi-LAT:2014ryh} for the isotropic emission), and $\sigma_X$ is the upper (lower) uncertainty of the reference when $F_X$ lies above (below) $F_X^{\rm ref}$. The energy bins are fitted independently, with five free normalizations per bin. We adopt flat priors $c_k\geq0$ for all normalizations and sample the posterior with 100 walkers and 1000 steps, discarding the first 400 steps as burn-in. The walkers are initialized uniformly in positive values. We take as the best fit the post-burn-in sample with the highest posterior probability, and quote the 16th and 84th percentiles of the marginalized posterior as the 68\% credible interval.

\subsection{Systematic Uncertainties}
The GC region contains unresolved point sources and substantial uncertainties in diffuse-emission modeling, both of which can affect the inferred GCE flux. In this section, we estimate empirical systematic uncertainties by shifting the ROI along the Galactic plane and examining the extent to which mismodeled emission can produce a GCE-like signal in these control regions. We also investigate how the inferred GCE flux varies across the 80 GDE models. While the empirical covariance is included in our DM interpretation, the additional model-to-model spread is presented for comparison and is not incorporated as a separate uncertainty contribution. The fitted GCE flux remains nonzero across all 80 models, indicating that a GCE component persists across the diffuse-emission models considered, although this observation alone does not establish its statistical significance.

\paragraph{Empirical uncertainty from fits at different Galactic longitudes}
We assess empirical systematic uncertainties in the 17.7-year analysis
by repeating the template-fitting analysis in control regions centered
at different Galactic longitudes along the Galactic plane.
Imperfect modeling of the Galactic diffuse emission can leave residuals
that are partly absorbed by the translated GCE-like component, affecting
its fitted normalization and spectrum. The variation of the recovered
spectra across these regions provides an empirical estimate of the
uncertainty associated with spatial mismodeling of the diffuse
background~\cite{Calore:2014xka,Cholis:2021rpp}.
The GC signal region is analyzed separately and is not included in the
empirical covariance estimate.

Following the control-region selection of Sec.~VI of
Ref.~\cite{Cholis:2021rpp}, we use 22 regions of
$40^\circ\times40^\circ$, centered at Galactic latitude $b=0^\circ$
and longitudes $20^\circ\leq|\ell|\leq70^\circ$ in steps of $5^\circ$.
The GC and regions centered at $|\ell|\leq15^\circ$ are excluded
to reduce contamination by the GCE itself.
For each control region, we use count and component maps centered on
that region, translate the GCE template to the region center, and
construct the FL16Y source mask using the same masking prescription
as for the GC. The diffuse background is modeled with GDE Model~X
in all control regions, rather than selecting a different GDE model
for each region.

The control-region fits use the same five template components as in
Sec.~\ref{sec:Template fitting}, with a dedicated control-region
fitting prescription. The $\pi^0$+bremsstrahlung and IC normalizations
are restricted to non-negative, as well as the bubble and isotropic
normalizations. The GCE-like normalization may take
either sign, subject to non-negative total expected counts in every
unmasked pixel. Each energy bin is fitted with sufficient numbers of MCMC walkers and steps. The reported control-region spectrum is evaluated at
the highest-posterior walker in the final ensemble state after the chains have converged.

We construct the empirical covariance from the fitted spectra
$F_{a,i}$ of the 22 control regions as
\begin{align}
\Sigma^{\rm sys}_{ij}
 &= \frac{1}{N}\sum_{a=1}^{N}
    (F_{a,i}-\bar F_i)(F_{a,j}-\bar F_j),
    \qquad N=22,\\
\bar F_i &= \frac{1}{N}\sum_{a=1}^{N}F_{a,i},
\end{align}
where $i$ and $j$ label the 14 energy bins and $F_{a,i}$ is the energy-weighted differential intensity in given energy bin and given control-region $a$, in units of GeV\,cm$^{-2}$\,s$^{-1}$\,sr$^{-1}$. Here, $\bar{F}_i$ is the average intensity over the 22 translated ROIs in the $i$th energy bin. We use the region-to-region scatter as an estimate of the systematic uncertainty in the GCE spectrum in the GC region. The control-region mean $\bar{F}_i$ is used only to construct the covariance matrix and is not subtracted from the measured GCE spectrum.
  
To characterize the dominant correlated modes of uncertainty, we perform a principal component analysis (PCA) of the empirical covariance matrix and examine the fraction of the total variance accounted for by its leading components.
The first principal component has the same sign in all 14 energy bins and accounts for $94.2\%$ of the total variance; the first three components together account for $98.95\%$. These fractions are reported as diagnostics only.
A possible regularization prescription is to retain only the leading principal components and discard smaller eigenmodes that may be affected by sampling noise, as in Ref.~\cite{Cholis:2021rpp}.
In our case, retaining the first three components would preserve approximately $99\%$ of the total variance. However, this high retained variance fraction does not by itself establish that the discarded modes are noise or that truncation leaves the spectral-fit results unchanged. Therefore in this analysis, we retain the full $14\times14$ empirical covariance matrix, which has numerical rank 14, without PCA truncation. 
Writing its eigendecomposition as
\begin{align}
\Sigma^{\rm sys}
 = \sum_{k=1}^{14}\lambda_k\,\boldsymbol{v}_k\boldsymbol{v}_k^{\mathsf T},
\end{align}
we retain all 14 eigenmodes rather than truncating the sum to the
three largest eigenvalues.
For the full $\chi^2$ definition, we use
\begin{align}
C_{ij} &= \Sigma^{\rm sys}_{ij}
          +\delta_{ij}\sigma_{{\rm stat},i}^{2},\\
\chi^2(\boldsymbol{\theta})
 &= [\boldsymbol{F}^{\rm GC}-\boldsymbol{F}^{\rm DM}(\boldsymbol{\theta})]^{\mathsf T}
    C^{-1}
    [\boldsymbol{F}^{\rm GC}-\boldsymbol{F}^{\rm DM}(\boldsymbol{\theta})],
\end{align}
where $\sigma_{{\rm stat},i}$ is the sample standard deviation of
the post-burn-in MCMC GCE-normalization samples in the main ROI, mapped to intensity. The statistical contribution is diagonal because the energy bins are fitted independently. For comparison with the measured energy-weighted intensities, we use the finite-bin prediction
\begin{equation}
F_i^{\rm DM}
=
\frac{E_i^2}{\Delta E_i}
\int_{E_{i,\min}}^{E_{i,\max}}
\mathrm{d}E_\gamma\,I_\gamma^{\rm DM}(E_\gamma),
\end{equation}
where
$I_\gamma^{\rm DM}(E_\gamma)
=(\Delta\Omega)^{-1}
\mathrm{d}\Phi_\gamma^{\rm DM}/\mathrm{d}E_\gamma$
is the ROI-averaged differential photon intensity,
$\mathrm{d}\Phi_\gamma^{\rm DM}/\mathrm{d}E_\gamma$
is the predicted differential photon flux integrated over the ROI,
and $\Delta\Omega$ is the ROI solid angle.
Here, $E_i=\sqrt{E_{i,\min}E_{i,\max}}$ is the geometric bin center
and $\Delta E_i=E_{i,\max}-E_{i,\min}$ is the bin width.

\paragraph{Uncertainty from cosmic-ray propagation modeling}

The diffusion coefficient and its rigidity dependence, the propagation halo height, convection, and reacceleration affect the spatial distributions and energy spectra of cosmic rays~\cite{Strong:2007nh}.
These effects, together with uncertainties in cosmic-ray sources and the interstellar environment, modify the predicted Galactic diffuse emission and consequently the flux attributed to the GCE~\cite{Calore:2014xka, Cholis:2021rpp}.

\begin{figure}[htbp]
  \centering
  \includegraphics[width=0.8\textwidth]{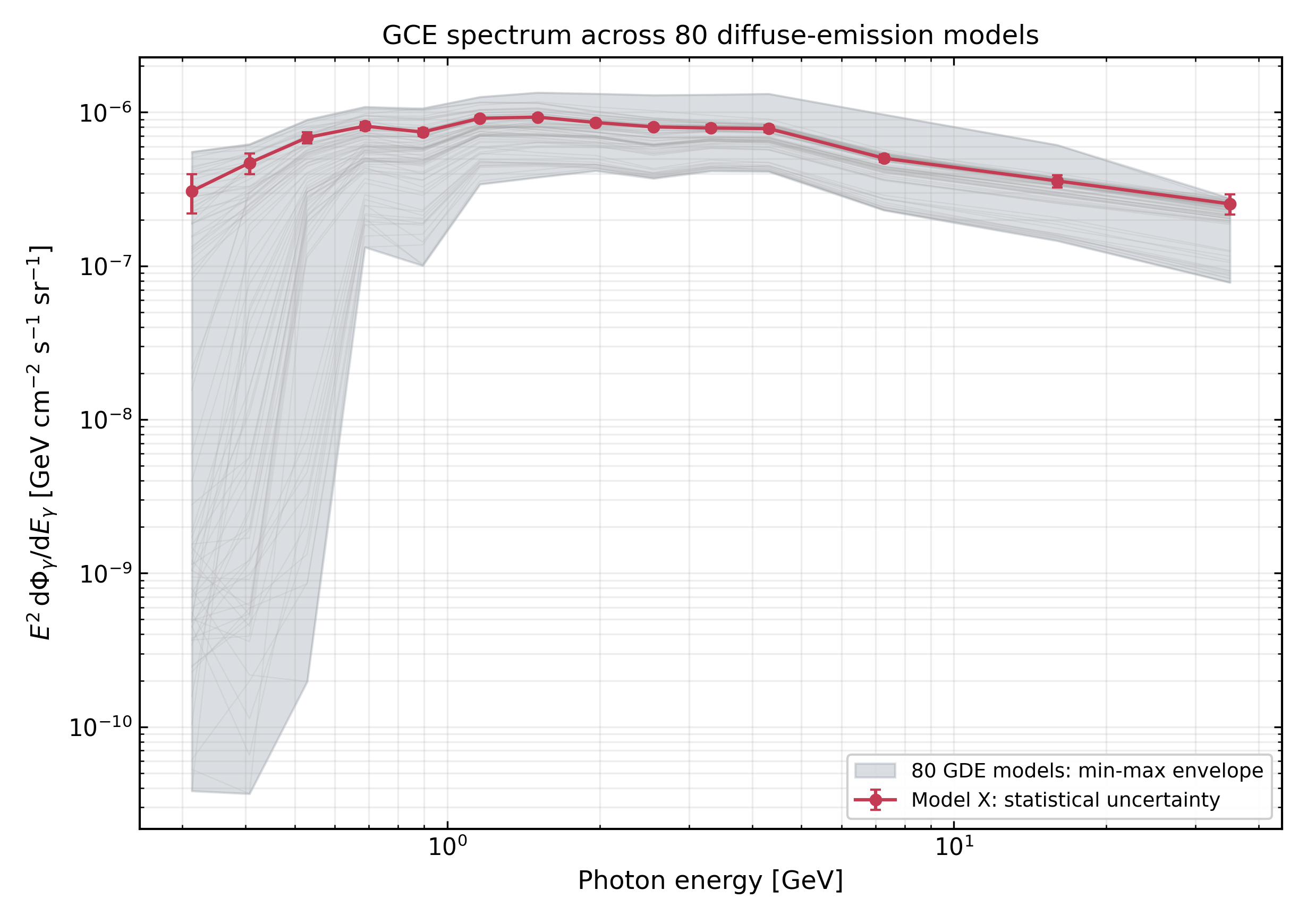}
  \caption{\justifying
GCE spectra from 17.7 years of Fermi-LAT data using
80 GDE models (gray curves).
The light blue band shows their per-bin minimum--maximum
envelope and has no assigned confidence level.
Red points show the result for the preferred Model~X
of Ref.~\cite{Cholis:2021rpp}, with statistical
68\% credible intervals.
The model spread is not added to the covariance used
in the DM fits.}
  \label{fig:gce_envelope80}
\end{figure}

We investigate this dependence by repeating the Galactic-center fit with the 80 \texttt{GALPROP} models described above~\cite{Cholis:2021rpp}.
We characterize this model dependence by the minimum--maximum envelope of the fitted GCE spectra across the 80 models; this spread is not added as a separate covariance contribution in the DM fits.
Figure~\ref{fig:gce_envelope80} shows the resulting spectra and their spread across the full model set.
The model dependence is particularly pronounced below 
$\sim 1$~GeV. Above $1.02$~GeV the envelope spans a factor of $3.2$--$4.2$ in each bin, whereas below $0.6$~GeV some models yield GCE intensities more than three orders of magnitude below that of Model~X.
Among the 80 GDE models considered, Model X provides the best fit to the 17.7-year FRONT-event data using the FL16Y catalog. The next five preferred models are LIII, XV, XLI, VII, and XLVII, with increases in the constrained fit statistic, $\Delta(-2\ln\mathcal{L})$, of 72.1, 149.7, 487.9, 517.5, and 638.2, respectively, relative to Model X.
Because the GDE models are not nested, their differences in $\Delta(-2\ln L)$ cannot be directly interpreted as squared Gaussian significances

\subsection{Dark Matter interpretation}

We now examine whether the inferred GCE spectrum can be explained by DM annihilation.
For self-conjugate DM, the prompt differential gamma-ray flux integrated over a region of solid angle $\Delta\Omega$ is
\begin{align}
\frac{d\Phi_\gamma}{dE_\gamma}
=
\frac{J_{\rm ann}(\Delta\Omega)}{8\pi m_\chi^2}
\sum_f \langle\sigma v\rangle_f
\frac{dN_\gamma^f}{dE_\gamma},
\label{eq:annihilation formula}
\end{align}
where $m_\chi$ is the DM mass, $\langle\sigma v\rangle_f$ is the velocity-averaged annihilation cross section into channel $f$, and $dN_\gamma^f/dE_\gamma$ is the photon yield per annihilation into that channel.
For the non-self-conjugate DM, extra 1/2 factor should be multiplied.
For the individual channels considered below, only the corresponding term is retained.
The astrophysical dependence is contained in the $J$-factor,
\begin{align}
J_{\rm ann}(\Delta\Omega)
\equiv
\int_{\Delta\Omega} d\Omega
\int_{\rm l.o.s.} ds\,\rho^2\left[R(s,\theta)\right],
\label{eq:J-factor}
\end{align}
where $\rho$ is the DM mass density and $s$ is the distance along the line of sight from the Sun.
The $J$-factor depends on the adopted density profile and its normalization, which can be constrained by dynamical observations.

For a spherically symmetric density profile, the Galactocentric distance along the line of sight $s$ is
\begin{align}
R(s,\theta)
=
\sqrt{r_0^2+s^2-2r_0s\cos\theta}\,,
\qquad
\cos\theta=\cos l\cos b\,,
\end{align}
where we adopt $r_0=8.5\,\mathrm{kpc}$ for the Sun--GC distance, and $l$ and $b$ are Galactic longitude and latitude.
We adopt the generalized NFW profile used for the GCE template~\cite{Navarro1996, Cholis:2021rpp}, $\rho(r)=\rho_s\,(r/r_s)^{-\gamma}\,(1+r/r_s)^{\gamma-3}$, with $\gamma=1.2$, $r_s=20$~kpc, and $\rho_s=0.271~\mathrm{GeV\,cm^{-3}}$, which gives a local density $\rho(r_0)=0.4~\mathrm{GeV\,cm^{-3}}$. For the fitting region, the $40^\circ\times40^\circ$ square with $|b|>2^\circ$ ($\Delta\Omega=0.429$~sr), this yields $J_{\rm ann}(\Delta\Omega)=8.86\times10^{22}~\mathrm{GeV^2\,cm^{-5}}$.

\begin{figure}[htbp]
  \centering
  \includegraphics[width=0.85\textwidth]{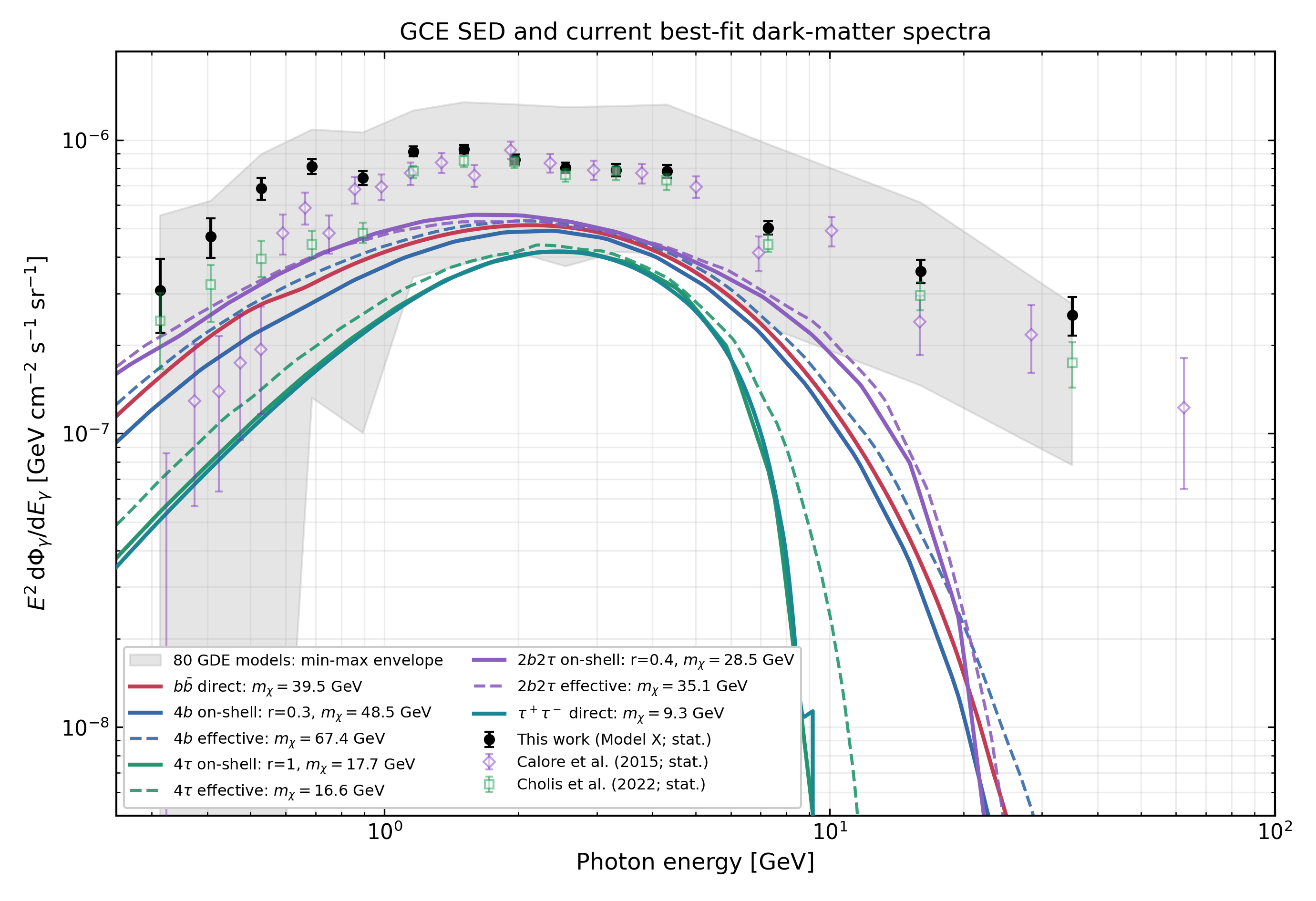}
  \caption{\justifying
GCE spectrum from 17.7 years of Fermi-LAT data
(Model~X, black points), compared with
Refs.~\cite{Calore:2014xka} (purple diamonds)
and~\cite{Cholis:2021rpp} (green squares).
Error bars are statistical.
The gray band shows the minimum--maximum envelope of
80 GDE models, with no assigned confidence level.
Best-fit DM curves show direct $b\bar b$ and $\tau^+\tau^-$
annihilation (\texttt{CosmiXs}) and the $4b$, $4\tau$,
and $2b2\tau$ channels.
Solid and dashed four-body curves denote on-shell cascades
and effective annihilation, respectively
(Sec.~\ref{sec:mediator}).
Fits use the empirical systematic covariance from 22 control
regions plus the diagonal statistical covariance;
the 80-model spread is not added.
}
  \label{fig:gce_sed_vs_channels}
\end{figure}

Figure~\ref{fig:gce_sed_vs_channels} compares the GCE spectrum inferred using GDE Model~X with the best-fit DM spectra for the direct $b\bar b$ channel and the four-body state channels $4b$, $4\tau$, and $2b2\tau$.
For each four-body channel, we compare an on-shell mediator cascade with an effective four-body prescription for direct DM annihilation into the same four-body final state, evaluated at their respective best-fit parameter values.
The mediator-to-DM mass ratio is defined as $r\equiv m_{h_2}/m_\chi$, where $m_{h_2}$ is the mediator mass.
The dependence on $r$ and the effective prescription are described in Sec.~\ref{sec:mediator}.

We compute the photon spectrum of the cascade channels with \textsc{MadGraph5\_\allowbreak aMC@NLO} v3.5.12~\cite{Alwall:2014hca}, using a UFO implementation of the simplified DM model\footnote{As in this case, we are only computing prompt photon spectrum therefore detailed DM model is irrelevant, and only the kinematics matter for the case of scalar-mediator decay prescription. This must be modified in different scenario, such as off-shell mediator cascade annihilation.}, and \textsc{Pythia}~8.312~\cite{Bierlich:2022pfr} for the parton shower, hadronization, and decays. We simulate the full tree-level $2\to4$ processes $\chi\bar\chi\to b\bar b\,b\bar b$, $\tau^+\tau^-\tau^+\tau^-$, and $b\bar b\,\tau^+\tau^-$, in which the scalar mediators appear as internal lines, without generator-level kinematic cuts and with multiparton interactions switched off. Because the final state is fixed, each spectrum corresponds to mediators decaying exclusively into the indicated fermion pairs. For each simulated value of $r$, we generate 100 DM masses spaced logarithmically from $2m_f/r$, above which the mediators can decay on shell into $f\bar f$, to $1$~TeV, with $5\times10^5$ events per mass.~\footnote{We set each beam energy to $1.01\,m_\chi$, i.e.\ $\sqrt{s}=2.02\,m_\chi$, slightly above threshold. This shifts the energy scale of the spectra by about 1\%.}For the case of effective four-body annihilation, we modified the interaction vertex as contact interaction, where the $r$ dependence is not involved.

\section{Gamma-ray constraints from dwarf spheroidal galaxies}
\label{sec:dsph-gamma-analysis}

We analyze a sample of 30 dSphs with $J$-factor estimates inferred from stellar kinematics.
The sample comprises Aquarius II, Bo{\"o}tes II, Canes Venatici I, Canes Venatici II, Carina, Carina II, Coma Berenices, Draco, Draco II, Eridanus II, Fornax, Grus I, Hercules, Horologium I, Hydrus I, Leo I, Leo II, Leo IV, Leo V, Pegasus III, Pisces II, Reticulum II, Sagittarius II, Segue 1, Sextans, Tucana II, Tucana IV, Ursa Major I, Ursa Major II, and Ursa Minor.
We adopt $J$-factors integrated within an angular radius of $0.5^\circ$ around each dwarf, following the aperture convention used in Ref.~\cite{Fermi-LAT:2015att}.

Our baseline analysis uses the batch-iterative residual-source procedure described below.
Unless otherwise stated, the results include a likelihood term accounting for the stellar-kinematic constraints on the $J$-factors.
For comparison, we also derive limits with the $J$-factors fixed at their adopted central values, using the same spectral-energy-distribution (SED) likelihood scans and DM parameter grids.

\subsection{Data selection and source modeling}
\label{subsec:data-selection}

We analyzed two observation periods spanning approximately 14.3 and 17.7 years data sets. Both periods began at Fermi mission elapsed time (MET) $t_{\min}=239500800~\mathrm{s}$ and ended at $t_{\max}=691545600~\mathrm{s}$ and $798163205~\mathrm{s}$, respectively. 
For each target, we selected events within $20^\circ$ of the dwarf position with reconstructed energies between $500~\mathrm{MeV}$ and $1~\mathrm{TeV}$. We used P8R3 \texttt{SOURCE} events (\texttt{evclass=128}), requiring \texttt{DATA\_QUAL>0} and \texttt{LAT\_CONFIG==1}, and excluded events with zenith angles greater than $100^\circ$. No additional ROI-based zenith-angle cut was applied when defining the good time intervals (\texttt{roicut=no}).

The binned likelihood analysis used a $10^\circ\times10^\circ$ ROI centered on each dwarf, with an Aitoff projection in equatorial coordinates, $0.08^\circ$ pixels, and eight logarithmic energy bins per decade. The data were divided into four PSF event types, PSF0, PSF1, PSF2, and PSF3, corresponding to \texttt{evtype} = 4, 8, 16, and 32, respectively, and treated as separate likelihood components. Each component used the corresponding \texttt{P8R3\_SOURCE\_V3} instrument response functions and isotropic diffuse template. Energy dispersion was included for all model components except the isotropic diffuse emission.

Galactic diffuse emission was modeled with the standard interstellar-emission template supplied for the Pass 8 analysis. 
The main comparison shows four catalog--duration configurations: 14.3-year and 17.7-year data, each analyzed with 4FGL-DR4 and FL16Y.
All other analysis settings and the residual-source procedure were kept unchanged across catalog choices.
Extended sources were modeled using the spatial templates associated with the corresponding catalog release. Each dwarf was represented by a point source at its optical position, initially assigned a power-law spectrum proportional to $E^{-2}$, with a reference energy of $1~\mathrm{GeV}$ and a differential normalization of $10^{-13}~\mathrm{cm}^{-2}\,\mathrm{s}^{-1}\,\mathrm{MeV}^{-1}$ at that energy.

For each target and observation period, catalog-independent products, including the selected-event files, livetime cubes, counts cubes, and exposure maps, were reused across catalog choices when the event selection, binning, and instrumental response were unchanged. 
Products specific to a PSF component were reused only within that component. 
Separate source-model files and source-map collections were maintained for each catalog--duration configuration and updated when residual sources were added.

We constructed the residual-source model using an iterative procedure in which candidate sources were selected and fitted in batches. 
The dwarf source was retained in the analysis configuration to define the target for subsequent SED extraction, but its component was removed from the ROI model during the initial fit and residual-source search. 
Thus, the local source model was constructed without a fitted dwarf component. 
The dwarf was restored before the final fit and SED extraction.

Before searching for additional sources, we performed an initial optimization of the ROI model with \texttt{Fermipy}, followed by a likelihood fit. 
For the latter fit, all model parameters were fixed except the Galactic diffuse normalization and spectral-tilt parameter, the isotropic normalization, and the normalizations of selected catalog sources. 
For selecting catalog sources whose normalizations were allowed to vary, we used the squared catalog significance, $T_{\rm sel}\equiv \texttt{Signif\_Avg}^{2}$, as a proxy for the catalog likelihood-ratio TS. A catalog source normalization was allowed to vary if either $T_{\rm sel}\geq25$ and $\Delta\theta\leq5^\circ$, or $T_{\rm sel}\geq500$ and $\Delta\theta\leq7^\circ$, where $\Delta\theta$ is the angular separation from the ROI center. This selection variable is an approximation to, rather than the exact value of, the catalog likelihood-ratio TS.

At each source-finding iteration, we computed a separate TS map for each of the five spectral templates used in the 4FGL source-finding analysis, listed in Table~\ref{tab:sourcefinding-templates}. 
The trial spectral shapes were fixed during map construction, while their normalizations were fitted at each trial position. 
We selected peaks satisfying $\sqrt{\mathrm{TS}}>4$, with a minimum separation of $1^\circ$ between peaks selected from the same map. 
Candidates from all five maps were then combined into connected groups by linking candidates separated by no more than $1^\circ$. 
The candidate with the highest TS in each group was retained as its representative. For example, in the region of Aquarius II, we found in total nine additional TS peaks, the closest to the target center was 0.428$\circ$ away. And for Draco, we found four additional peaks, and the closest one was $2.073^\circ$ away. The $1^\circ$ separation criterion applies only between additional peaks, not between a peak and the target center.

Up to four representatives were added as point sources in each iteration. 
The newly added sources were fitted together in a single likelihood fit, after which their parameters were fixed. 
The search was repeated for at most five iterations and stopped earlier if no new sources were added. 
Selecting candidates from all five maps before fitting them together was intended to reduce dependence on the order in which sources were added.

\begin{table}[t]
\centering
\begin{tabular}{lccc}
\hline
Template & $\alpha$ & $\beta$ & $E_b$ [GeV] \\
\hline
Hard         & 1.7 & 0.0 & 50.0 \\
Intermediate & 2.2 & 0.0 & 1.0  \\
Soft         & 2.7 & 0.0 & 0.25 \\
Peaked       & 2.0 & 0.5 & 2.0  \\
Pulsar-like  & 2.0 & 0.3 & 1.0  \\
\hline
\end{tabular}
\caption{\justifying
Spectral templates used to construct residual-source TS maps, following the template choices in the 4FGL source-finding analysis. Each template represents a point source with a \texttt{LogParabola} spectrum,
$dN_\gamma/dE=N_0(E/E_b)^{-\alpha-\beta\ln(E/E_b)}$.
The initial differential normalization at $E_b$ was
$N_0=10^{-13}~\mathrm{cm}^{-2}\,\mathrm{s}^{-1}\,\mathrm{MeV}^{-1}$.
Templates with $\beta=0$ reduce to power laws.}
\label{tab:sourcefinding-templates}
\end{table}

After residual-source finding, the dwarf component was restored to the ROI model. 
Then we fitted its normalization and spectral index jointly with the normalizations of the added residual sources and the selected background parameters.
We allowed the Galactic diffuse normalization and spectral-tilt parameter, the isotropic normalization, the dwarf normalization and power-law index, the normalizations of all newly added residual sources, and the catalog-source normalizations selected by the criteria above to vary.~\footnote{A residual-source candidate near the dwarf need not represent a separate physical source. If the components have similar spatial and spectral distributions, their individual contributions may be poorly constrained even when their combined emission is well determined.} 
A final likelihood fit was performed before SED extraction.

\subsection{SED extraction and dark-matter likelihood construction}
\label{subsec:sed-extraction}

For each dwarf, we extracted the SED with \texttt{Fermipy} using the
restored dwarf source as the target. 
Immediately before SED
extraction, all model parameters were fixed except the Galactic
diffuse prefactor and isotropic normalization. 
With
\texttt{free\_background=True}, these two normalizations were
refitted in each energy bin, while the parameters of the catalog
and added residual sources remained fixed. 
No additional
background sources were freed through a spatial-radius selection.

The target normalization was fitted independently in each SED
bin. 
We set \texttt{bin\_index=2.0} and enabled
\texttt{use\_local\_index=True}. With the local-index option
enabled, the spectral index in each bin was fixed to the local
slope of the global target spectrum, superseding the constant
index specified by \texttt{bin\_index}. We used
\texttt{cov\_scale=3.0}, which sets the Gaussian constraint widths
on the free diffuse normalizations to three times their
uncertainties from Fermipy's broadband background fit, and is the default setting of \texttt{Fermipy}.
For the subsequent DM analysis, we retained the full per-bin
normalization and log-likelihood scans, together with the
reference differential fluxes and reference energies.

For each annihilation channel, dark-matter mass, and annihilation
cross section, we evaluated the predicted differential photon
flux at the reference energy \(E_\gamma\) of each SED bin:
\begin{align}
  \frac{d\Phi_\gamma}{dE_\gamma}(E_\gamma)
  =
  \frac{\langle\sigma v\rangle}{8\pi m_\chi^2}
  J
  \frac{dN_\gamma}{dE_\gamma}(E_\gamma;m_\chi)\,,
  \label{eq:dm-flux}
\end{align}
where \(dN_\gamma/dE_\gamma\) is the photon yield per annihilation and
\(J\) is the integral of the squared dark-matter density along
the line of sight and over the adopted angular aperture.
This expression assumes self-conjugate dark matter.
For the \(b\bar b\) channel, we used the prompt photon spectra
from CosmiXS.
For each target and each point on the mass--cross-section grid,
we evaluated the predicted differential fluxes against the SED
likelihood scans. 
In each energy bin, the flux coordinate of
the scan was obtained by multiplying the trial normalization
by the reference differential flux. 
Denoting the predicted
differential flux in bin \(k\) by \(F_k^{\rm DM}\), we constructed
the target gamma-ray log likelihood as
\begin{align}
  \log\mathcal L_\gamma
  (m_\chi,\langle\sigma v\rangle,J)
  =
  \sum_k
  \log\mathcal L_k\!\left(F_k^{\rm DM}\right)\,,
  \label{eq:dm-bin-likelihood}
\end{align}
where \(\mathcal L_k\) is the likelihood in bin \(k\), evaluated
from its tabulated log-likelihood scan.
We used linear interpolation between scan points. 
Outside the
scanned flux range, the implementation assigned the log-likelihood
value at the nearest endpoint.
The DM limits were derived from these likelihood scans rather
than from the best-fit SED flux points.

We introduced \(J\)-factor prior by their measured values and uncertainty from Ref. \cite{McDaniel:2023bju}. For the prior, we wrote \(J=sJ_0\), where \(s\) is
an energy-independent normalization parameter for each target.
Varying \(s\) rescales the entire predicted DM spectrum of that
target without changing its spectral shape.
We added the Gaussian penalty
\begin{align}
  \log\mathcal L_J
  =
  -\frac{(\log_{10}s)^2}{2\sigma_J^2}
  \label{eq:j-factor-penalty}
\end{align}
to the gamma-ray log likelihood summed over the SED bins and
profiled the resulting likelihood over
\(\log_{10}s\in[-3,3]\). 
Here, \(\sigma_J\) is the quoted
uncertainty in \(\log_{10}J\). Our dSphs limits are then obtained after profiling over this $J$ prior. As this prior is determined from stellar-kinematics, our DM limit is also constrained by such kind of observation.

At each mass, we identified the maximum of the stored likelihood
profile, \(\log\mathcal L_{\max}(m_\chi)\). The approximate
one-sided 95\% confidence-level upper limit was determined on
the upper-cross-section side of this maximum using
\begin{align}
  2\left[
    \log\mathcal L_{\max}(m_\chi)
    -
    \log\mathcal L
    \left(m_\chi,\langle\sigma v\rangle_{95}\right)
  \right]
  =
  2.71\,.
  \label{eq:dm-upper-limit}
\end{align}
This corresponds to a decrease of \(2.71\) in
\(\mathrm{TS}\), or \(1.355\) in the log likelihood,
and is independent of the absolute value of the likelihood, but only relative difference. For the two-body state, numerical grid contained 40 logarithmically spaced masses
from \(1\) GeV to \(10\) TeV and 60 logarithmically spaced cross
sections from \(10^{-28}\) to
\(10^{-22}\,\mathrm{cm^3\,s^{-1}}\).

For nonrelativistic DM annihilation into $b\bar b$, the partonic threshold is
$m_\chi\geq m_b$, since the available center-of-mass energy is approximately $2m_\chi$.
The calculated limits are additionally restricted to the supported domain of the
adopted photon-spectrum tables and mass grid.

In addition to the \(b\bar b\) channel modeled with CosmiXS,
we derived annihilation limits using a custom effective
four-body spectrum, denoted effective-\(4b,4\tau,2b2tau\).
We evaluated the existing SED likelihood scans with the predicted
fluxes for this alternative prompt spectrum.

\subsection{Combined limits from dSphs}
\label{subsec:dSphsBounds}

\begin{figure}[htbp]
  \centering
    \includegraphics[width=0.8\textwidth]{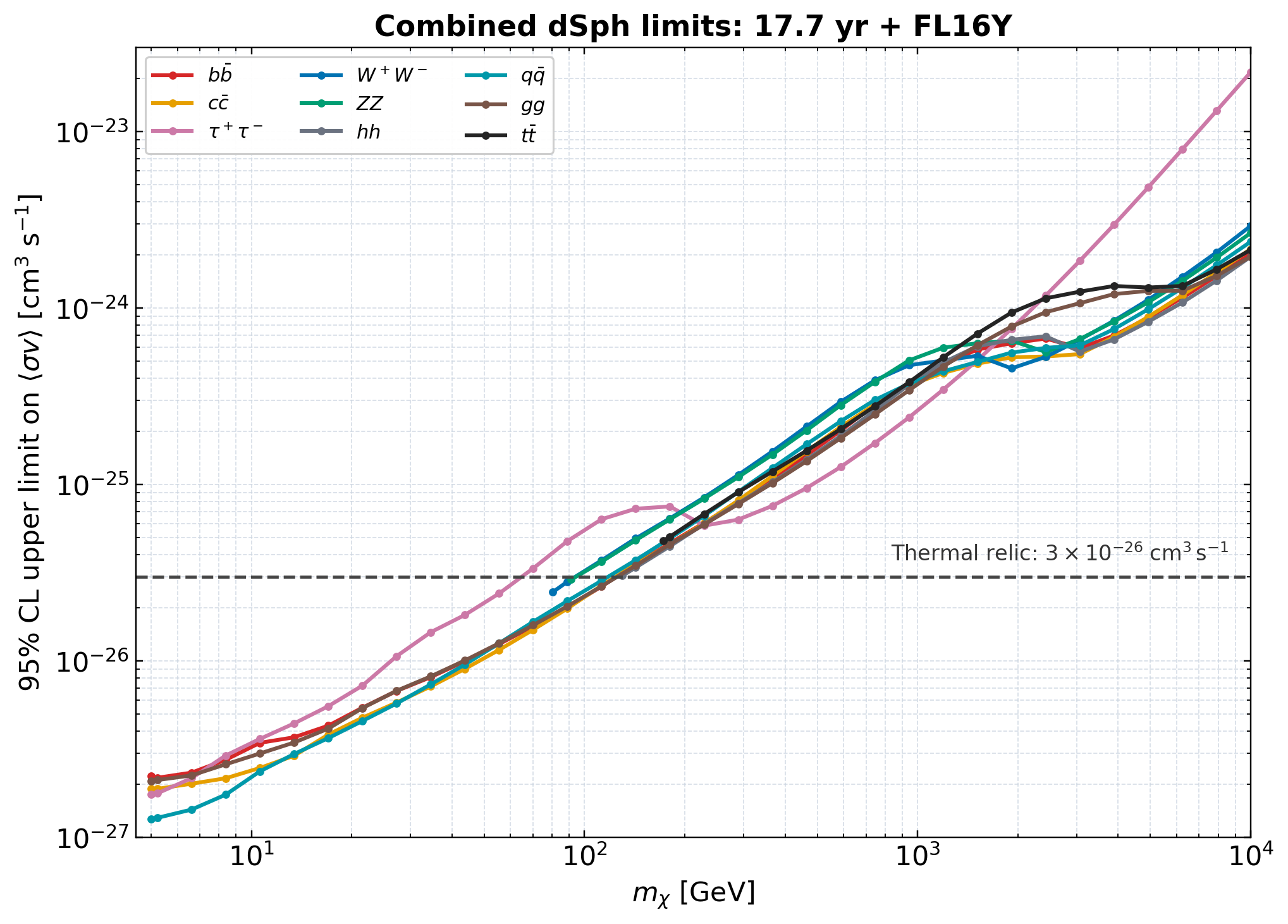}
  \caption{\justifying
  The dependence of dSphs limit on the two-body annihilation channel. Each lines represent the 95$\%$ CL upper limit from 30 measured dSphs joint likelihood with $J$-factor prior, with each different annihilation channel. $q\bar{q}$ represents light quarks, $u,\,d,\,s$.}
  \label{fig:dSphs-twobody-all}
\end{figure}

We derive combined limits from the joint likelihood of the 30 dSphs with stellar-kinematic $J$-factor measurements, profiling over the individual $J$ factors with their
corresponding uncertainties.
We assume self-conjugate DM that constitutes the entire DM abundance and consider each annihilation channel separately with a branching fraction of unity.
Figure~\ref{fig:dSphs-twobody-all} shows the combined 95\% CL
upper limits on $\langle\sigma v\rangle$ for the two-body states $b\bar b$, $c\bar c$, $\tau^+\tau^-$,
$W^+W^-$, $ZZ$, $hh$, $q\bar q$, $gg$, and $t\bar t$,
using 17.7 years of Fermi-LAT data and the FL16Y source catalog.
Here, $q$ denotes the light quarks ($u,d,s$), and $h$ is
the Standard Model Higgs boson. The prompt photon spectra are taken from \texttt{CosmiXs}~\cite{Arina:2023eic}.

For comparison, we adopt $\langle\sigma v\rangle =
3\times10^{-26}\,\mathrm{cm^3\,s^{-1}}$ as a conventional reference value for thermal relic DM with $s$-wave dominated annihilation.
The limits for the $b\bar b$, $c\bar c$, $q\bar q$, and $gg$ channels fall below this reference value up to
masses of approximately $100$--$150$~GeV, depending on the channel.
For $\tau^+\tau^-$, the corresponding mass reach is approximately $70$~GeV.
The relative strength of the limits depends on the DM mass: the $\tau^+\tau^-$ limits are generally weaker than those for the quark and gluon channels at $m_\chi\lesssim150$~GeV and $m_\chi\gtrsim2$~TeV, but are among the strongest at intermediate masses.

\begin{figure}[htbp]
\centering
\includegraphics[width=0.78\linewidth]{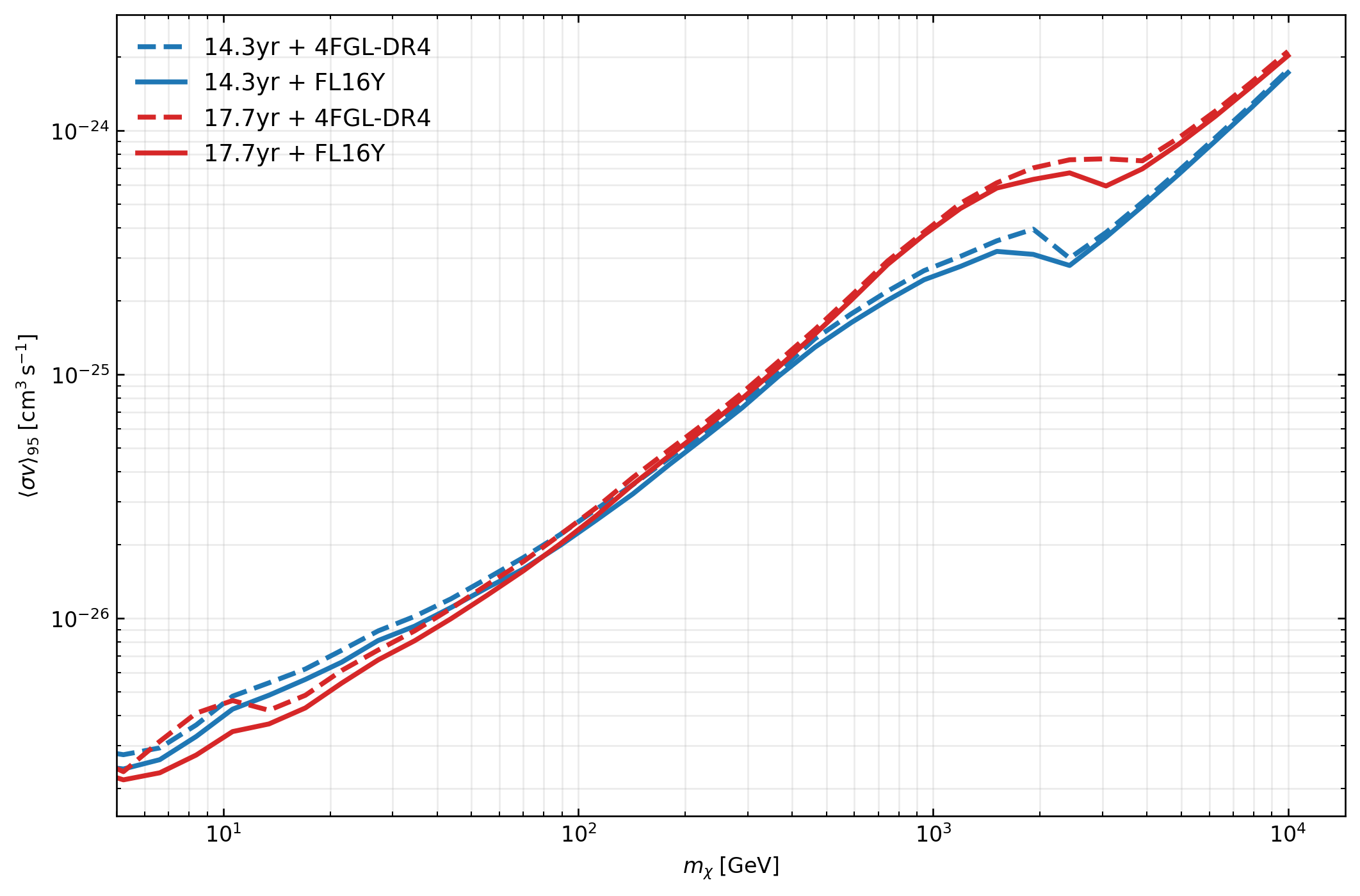}
\caption{\justifying
Combined 95\% CL upper limits on $\langle\sigma v\rangle$
for $b\bar b$ annihilation from 30 dSphs, including $J$-factor uncertainties. Blue and red curves show the 14.3-year and 17.7-year results; solid and dashed curves use FL16Y and 4FGL-DR4, respectively.
All configurations use the same residual-source search
and likelihood procedures.
}
\label{fig:matrix_comparison_bb}
\end{figure}

To examine the dependence of the combined constraints on the
observation interval and source catalog, we compare the four combinations of two observation intervals, 14.3 and 17.7 years,
and two source catalogs, 4FGL-DR4 and FL16Y.
The resulting combined limits for $\chi\chi\to b\bar b$
are shown in Fig.~\ref{fig:matrix_comparison_bb}.

For the FL16Y source model, extending the observation interval
from 14.3 years (blue solid curve) to 17.7 years (red solid curve) strengthens the limits at low DM masses but weakens them at high masses.
The weakening is particularly apparent at $m_\chi\gtrsim500$~GeV.
A similar trend is found for 4FGL-DR4, although the longer observation interval also gives weaker limits over a limited low-mass range, approximately $7$--$15$~GeV.

For the 14.3-year analysis, FL16Y generally gives the stronger limit than 4FGL-DR4.
For the 17.7-year analysis, FL16Y gives the stronger
limits at low masses, whereas its limits are weaker than that obtained with 4FGL-DR4 at high masses.~\footnote{Although not shown here, the FL16Y results show
behavior similar to that of the 4FGL-DR3 results presented
in Appendix~\ref{app:validation}. As well as the 4FGL-DR3 comparision, we checked the validation of our analysis with the previous results in Ref. \cite{McDaniel:2023bju}.}
Extending the observation interval therefore does not uniformly strengthen the observed limits; the effect depends on the DM mass and source catalog.

The high-mass weakening occurs for all three catalogs and is therefore not specific to the FL16Y source model.
One possible explanation is that the additional data produce positive residuals relative to the fitted backgrounds in energy bins contributing to the high-mass $b\bar b$ fits, allowing a larger DM contribution despite the increased exposure.~\footnote{Between 14.3 yr and 17.7 yr duration, we additionally identified extra photon signal at high energy, that might possibly alter the result, especially for high $m_\chi$ region. See Appendix~\ref{app:dSphs_weak} for more details.}
Statistical fluctuations, background mismodeling, and
source variability may contribute to these residuals, although their individual effects have not been isolated.
Determining the origin of the weakening requires examining the likelihood contributions from individual dwarfs and energy bins.
The same likelihood and limit-setting procedure is used in all six configurations.

As a validation check, we compare the combined $b \bar b$ limits from our 14.3-year analysis using 4FGL-DR3 with those derived from the published likelihood products of
Ref.~\cite{McDaniel:2023bju}.
Details of this comparison are given in Appendix~\ref{app:validation}.

\section{Constraints from AMS-02 antiproton measurements}
\label{sec:AMS-02}

We constrain the DM annihilation cross section by comparing the
AMS-02 antiproton flux measurements~\cite{AMS:2016oqu} with the predicted secondary antiproton flux and a possible primary contribution from DM annihilation.
Secondary antiprotons are produced mainly through collisions of cosmic-ray protons and helium nuclei with interstellar gas.
Our calculation follows the propagation framework of
Ref.~\cite{Calore:2022stf}, which uses the semi-analytical code
\texttt{USINE}~\cite{Maurin:2018rmm}.
We adopt the \texttt{BIG} propagation benchmark of
Ref.~\cite{Genolini:2019ewc}, which includes spatial diffusion
with low- and high-rigidity breaks, convection, and reacceleration.

The spatial diffusion coefficient is parameterized as
\begin{equation}
\begin{aligned}
K(R,L)
={}&
K_0(L)\,\beta^\eta
\left(\frac{R}{1\,\mathrm{GV}}\right)^{\delta(L)}
\\
&\times
\left[
1+
\left(\frac{R}{R_l(L)}\right)^{
[\delta_l(L)-\delta(L)]/s_l}
\right]^{s_l}
\left[
1+
\left(\frac{R}{R_h(L)}\right)^{
\Delta_h(L)/s_h(L)}
\right]^{-s_h(L)} ,
\end{aligned}
\label{eq:transport}
\end{equation}
where $R$ is the absolute rigidity, $\beta=v/c$, and $K_0$
sets the diffusion normalization.
The parameters $R_l$ and $R_h$ specify the break rigidities,
while $s_l$ and $s_h$ control the smoothness of the transitions.
The rigidity indices below and between the breaks are
$\delta_l$ and $\delta$, respectively, apart from the
$\beta^\eta$ dependence.
The high-rigidity index is $\delta_h=\delta-\Delta_h$.

The parameter $L$ is the half-height of the Galactic diffusive
halo and, together with the transport coefficients, determines
the escape of cosmic rays from the propagation region.
The transport parameters inferred from cosmic-ray nuclei data
are correlated with $L$.
For the parameters tabulated in Ref.~\cite{Weinrich:2020ftb},
we account for this dependence using
\begin{equation}
\lambda_i(L)
=
A_i
\left(\frac{L}{5\,\mathrm{kpc}}\right)^{B_i},
\label{eq:scaling_relation_lambda}
\end{equation}
where $A_i$ is the value of the parameter $\lambda_i$ at the
reference halo half-height of $5\,\mathrm{kpc}$, and $B_i$
describes its dependence on $L$.
We take these coefficients from the \texttt{BIG} column for
the combined fit to the AMS-02 Li/C, Be/B, and B/C flux ratios,
given on the right-hand side of Table~A.1 of
Ref.~\cite{Weinrich:2020ftb}.

Although the theoretical uncertainties in the predicted
antiproton fluxes are sizable, we include only the experimental
covariance matrix $C_{\mathrm{data}}$ in the flux comparison.
We construct this matrix using the uncertainties reported by
AMS-02~\cite{AMS:2016oqu}, separating the systematic contributions
according to their origin and estimating their correlations
following Ref.~\cite{Heisig:2020nse}.
The uncertainty in the halo half-height $L$ is included
separately through the penalty term introduced below.

The contribution from the antiproton flux measurements is
\begin{equation}
\chi^2_{\bar p}
=
(\mathbf{f}-\boldsymbol{\mu})^{T}
C_{\mathrm{data}}^{-1}
(\mathbf{f}-\boldsymbol{\mu}),
\label{eq:chi2-antiproton}
\end{equation}
where $\mathbf{f}$ contains the measured fluxes and
$\boldsymbol{\mu}$ contains the predicted fluxes in the
corresponding rigidity bins, including both the secondary
and DM contributions.

We include the external constraint on the halo half-height
as a Gaussian penalty in $\log_{10}L$.
The resulting likelihood is
\begin{equation}
-2\ln\mathcal{L}_{\bar p}
(m_\chi,\langle\sigma v\rangle,L)
=
\chi^2_{\bar p}
+
\left[
\frac{\log_{10}(L/\widehat{L})}
{\sigma_{\log_{10}L}}
\right]^2
+
\mathrm{const.},
\label{eq:likelihood-antiproton}
\end{equation}
where $\widehat{L}$ is the preferred halo half-height from
the cosmic-ray nuclei analysis and $\sigma_{\log_{10}L}$
is the corresponding uncertainty in
$\log_{10}(L/\mathrm{kpc})$.
We adopt $\widehat{L}=4.96\,\mathrm{kpc}$ and
$\sigma_{\log_{10}L}=0.197$, and profile over
$1\leq L/\mathrm{kpc}\leq12$.
The propagated fluxes are interpolated between ten halo-height
grid points, allowing continuous profiling in $L$.
For each annihilation channel, we determine the upper limit
at fixed $m_\chi$ and, where applicable, fixed mediator mass ratio.
Defining $s\equiv\langle\sigma v\rangle$, we denote $-2\ln\mathcal{L}_{\bar p}(m_\chi,s,L)$ in Eq.~\eqref{eq:likelihood-antiproton} by $Q(m_\chi,s,L)$. For each trial annihilation rate, we minimize $Q$ over given $L$, keeping the DM mass, annihilation channel, and mediator mass ratio, where applicable, fixed. We define $\Delta Q$ as the increase in this minimized $-2\ln\mathcal{L}_{\bar p}$ relative to its minimum over non-negative annihilation rates at the same fixed mass. The dependence on the annihilation channel and mediator mass ratio is suppressed in this notation.
The reported upper limit is the upper endpoint of the
$\Delta Q\leq3.84$ region:
\begin{equation}
\Delta Q(s_{95};m_\chi)=3.84,
\qquad
s_{95}\geq\widehat{s}(m_\chi),
\label{eq:antiproton-upper-limit}
\end{equation}
where $\widehat{s}(m_\chi)$ is the best-fitting non-negative
annihilation rate at that mass.
We use the nominal 95\% likelihood-ratio interval threshold
of $3.84$ for one parameter of interest and report its upper
endpoint; this differs from the commonly used one-sided
95\% upper-limit prescription with threshold $2.71$.

\section{Benchmark dark matter interpretations of the GCE}
\label{sec:DMinterp}

We examine the DM interpretation of the GCE using three sets
of annihilation benchmarks.
The first two subsections consider direct annihilation into
$b\bar b$ and $\tau^+\tau^-$ and effective annihilation into
$4b$, $4\tau$, and $2b2\tau$, respectively.
For these benchmarks, any unstable particles produced are
assumed to decay over distances negligible on the relevant
astrophysical scales.
In the third subsection, we study cascade annihilation through
on-shell mediators into the same four-body final states,
examining how the mediator masses and decay kinematics affect
the predicted spectra.
We assume Majorana DM and treat the DM mass and
velocity-averaged annihilation cross section as free parameters
for each benchmark.
For the cascade benchmarks, we also vary the mediator-to-DM
mass ratio.
We compare the GCE-preferred regions with constraints from
dSphs and AMS-02 antiproton measurements.

\subsection{Interpretation with the two-body channels}
\label{subsec:two-body}

For the two-body channels, $\chi\chi \to b\bar b$ and
$\chi\chi \to \tau^+\tau^-$, we use the prompt photon spectra
from CosmiXs.
These spectra are fitted to the GCE spectrum extracted from
17.7 years of Fermi-LAT data using our preferred GDE model, Model~X, with the covariance matrix explained in Sec.~\ref{sec:GCE}.
We then compare the GCE preferred parameter regions with the dSphs constraints. 

\begin{figure*}[h]
    \centering
    \includegraphics[width=0.48\textwidth]{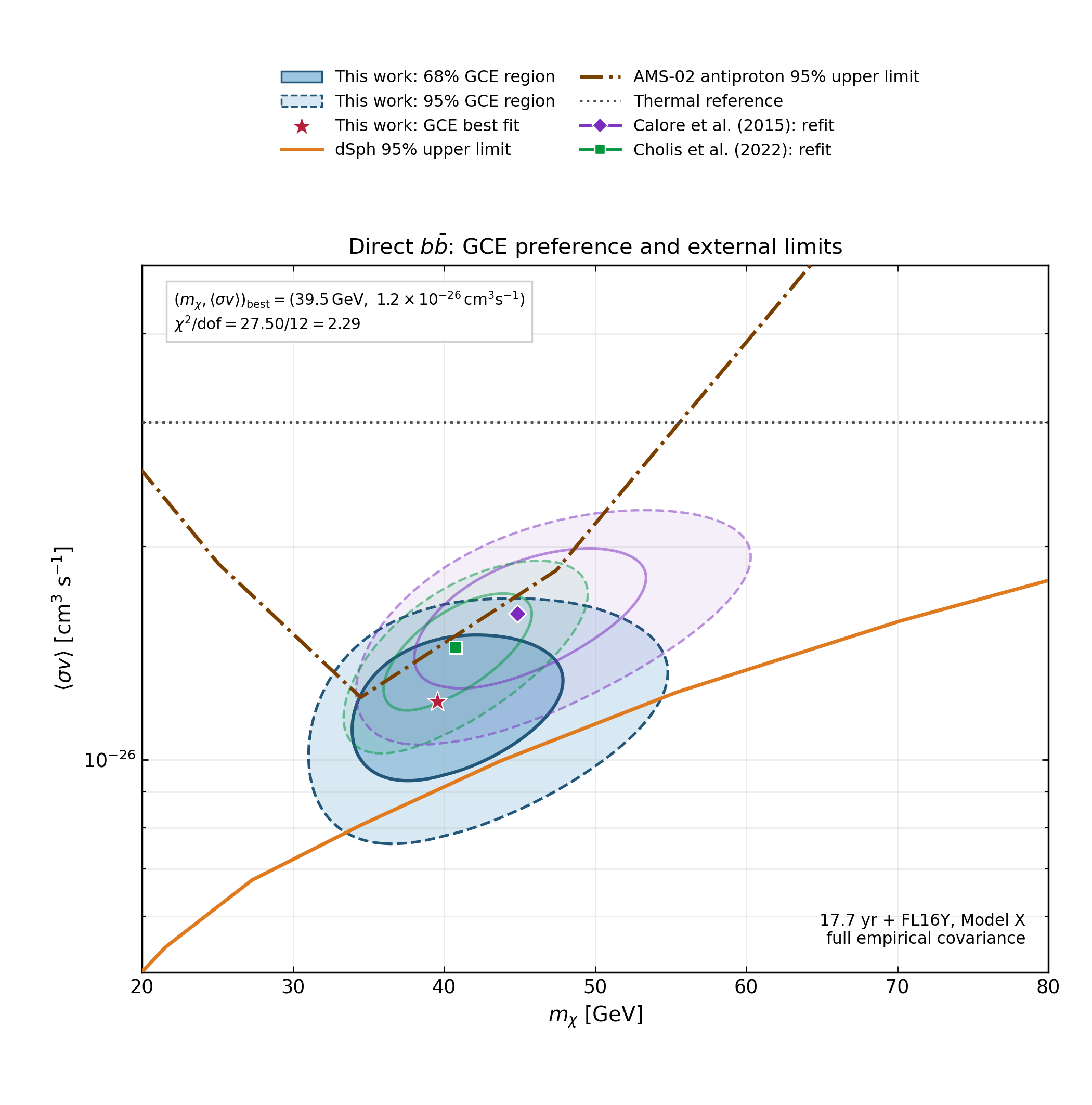}
    \hfill
    \includegraphics[width=0.48\textwidth]{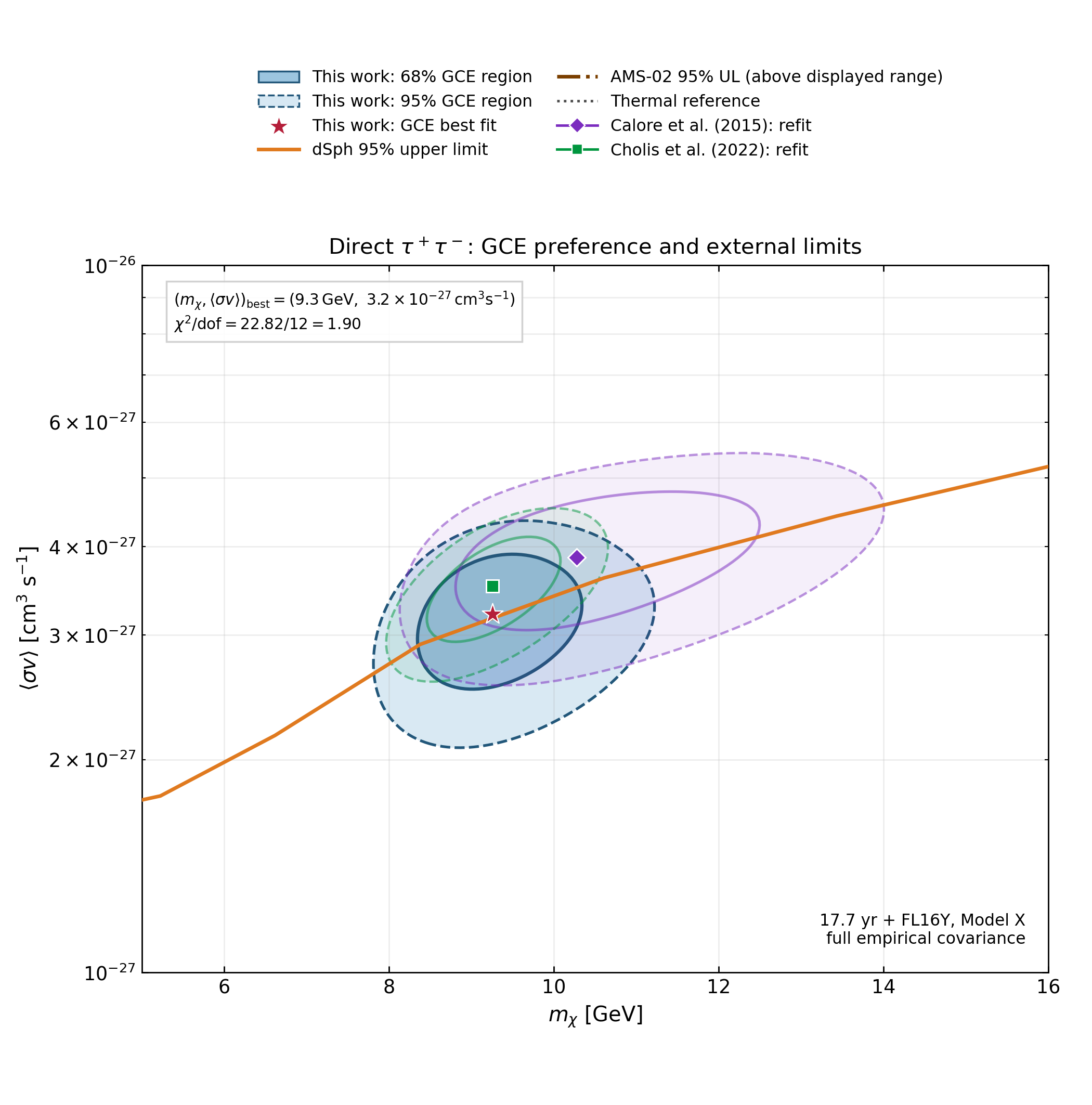}

    \caption{\justifying
    GCE-preferred regions for Majorana DM annihilation into
$b\bar b$ (left) and $\tau^+\tau^-$ (right).
Dark and light blue indicate the joint 68\% and 95\% confidence
regions from the 17.7-year Model~X GCE spectrum and adopted
covariance, with best-fit points marked by red stars.
The 95\% CL upper limits are from combined dSph observations
(orange solid; 17.7 years, FL16Y, including $J$-factor
uncertainties) and AMS-02 antiprotons (brown dot-dashed, left).
Cross sections above these curves are excluded under the
adopted assumptions; antiprotons do not constrain the displayed
$\tau^+\tau^-$ region.
The black dotted line marks the conventional thermal relic
reference, $\langle\sigma v\rangle=3\times10^{-26}
~\mathrm{cm^3\,s^{-1}}$.
Green and purple contours and markers show our refits to the
GCE spectra obtained with the preferred GDE models in
Refs.~\cite{Cholis:2021rpp} and~\cite{Calore:2014xka},
respectively.
}
\label{fig:gce_contour_dsph_pbar}
\end{figure*}

Figure~\ref{fig:gce_contour_dsph_pbar} shows the joint 68\% and 95\% GCE confidence regions in the $(m_\chi, \langle\sigma v\rangle)$ plane for the $b\bar b$ and $\tau^+\tau^-$ channels with the best fit points marked as red stars.
For comparison, we also show the previous results in Ref.~\cite{Calore:2014xka} with light purple and Ref.~\cite{Cholis:2021rpp} with light green shaded regions.
The orange solid curves show the combined dSph 95\% CL upper
limits obtained in this work.
The brown dot-dashed curve in the left panel shows the
95\% CL upper limit derived from the AMS-02 antiproton data.

For the $b\bar b$ channel, the best-fit parameters are
\begin{align}
(m_\chi,\langle\sigma v\rangle)_{\rm best}^{b\bar b}
=
\left(
39.5~\mathrm{GeV},
1.2\times10^{-26}~\mathrm{cm^3\,s^{-1}}
\right),
\end{align}
with $\chi^2_{\min}=27.50$ for 12 degrees of freedom.
The best-fit mass is somewhat lower than that reported in Refs.~\cite{Calore:2014xka,Cholis:2021rpp}, as shown in the figure.
The enhanced low-energy GCE emission discussed in Sec.~\ref{sec:GCE} may
contribute to this shift.
The preferred cross sections lie below the conventional thermal relic reference value of $3\times10^{-26}~\mathrm{cm^3\,s^{-1}}$ for the $s$-wave dominant annihilation, shown by the black dotted line.
The entire 68\% GCE-preferred region, including the best-fit
point, lies above the dSph 95\% CL upper limit.
The AMS-02 antiproton constraint is less restrictive near the best-fit mass, leaving most of this region, including the best-fit point, unexcluded.~\footnote{Our antiproton analysis differs from that of Ref.~\cite{DiMauro:2021qcf} in its propagation model, secondary-background treatment, and uncertainty prescription.
We use the \texttt{BIG} model with an external constraint on
the diffusion-halo half-height $L$, whereas that work adopts
a \texttt{QUAINT}-type setup and presents limits at fixed values of $L$. These choices affect both the propagated DM flux and the additional DM contribution allowed by the background fit, and can therefore lead to different upper
limits~\cite{Heisig:2020nse, Calore:2022stf}.}

For the $\tau^+\tau^-$ channel, the best-fit parameters are
\begin{align}
(m_\chi, \langle\sigma v\rangle)_{\rm best}^{\tau^+ \tau^-}
=
\left(
9.3~\mathrm{GeV},
3.2\times10^{-27}~\mathrm{cm^3\,s^{-1}}
\right)\,,
\end{align}
with $\chi^2_{\min}=22.82$ for 12 degrees of freedom.
This channel gives a lower minimum $\chi^2$ than the $b\bar b$ channel.
A substantial part of the 68\% GCE-preferred region remains below the dSph
95\% CL upper limit, while the best-fit value is above it by 0.67\%.
The AMS-02 antiproton analysis does not constrain the GCE-preferred region for this channel.

\begin{figure}[htbp]
  \centering    \includegraphics[width=\textwidth]{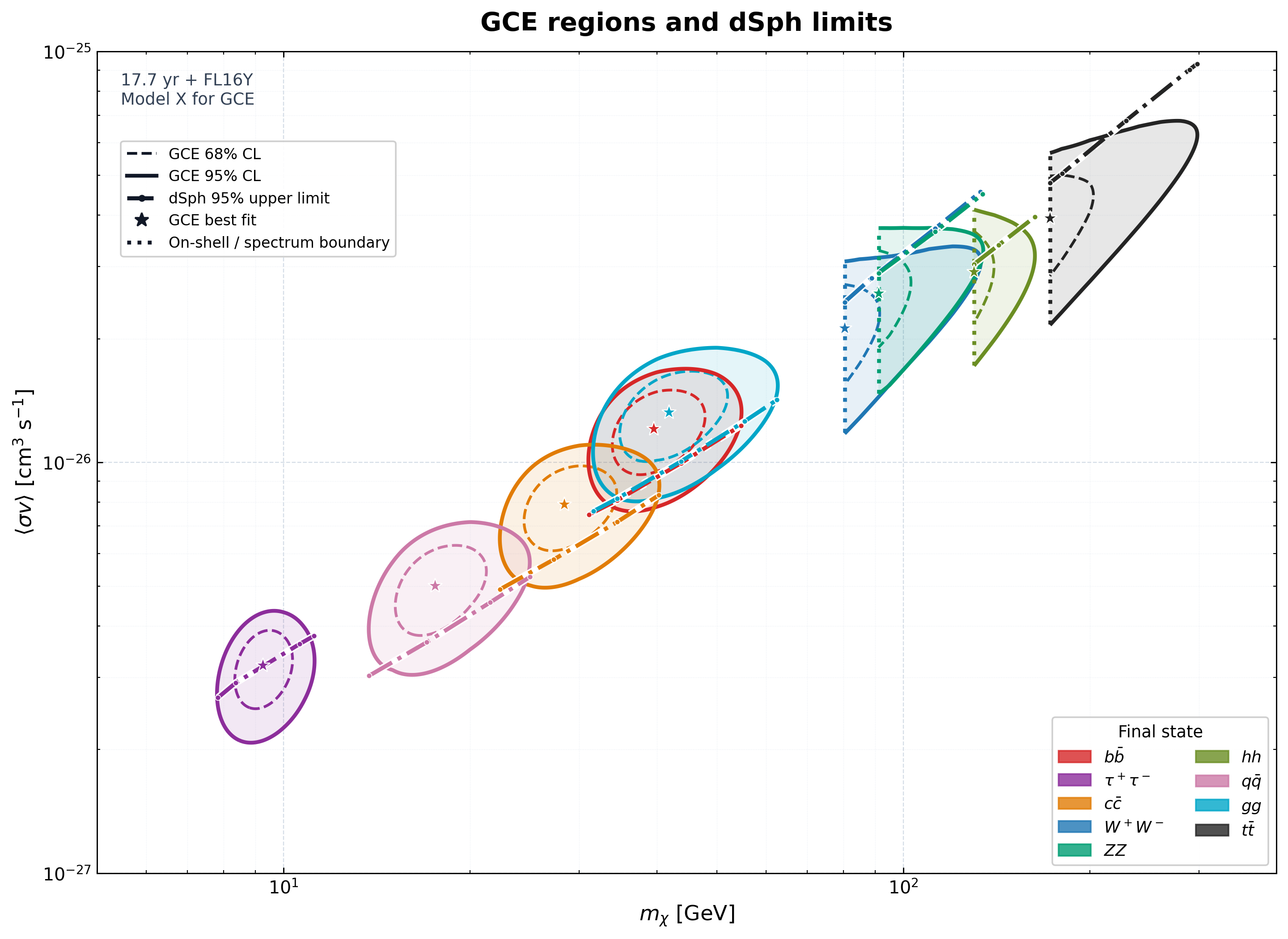}
  \caption{\justifying
  Comparison of the GCE parameter regions obtained
with Model~X and the combined dwarf-spheroidal constraints
for two-body DM annihilation channels, using CosmiXs
prompt photon spectra.
Stars mark the GCE best-fit points, and dashed and solid
contours indicate the 68\% and 95\% confidence regions,
respectively.
Curves with circular markers show the corresponding
95\% confidence-level upper limits from the 17.7-year
FL16Y analysis of 30 dwarfs with $J$-factor constraints.
For each channel, parameter points above its dwarf
upper-limit curve are excluded by that constraint.
Dotted segments indicate the lower mass boundaries
adopted in the GCE fits.
The light-quark channels $u\bar u$, $d\bar d$, and
$s\bar s$ are grouped under the label $q\bar q$;
$h$ denotes the Standard Model Higgs boson.
Each channel is considered separately, assuming
self-conjugate DM and unit branching fraction.
  \label{fig:gce-dsph-two-body}
  }
\end{figure}

Figure~\ref{fig:gce-dsph-two-body} compares the joint 68\% and 95\% CL GCE-preferred parameter regions for the two-body final states $b\bar b$, $c\bar c$, $\tau^+\tau^-$,
$W^+W^-$, $ZZ$, $hh$, $q\bar q$, $gg$, and $t\bar t$ with the corresponding combined 95\% CL upper limits from dSphs.
Similar comparisons of the GCE-preferred parameter regions
for different two-body annihilation channels were presented
in Refs.~\cite{Agrawal:2014oha, Calore:2014nla}.
We omit the AMS-02 antiproton limits, which are weaker in the relevant parameter regions under the propagation and background assumptions adopted in this work.

At $m_\chi\lesssim70$~GeV, most of the 68\% CL GCE-preferred regions for the $b\bar b$, $c\bar c$,
$q\bar q$, and $gg$ channels lie above their corresponding dSph upper limits.
For the $\tau^+\tau^-$ channel, part of the 68\% CL GCE-preferred region remains below the dSph limit.
The $W^+W^-$, $ZZ$, $hh$, and $t\bar t$ channels require larger DM masses because of their on-shell production thresholds.
Parts of their GCE-preferred regions also remain compatible with the corresponding dSph limits.

\subsection{Interpretation with the effective four-body channels}
\label{subsec:effective-4body}

\begin{figure}[htbp]
    \centering    \includegraphics[width=0.5\textwidth]{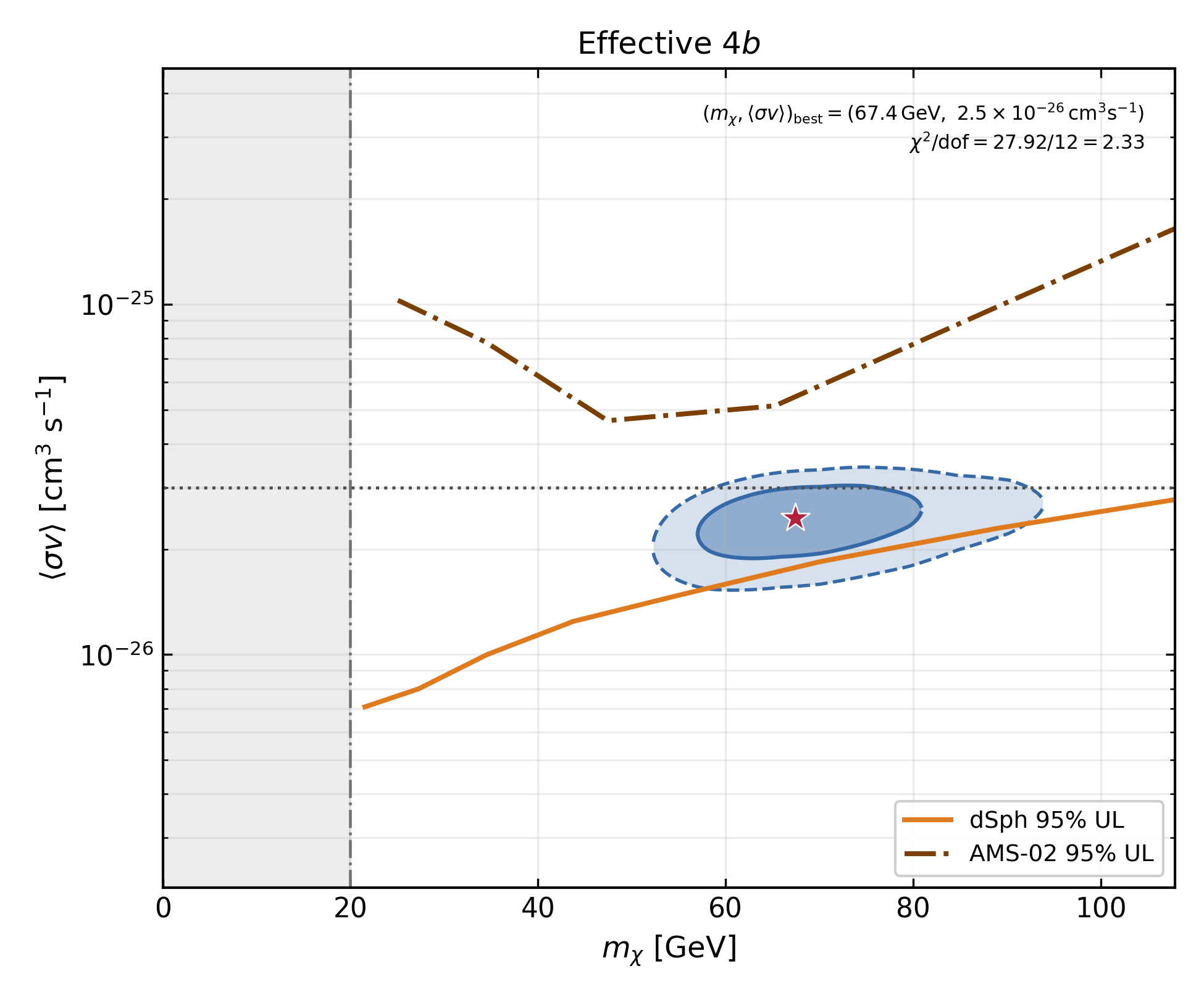}\hfill%
    \includegraphics[width=0.5\textwidth]{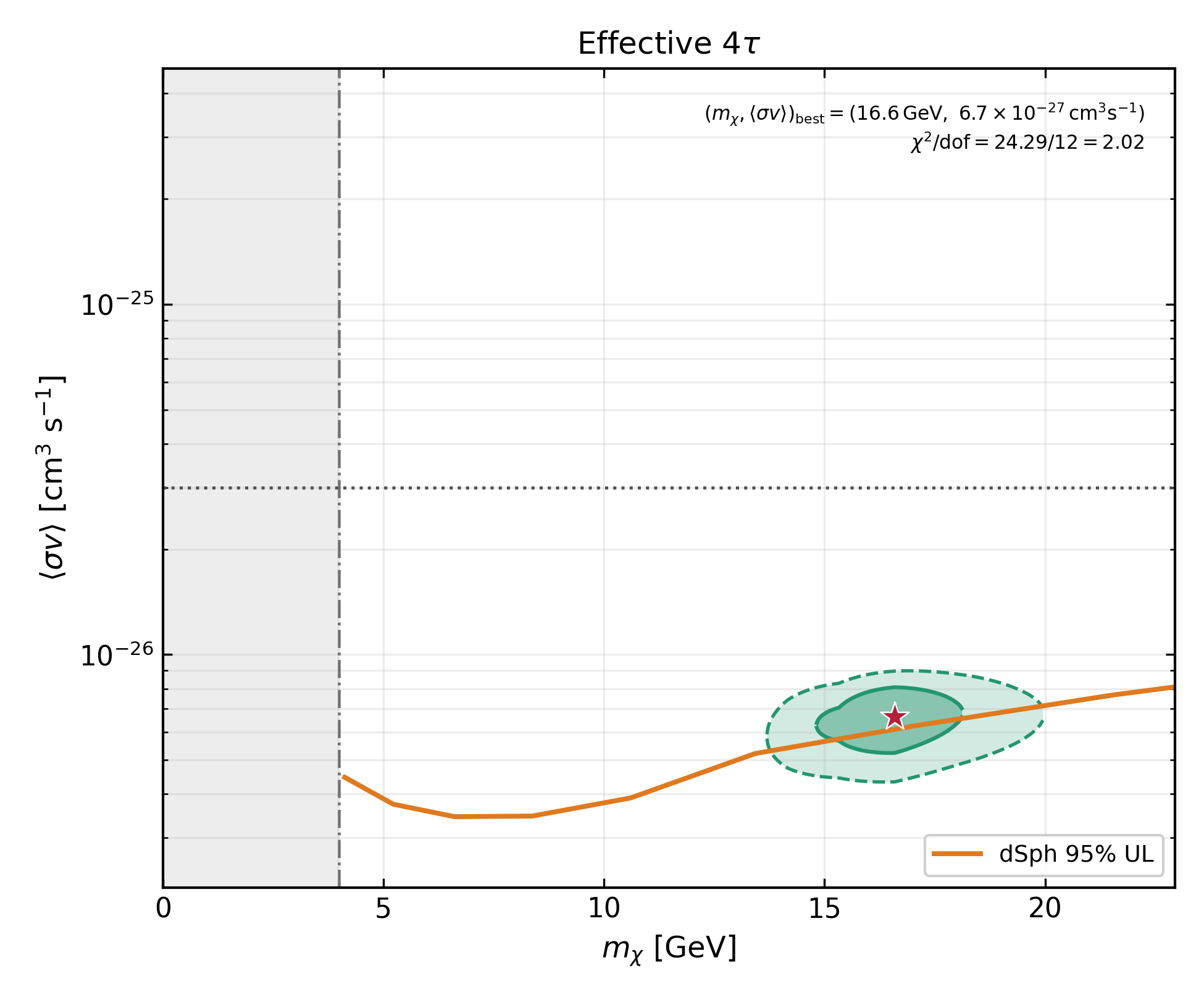}\hfill%
    \includegraphics[width=0.5\textwidth]{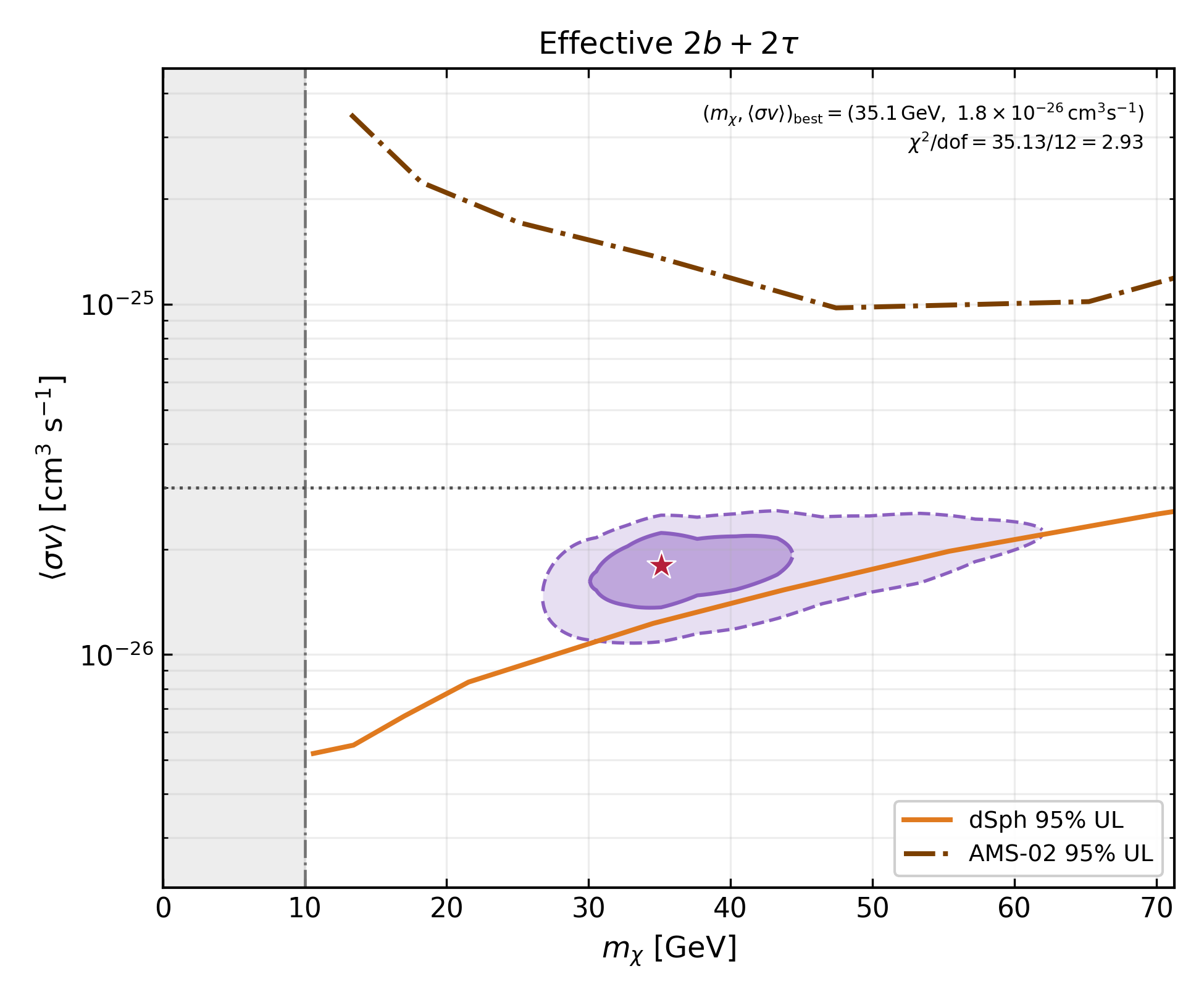}
    \\[4pt]
    \includegraphics[width=0.72\textwidth]{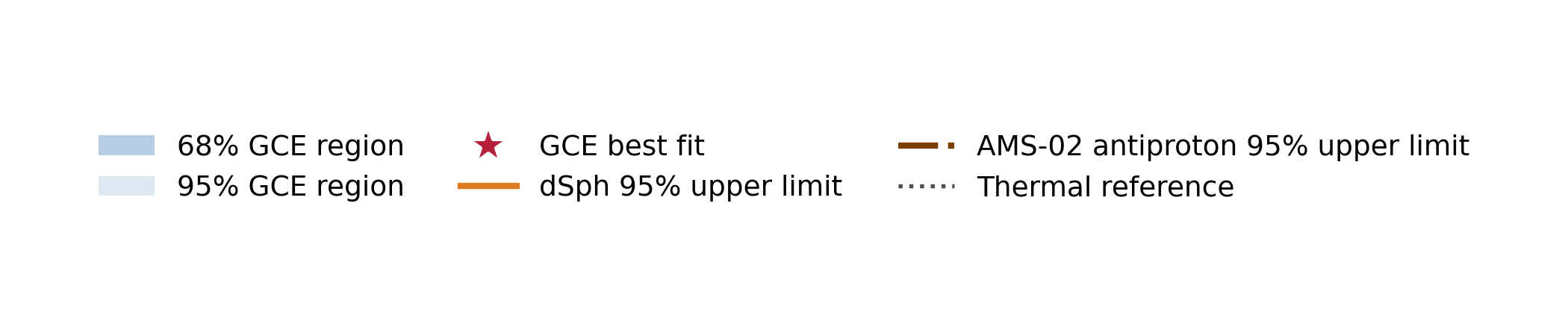}
    \caption{\justifying
    GCE-preferred regions for effective $4b$ (top left),
$4\tau$ (top right), and $2b2\tau$ (bottom) annihilation,
assuming Majorana DM.
Fits use the same Model~X spectrum and covariance as the
two-body analysis.
Dark and light shading show the joint 68\% and 95\%
confidence regions, with solid and dashed boundaries,
respectively; red stars mark the best-fit points.
Orange solid curves show the combined dSph 95\% CL upper
limits, including $J$-factor uncertainties.
Brown dot-dashed curves show the AMS-02 antiproton limits
at the same confidence level for $4b$ and $2b2\tau$.
Cross sections above these curves are excluded under the
adopted assumptions.
The black dotted line marks the conventional thermal relic
reference, $\langle\sigma v\rangle = 3 \times 10^{-26}
~\mathrm{cm^3\,s^{-1}}$.
Vertical gray lines mark the lower mass boundaries;
omitted ranges are shaded gray.
    }
    \label{fig:chi2_effective_4body}
\end{figure}

We extend the analysis to the effective four-body annihilation
channels $\chi\chi\to4b$, $4\tau$, and $2b2\tau$, using the
spectral prescription defined above without an explicit
on-shell mediator.
We generate the hard annihilation events with
\texttt{MadGraph5}~\cite{Alwall:2011uj} and use
\texttt{PYTHIA8}~\cite{Sjostrand:2014zea} for parton showering,
hadronization, and unstable-particle decays.
Figure~\ref{fig:chi2_effective_4body} shows the joint 68\% and 95\% GCE confidence regions in the $(m_\chi,\langle\sigma v\rangle)$ plane for the $4b$ (top left), $4\tau$ (top right), and $2b2\tau$ (bottom) channels.
As in Fig.~\ref{fig:gce_contour_dsph_pbar}, the orange solid
curves show the combined dSph 95\% CL upper limits.
The brown dot-dashed curves show the AMS-02 antiproton 95\% CL upper limits for the $4b$ and $2b2\tau$ channels.

The effective $4b$ and $4\tau$ channels favor larger DM masses
than the corresponding two-body channels.
This is consistent with the annihilation energy being shared
among four primary fermions.
The best-fit parameters are
\begin{align}
(m_\chi,\langle\sigma v\rangle)_{\rm best}^{4b}
&=
\left(
67.4~\mathrm{GeV},
2.5\times10^{-26}~\mathrm{cm^3\,s^{-1}}
\right), \\
(m_\chi,\langle\sigma v\rangle)_{\rm best}^{4\tau}
&=
\left(
16.6~\mathrm{GeV},
6.7\times10^{-27}~\mathrm{cm^3\,s^{-1}}
\right), \\
(m_\chi,\langle\sigma v\rangle)_{\rm best}^{2b2\tau}
&=
\left(
35.1~\mathrm{GeV},
1.8\times10^{-26}~\mathrm{cm^3\,s^{-1}}
\right).
\end{align}
The corresponding minimum $\chi^2$ values are
$27.92$, $24.29$, and $35.13$, respectively,
with 12 degrees of freedom in each case.

The AMS-02 antiproton analysis does not exclude the displayed GCE-preferred regions for these benchmarks.
The dSph constraints are more restrictive, excluding the entire 68\% GCE-preferred region for the effective $4b$ channel and the entire 68\% region for the effective
$2b2\tau$ channel.
Parts of their 95\% GCE-preferred regions nevertheless remain below the dSph upper limits.
The effective $2b2\tau$ benchmark gives a poorer GCE fit than the other two channels and is strongly constrained by the dSph observations under the adopted assumptions.
The effective $4\tau$ channel gives the lowest minimum $\chi^2$ among these three benchmarks, and part of its 68\% GCE-preferred region remains unexcluded by the dSph limits, while the best-fit value is above the limit by 8.7\%.

The effective $4b$ and $4\tau$ benchmarks give minimum $\chi^2$ values similar to, but slightly higher than, those of the corresponding two-body channels.
The effective $4b$ benchmark also remains strongly constrained by the dSph observations.
These conclusions are based on the effective four-body spectra used in this section.
Annihilation into on-shell mediators followed by their decays (cascade annihilation) can produce different photon and antiproton spectra for the same four-body final states,
depending on the mediator masses and decay kinematics.
The resulting changes in the photon and antiproton spectra can affect both the GCE-preferred regions and the complementary constraints.
In the following section, we investigate these effects using
the singlet fermionic dark matter model as a reference framework.

\subsection{Mediator effects in dark matter cascade annihilation}
\label{sec:mediator}

We investigate how the mediator mass and decay channels affect the DM interpretation of the GCE and its compatibility with dSph and antiproton constraints.
We use the singlet fermionic dark matter model as a reference framework and treat the mediator branching fractions as independent benchmark assumptions.
The spectral results can also be applied to other models with the same cascade kinematics and decay distributions.
We distinguish these kinematic effects from those that depend on the couplings and branching fractions of a specific model.
For the mixed $2b2\tau$ channel, we consider $\chi\bar{\chi}\to \phi_b \phi_\tau$, where $\phi_b$
and $\phi_\tau$ are distinct mediators with $\mathrm{Br}(\phi_b\to b\bar b)=
\mathrm{Br}(\phi_\tau\to\tau^+\tau^-)=1$; we assume equal mediator masses,
$m_{\phi_b}=m_{\phi_\tau}\equiv m_\phi$,
and use the common mass ratio $r=m_\phi/m_\chi$.

In the cascade $\chi\bar{\chi}\to\phi\phi\to
b\bar{b}b\bar{b}$, the energy distribution of the bottom quarks,
and hence the resulting photon and antiproton spectra, depends on the mediator mass $m_\phi$.
Defining $r \equiv m_\phi/m_\chi$, our $4b$ analysis gives the best GCE fit near $r\simeq0.3$, although values in
the range $0.3\lesssim r\lesssim 0.5$ give comparable fit quality, as clearly seen in Table~\ref{tab:fixed-r-highlight-all}.
This range includes the value $r \simeq 0.5$ favored in earlier studies~\cite{Dutta:2015ysa, Kim:2016csm}.
For the pure $4b$ benchmark, we assume
$\mathrm{Br}(\phi\to b\bar{b})=1$.
Additional mediator decay channels can contribute in a specific model, changing both the GCE-preferred
parameter region and the constraints from other observations, which will be discussed in the next section.

\begin{table}[tpbh]
\centering
\small
\begin{tabular}{lrrrll}
\hline
Channel & $r$ & $m_{\chi,\rm best}$ [GeV] & $\langle\sigma v\rangle_{\rm best}$ [cm$^3$ s$^{-1}$] & $\chi^2/{\rm dof}$ & dSphs \\
\hline
$4b$ & $0.10^{*}$ & 114.0 & $1.5\times10^{-26}$ & $55.34/12=4.61$ & Pass \\
$4b$ & $0.20$ & 57.0 & $1.5\times10^{-26}$ & $29.39/12=2.45$ & Fail \\
\textbf{\boldmath $4b$} & \textbf{\boldmath $0.30$} & \textbf{\boldmath 48.5} & \textbf{\boldmath $1.5\times10^{-26}$} & \textbf{\boldmath $26.26/12=2.19$} & \textbf{\boldmath Fail} \\
$4b$ & $0.40$ & 48.3 & $1.6\times10^{-26}$ & $26.71/12=2.23$ & Fail \\
$4b$ & $0.50$ & 52.5 & $1.8\times10^{-26}$ & $27.00/12=2.25$ & Fail \\
$4b$ & $0.60$ & 58.7 & $2.1\times10^{-26}$ & $27.09/12=2.26$ & Fail \\
$4b$ & $0.70$ & 64.4 & $2.3\times10^{-26}$ & $27.19/12=2.27$ & Fail \\
$4b$ & $0.80$ & 69.5 & $2.4\times10^{-26}$ & $27.50/12=2.29$ & Fail \\
$4b$ & $0.90$ & 69.7 & $2.5\times10^{-26}$ & $27.64/12=2.30$ & Fail \\
$4b$ & $1.00$ & 74.8 & $2.6\times10^{-26}$ & $27.81/12=2.32$ & Fail \\
\hline
$4\tau$ & $0.10^{*}$ & 36.8 & $5.9\times10^{-27}$ & $57.30/12=4.77$ & Pass \\
$4\tau$ & $0.20$ & 18.5 & $6.6\times10^{-27}$ & $23.42/12=1.95$ & Fail \\
$4\tau$ & $0.30$ & 15.6 & $6.7\times10^{-27}$ & $24.78/12=2.06$ & Fail \\
$4\tau$ & $0.40$ & 14.3 & $6.6\times10^{-27}$ & $25.45/12=2.12$ & Fail \\
$4\tau$ & $0.50$ & 13.6 & $6.4\times10^{-27}$ & $25.94/12=2.16$ & Fail \\
$4\tau$ & $0.60$ & 14.3 & $6.5\times10^{-27}$ & $25.73/12=2.14$ & Fail \\
$4\tau$ & $0.70$ & 15.0 & $6.5\times10^{-27}$ & $24.87/12=2.07$ & Fail \\
$4\tau$ & $0.80$ & 16.4 & $6.6\times10^{-27}$ & $24.28/12=2.02$ & Fail \\
$4\tau$ & $0.90$ & 17.2 & $6.5\times10^{-27}$ & $23.56/12=1.96$ & Fail \\
\textbf{\boldmath $4\tau$} & \textbf{\boldmath $1.00$} & \textbf{\boldmath 17.7} & \textbf{\boldmath $6.3\times10^{-27}$} & \textbf{\boldmath $22.69/12=1.89$} & \textbf{\boldmath Fail} \\
\hline
$2b2\tau$ & $0.10^{*}$ & 114.0 & $2.4\times10^{-26}$ & $53.73/12=4.48$ & Pass \\
$2b2\tau$ & $0.20^{*}$ & 57.0 & $1.7\times10^{-26}$ & $40.69/12=3.39$ & Pass \\
$2b2\tau$ & $0.30$ & 38.0 & $1.5\times10^{-26}$ & $32.95/12=2.75$ & Fail \\
\textbf{\boldmath $2b2\tau$} & \textbf{\boldmath $0.40$} & \textbf{\boldmath 28.5} & \textbf{\boldmath $1.5\times10^{-26}$} & \textbf{\boldmath $31.95/12=2.66$} & \textbf{\boldmath Fail} \\
$2b2\tau$ & $0.50$ & 30.9 & $1.5\times10^{-26}$ & $32.85/12=2.74$ & Fail \\
$2b2\tau$ & $0.60$ & 31.3 & $1.6\times10^{-26}$ & $34.03/12=2.84$ & Fail \\
$2b2\tau$ & $0.70$ & 32.6 & $1.6\times10^{-26}$ & $35.03/12=2.92$ & Fail \\
$2b2\tau$ & $0.80^{*}$ & 34.2 & $1.7\times10^{-26}$ & $35.77/12=2.98$ & Fail \\
$2b2\tau$ & $0.90^{*}$ & 34.7 & $1.7\times10^{-26}$ & $37.92/12=3.16$ & Fail \\
$2b2\tau$ & $1.00^{*}$ & 38.2 & $1.7\times10^{-26}$ & $40.23/12=3.35$ & Fail \\
\hline
\end{tabular}
\caption{Conditional GCE best fits at fixed $r=m_{\rm med}/m_\chi$, with ${\rm dof}=12$. An asterisk marks a conditional best-fit point outside the channel-global joint $1\sigma$ volume, defined by $\Delta\chi^2>3.52674038$ for three parameters. Within each channel, the row with reduced $\chi^2$ closest to unity is shown in bold, using unrounded values; this highlighting criterion is distinct from minimizing $\chi^2$ over $r$. Pass/Fail compares the unrounded rate against the interpolated dSph 95\% upper limit at the same mass and $r$.}
\label{tab:fixed-r-highlight-all}
\end{table}

For the $4b$ cascade through on-shell mediators, the GCE-preferred regions in the $(m_\chi,\langle\sigma v\rangle)$ plane generally shift toward smaller masses and annihilation cross sections
as $r$ decreases, as discussed below.
Neglecting the initial DM kinetic energy, the partonic cascade requires $2m_b < m_\phi < m_\chi$.
At fixed $r$, this imposes the lower bound $m_\chi>2m_b/r$, which increases as $r$ decreases and can truncate the low-mass part of the preferred region.
Near the bottom-hadron production threshold, the photon spectrum also depends sensitively on the treatment of hadronization and hadronic thresholds.
The continuum approximation underlying Lund string fragmentation used in \texttt{pythia} has limitations in this region~\cite{Sjostrand:2014zea}.
The spectral behavior near threshold therefore cannot be explained by the mediator boost alone.

\begin{figure}[h]
  \centering
  \setlength{\hkpanelw}{0.31\textwidth}%
  \includegraphics[width=1\textwidth]{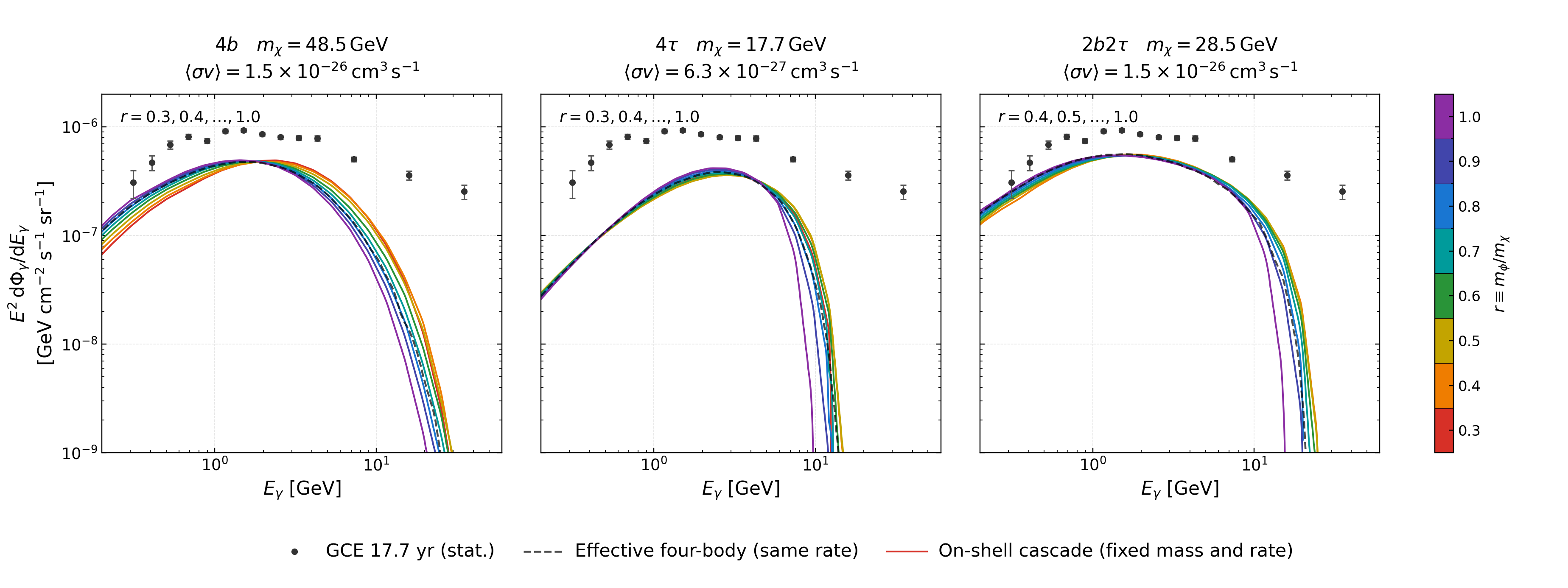}
  \caption{\justifying
  Photon intensity spectra for $4b$ (left), $4\tau$ (center),
and $2b2\tau$ (right).
All curves in each panel use the channel's global cascade
best-fit mass and annihilation cross section.
Colored curves vary $r=m_\phi/m_\chi$ without refitting:
$r=0.3,0.4,\ldots,1.0$ for $4b$ and $4\tau$, and
$r=0.4,0.5,\ldots,1.0$ for $2b2\tau$.
Omitted ratios lie outside the adopted kinematic or
spectrum-support domain, or they are not allowed for given best-fit $m_\chi$ \footnote{For example, $4\tau$ with $m_\chi = 17.7$ GeV does not allow $r=0.2$ due to the on-shell requirement.}.
Black dashed curves show the effective four-body
contact-interaction benchmarks.
Black points show the 17.7-year Fermi-LAT GCE spectrum
with per-bin uncertainties.
For $2b2\tau$, two distinct mediators of equal mass decay
exclusively into $b\bar b$ and $\tau^+\tau^-$, respectively.
  }  \label{fig:cascade_photon_spectra}
\end{figure}

Figure~\ref{fig:cascade_photon_spectra} shows the photon intensity spectra obtained
by fixing $m_\chi$ and $\langle\sigma v\rangle$ to the global best-fit values of each
channel and varying only $r$. The fixed masses are $48.5$, $17.7$, and
$28.5~\mathrm{GeV}$ for $4b$, $4\tau$, and $2b2\tau$, respectively; the corresponding
rates are $1.5\times10^{-26}$, $6.3\times10^{-27}$, and
$1.5\times10^{-26}~\mathrm{cm^3\,s^{-1}}$. Only ratios supported at each fixed mass are
shown: $r=0.3,0.4,\ldots,1.0$ for $4b$ and $4\tau$, and $r=0.4,0.5,\ldots,1.0$ for
$2b2\tau$. The $r=1$ spectrum is retained as the at-rest-mediator limiting template.
The effective four-body comparison uses the same fixed mass and annihilation rate in
each panel. Thus, differences between curves within a panel reflect changes in
spectral shape, especially by changing $r$.

Decreasing $r$ generally extends the high-energy tail
of the photon spectrum in the examples shown.
Part of this dependence can be understood from the
decay kinematics.
For $\phi\to X \bar X$, where $X$ is a SM particle
of mass $m_X$, the energy of either daughter in the
mediator rest frame is $E_X^\ast=m_\phi/2$.
Neglecting the initial DM velocities, each mediator has
energy $E_\phi=m_\chi$ in the annihilation center-of-mass
frame.
Its Lorentz factor and speed are therefore
$\gamma_\phi=m_\chi/m_\phi=1/r$ and
$\beta_\phi=\sqrt{1-r^2}$.
The daughter energy in this frame is
\begin{align}
E_X = E_X^\ast \left( \gamma_\phi + \sqrt{\gamma_\phi^2 - 1} \sqrt{1 - \frac{4 m_X^2}{m_\phi^2}} \cos \theta^\ast \right)\,,
\end{align}
where $\theta^\ast$ is the angle of the $X$-particle emission from the boost direction of $\phi$, measured in the mediator rest frame. For a scalar mediator, the decay is isotropic, so $\cos\theta^\ast$ is uniformly distributed.
Consequently, $dN_X/dE_X$ is constant between the endpoints
\begin{align}
E_X^\pm
=
\frac{m_\chi}{2}
\left[
1\pm
\sqrt{1-r^2}
\sqrt{1-\frac{4m_X^2}{r^2m_\chi^2}}
\right]\,. 
\label{eq:Erange}
\end{align}
The mean daughter energy is
$\langle E_X\rangle=m_\chi/2$, independent of $r$.
When $m_\phi\gg2m_X$, decreasing $r$ broadens this distribution toward both lower and higher energies.
Weighting by $E_X^2$ emphasizes its high-energy part but does not shift the underlying box distribution.
Near the decay threshold, the phase space factor decreases, and the energy range can narrow again.

These relations describe the kinematic contribution to the spectral changes in
Fig.~\ref{fig:cascade_photon_spectra}.
The photon spectra also depend on radiation, hadronization, and particle decays in the mediator rest frame.
They therefore require the full simulation and cannot be inferred from the daughter-energy distribution alone.

\begin{figure}[htbp]                 
      \centering                    
      \subfloat[]{%
          \includegraphics[
              width=0.45\linewidth,
              trim=20bp 35bp 35bp 15bp,
              clip
          ]{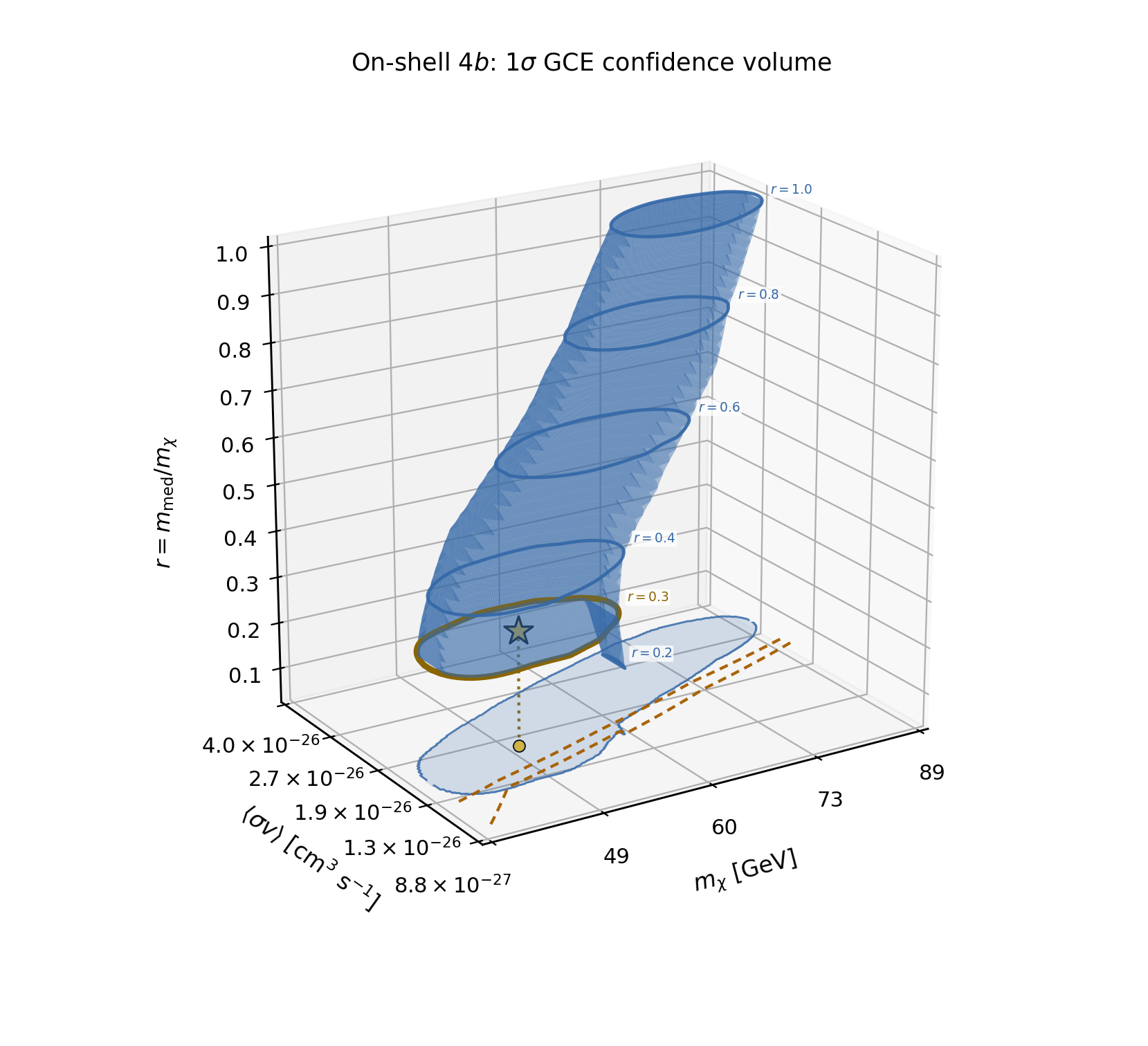}%
      }%
      \subfloat[]{%
          \includegraphics[
              width=0.45\linewidth,
              trim=20bp 35bp 35bp 15bp,
              clip
          ]{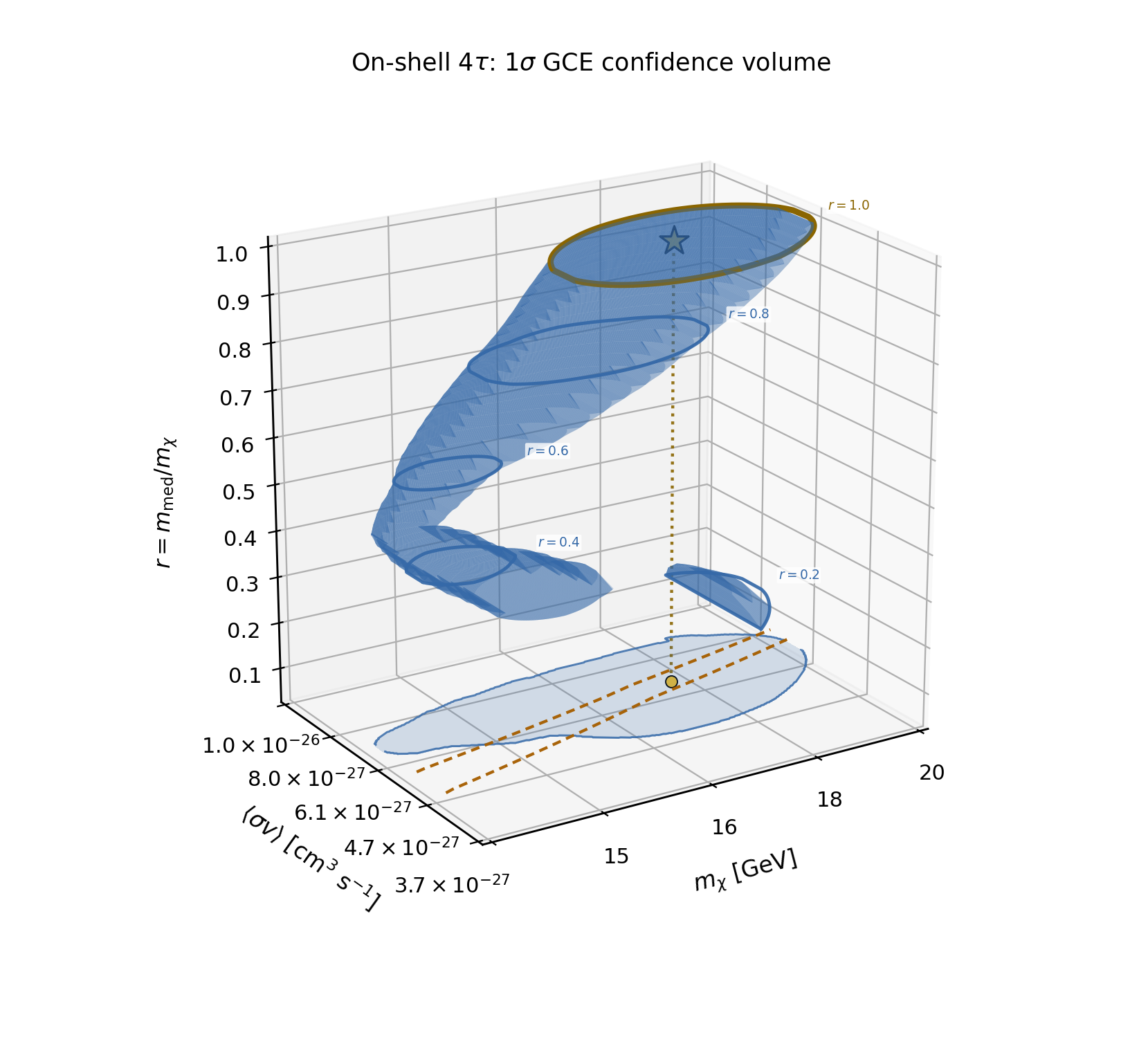}%
      }%
            \vspace{0.5cm} 
            
      \subfloat[]{%
          \includegraphics[
              width=0.45\linewidth,
              trim=20bp 35bp 35bp 15bp,
              clip
          ]{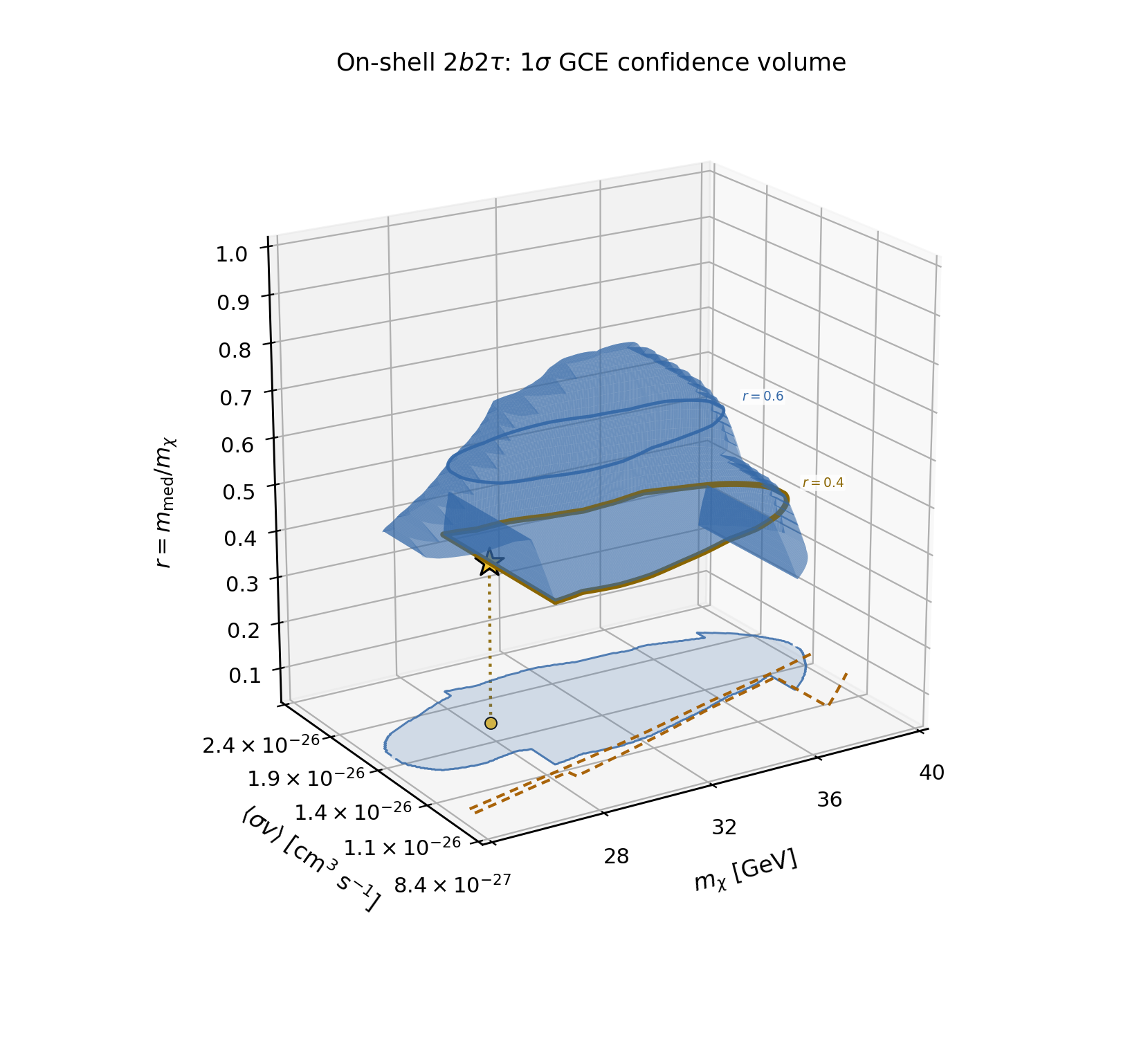}%
      }%
      \par\vspace{4pt}

      \includegraphics[width=0.85\linewidth]{%
          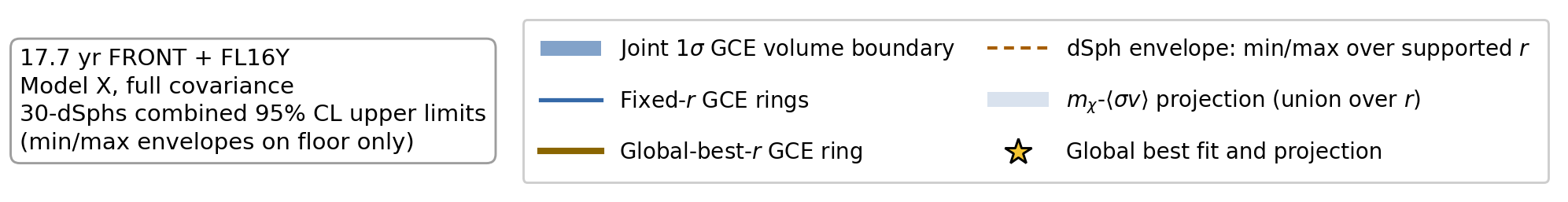}
    \caption{\justifying
        Joint $1\sigma$ GCE volumes in
$(m_\chi,\langle\sigma v\rangle,r)$ for
(a) $4b$, (b) $4\tau$, and (c) $2b2\tau$.
Fits use Model~X and the full covariance.
Colored curves show fixed-$r$ slices; gold curves and stars
mark the best-fit $r$ slices and global best-fit points.
Shaded projections are unions over $r$, not separate
two-dimensional confidence regions.
Brown dashed curves show the minimum and maximum combined dSph
95\% CL upper limits over supported $r$ at each mass.
Rates below the minimum satisfy all these limits;
rates above the maximum are excluded for all supported r.
Intermediate rates require comparison at the same $r$.
    }
    \label{fig:mediator_region_3d}
\end{figure}

Allowing the mediator mass to vary, we describe the GCE-preferred
region in the three-dimensional parameter space
$(m_\chi,\langle\sigma v\rangle,r)$.
Figure~\ref{fig:mediator_region_3d} shows the joint $1\sigma$
confidence regions for the $4b$, $4\tau$, and $2b2\tau$ channels.
For each channel, the region is defined by
\begin{align}
    \Delta\chi^2
\equiv
\chi^2(m_\chi,\langle\sigma v\rangle,r)
-\chi^2_{\min}
\leq 3.53\,,
\end{align}
where $\chi^2_{\min}$ is the global minimum over the parameter
range considered.
The photon spectra are simulated at
$r=0.1,0.2,\ldots,1.0$ and linearly interpolated at fixed energy
between adjacent values of $r$.
For the $4b$ and $2b2\tau$ channels, we restrict the fit to
$m_\phi>11~\mathrm{GeV}$, our adopted lower cutoff for the
hadronic spectra.
For the $4\tau$ channel, we require
$m_\phi>2m_\tau$, or equivalently $m_\chi>2m_\tau/r$.

For the $4b$ channel in the top-left panel, the GCE-preferred
region has the approximate shape of a tilted elliptic cylinder
in the $(m_\chi,\langle\sigma v\rangle,r)$ parameter space.
Over most of the displayed range, decreasing $r$ shifts the
preferred masses and annihilation cross sections toward smaller
values.
This trend is consistent with the changes in the photon spectra
shown in the left panel of Fig.~\ref{fig:cascade_photon_spectra}.
As $r$ decreases, the mediator boost $\gamma_\phi=1/r$ increases, modifying the decay-product energy
distribution and the resulting photon spectrum.
The harder photon spectra obtained over this range can favor
smaller DM masses when fitting the GCE.
At small $r$, the mediator-mass cut also restricts the available
parameter space: $m_\phi>11~\mathrm{GeV}$ requires
$m_\chi>11~\mathrm{GeV}/r$.
For example, at $r=0.2$, only masses above $57~\mathrm{GeV}$
are included in the fit, removing the lower-mass part of the
parameter space.
The global best fit of the $\chi \chi \to 4b$ channel is $(m_\chi, \langle \sigma v \rangle, r) = (48.5\,{\rm GeV}, 1.5 \times 10^{-26}\,{\rm cm}^3{\rm s}^{-1},0.3)$ with $\chi^2$/d.o.f. = 2.39.

The light-blue shaded region shows the projection of the
three-dimensional GCE-preferred region onto the $(m_\chi,\langle \sigma v\rangle)$ plane, taking the union over $r$.
The dashed brown curves show the minimum
and maximum of the dSph $95\%$ confidence level (CL) upper limits over
the supported values of $r$ at each DM mass.
Points above the dashed upper curve are excluded for all supported
values of $r$, whereas points below the dashed lower curve are not
excluded for any of them.
Between the two curves, exclusion depends on $r$.
Compatibility with the GCE must therefore be assessed by comparing
the GCE-preferred region and the dSph limit at the same value of
$r$; the projection alone does not retain this information.
For some values of $r \lesssim 0.3$, part of the joint $1\sigma$
GCE-preferred region for the $4b$ channel remains unexcluded by the dSph $95\%$ CL upper limits evaluated
at the same $r$.
This contrasts with the effective $\chi\chi\to4b$ benchmark,
for which the entire $1\sigma$ GCE-preferred region is excluded
by the dSph $95\%$ CL limits.

For the $4\tau$ channel in the top-right panel, the GCE-preferred DM mass decreases as $r$ is reduced to approximately $0.5$, but increases again at smaller $r$.
At intermediate values of $r$, the mediated cascade favors smaller masses than the effective $\chi\chi \to 4\tau$ benchmark.
The initial shift toward smaller masses is consistent with
the changes in the photon spectrum induced by the mediator boost.

The change in direction may be related to the competing effects of the mediator boost and the finite tau mass.
At fixed $m_\chi$, the tau-energy range in Eq.~\eqref{eq:Erange} is widest at $r=\sqrt{2m_\tau/m_\chi}$.
For $m_\chi \simeq 15~\mathrm{GeV}$, this corresponds to
$r \simeq 0.49$, close to the value at which the preferred region turns toward larger masses.~\footnote{For the $4b$ channel, the slight bend near $r \simeq 0.4$ may have a similar kinematic origin. At fixed $m_\chi$, the primary bottom-quark energy
distribution has its maximum width at
$r=\sqrt{2m_b/m_\chi}$, which lies near this range for the masses favored here. The connection with the GCE fit remains qualitative because the photon spectrum also depends on parton showering, hadronization, and subsequent decays.}
Subsequent $\tau$ lepton decays distribute the energy among neutrinos
and visible particles.
The predicted photon spectrum therefore depends on both the
$\tau$-energy distribution from mediator decay and the subsequent $\tau$-decay kinematics.
These decay effects are included at all values of $r$.
The global best fit is $(m_\chi, \langle \sigma v \rangle, r) = (17.7\,{\rm GeV}, 6.3 \times 10^{-27}\,{\rm cm}^3{\rm s}^{-1},1.0)$ with $\chi^2$/d.o.f. = 2.06 which is better than the $4b$ channel.
Notably, a substantial part of the joint $1 \sigma$ GCE-preferred region is not excluded by the dSph $95\%$ CL
limits evaluated at the same $r$, whereas the best-fit value is above the same-$r$ limit by 5.8\%.

For the $2b2\tau$ channel in the bottom panel, the joint
$1\sigma$ GCE-preferred region spans approximately
$0.3 \lesssim r \lesssim 0.7$.
The minimum $\chi^2$ is $31.954$ for 11 degrees of freedom,
indicating a poorer fit than for the $4b$ and $4\tau$ channels.
The best-fit point has $(m_\chi, \langle \sigma v \rangle, r) = (28.5\,{\rm GeV}, 1.5 \times 10^{-26}\,{\rm cm}^3{\rm s}^{-1}, 0.4)$ with $\chi^2$/d.o.f. = 2.90,
corresponding to $m_\phi=11.4$~GeV, close to the adopted
lower cut of $11$~GeV.
The extent of the preferred region is therefore also affected
by this cut.
The entire joint $1\sigma$ GCE-preferred region is excluded
by the dSph $95\%$ CL limits evaluated at
the corresponding values of $r$.

\begin{figure}[htbp]
    \centering  
    \subfloat[]{
        \includegraphics[width=0.45\linewidth]{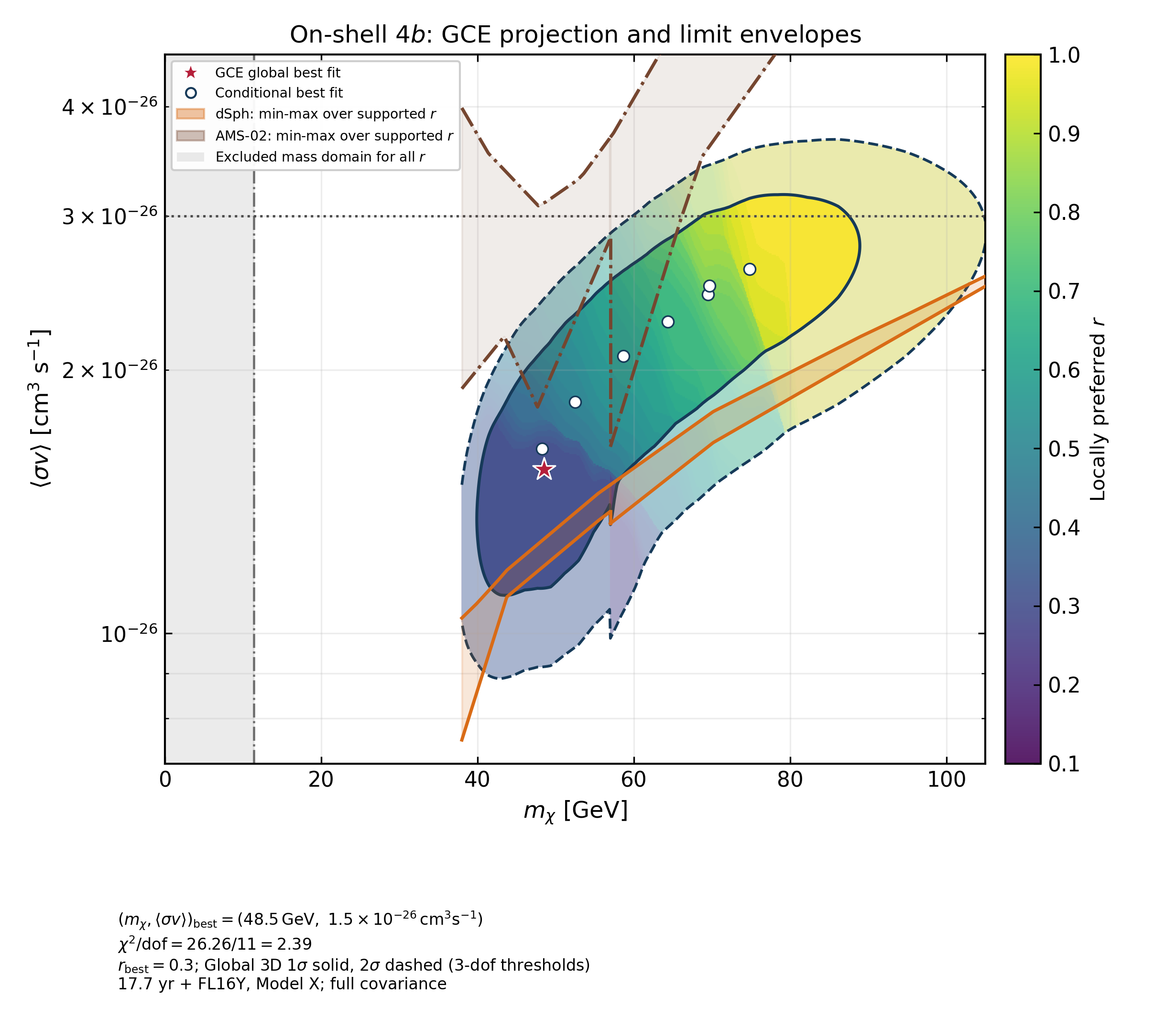}
    }
    \hfill
    \subfloat[]{
        \includegraphics[width=0.45\linewidth]{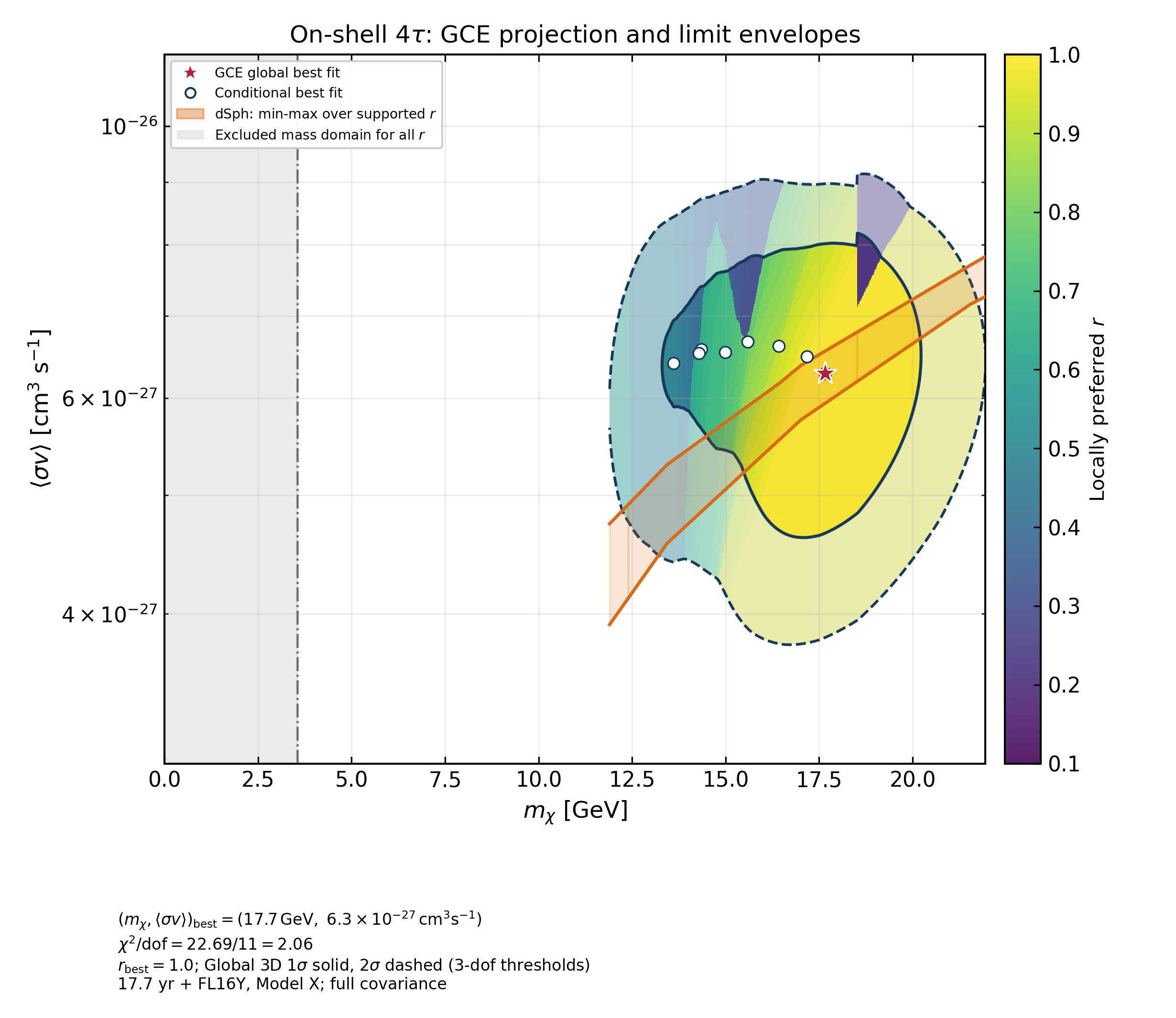}
    }
    
    \vspace{0.5cm} 
    \subfloat[]{
        \includegraphics[width=0.45\linewidth]{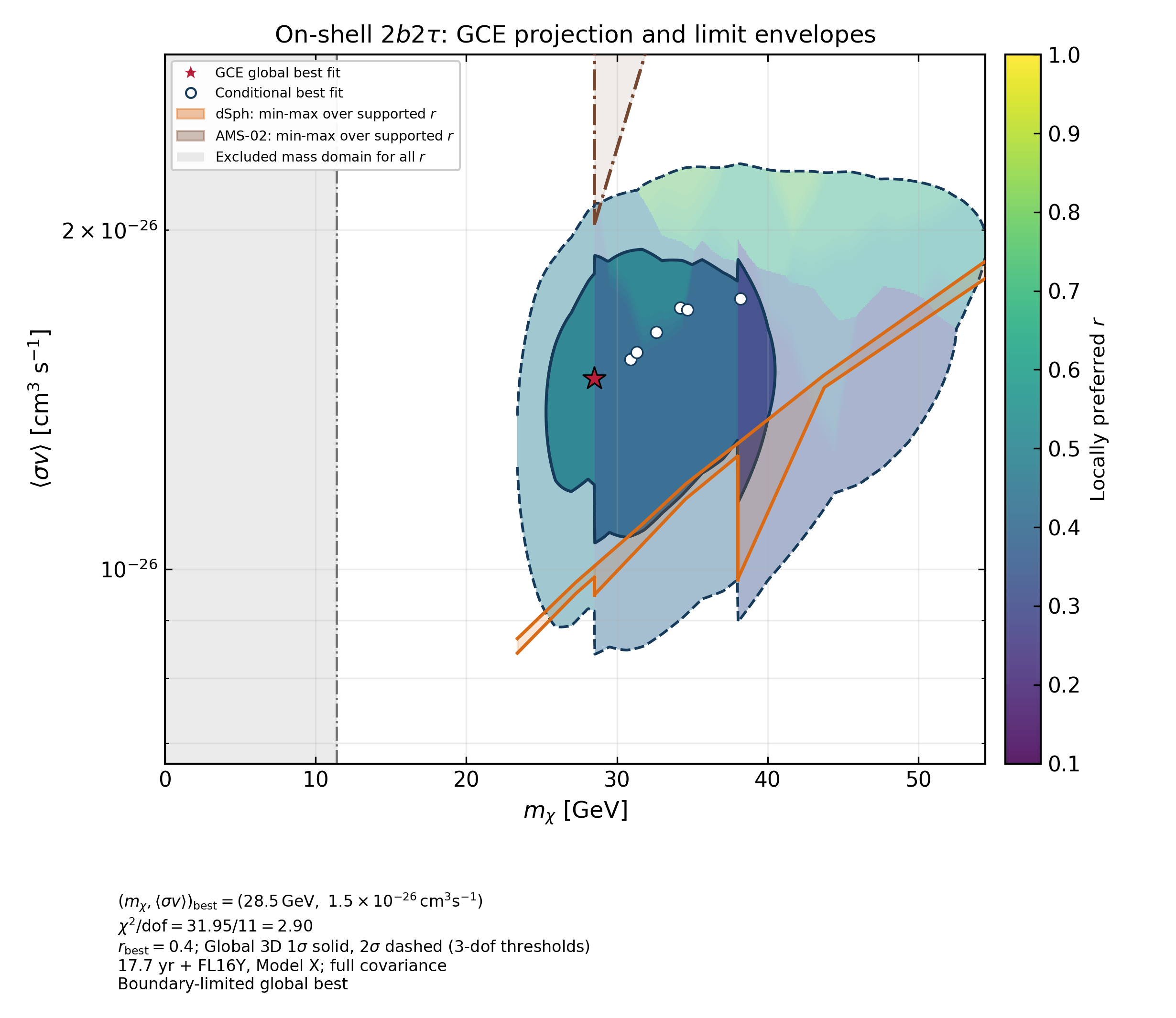}
    }
    
    \caption{\justifying
    Projections of the joint three-parameter $1\sigma$ (solid
dark blue) and $2\sigma$ (dashed dark blue) GCE regions onto
the $(m_\chi,\langle\sigma v\rangle)$ plane for $4b$,
$4\tau$, and $2b2\tau$, using Model~X and the full covariance.
Color shows the locally preferred $r=m_\phi/m_\chi$;
red stars and open circles mark global and fixed-$r$
best fits, respectively.
Orange and brown bands span the minimum--maximum dSph and
AMS-02 antiproton 95\% CL upper limits over available $r$
at each mass, with solid and dot-dashed boundaries.
Exclusion requires comparison with the limit at the same $r$.
No antiproton limit is shown for $4\tau$.
Gray shading marks masses outside the analysis domain for
all $r$.
The horizontal dotted line marks the thermal relic reference,
$\langle\sigma v\rangle=3\times10^{-26}\,
\mathrm{cm^3\,s^{-1}}$.
    }  \label{fig:mediator_msv_projection}
\end{figure}

Figure~\ref{fig:mediator_msv_projection} shows the projections of the joint 68\% and 95\% GCE-preferred regions in $(m_\chi,\langle\sigma v\rangle,r)$ onto the $(m_\chi,\langle\sigma v\rangle)$ plane.
The solid and dashed dark-blue contours correspond to the
68\% and 95\% regions, respectively.
The color indicates the locally preferred value of $r$, obtained by interpolating between the simulated mediator mass ratios.
The orange and brown bands span the minimum and maximum 95\% CL upper limits from dSphs and AMS-02 antiproton observations, respectively, over the available values of $r$ at each DM mass.
Their boundaries are shown as orange solid and brown dot-dashed curves.
For points between the two boundaries of an envelope, whether the point is excluded depends on $r$.
The GCE fit and the corresponding upper limit must therefore be compared at the same value of $r$.

For the $4b$ and $4\tau$ channels, parts of the projected 68\% GCE-preferred regions lie below even the strongest dSph limits represented by the envelopes.
Most of the projected 68\% region for the $2b2\tau$ channel, however, lies above the weakest dSph limit
in its envelope.
The antiproton envelope overlaps the projected 68\% GCE-preferred region for the $4b$ channel, whereas it does not intersect the corresponding region for $2b2\tau$.
No antiproton limit is shown for $4\tau$, whose antiproton yield is negligible in the mass range considered.

\begin{figure}[htbp]
  \centering \includegraphics[width=0.32\textwidth]{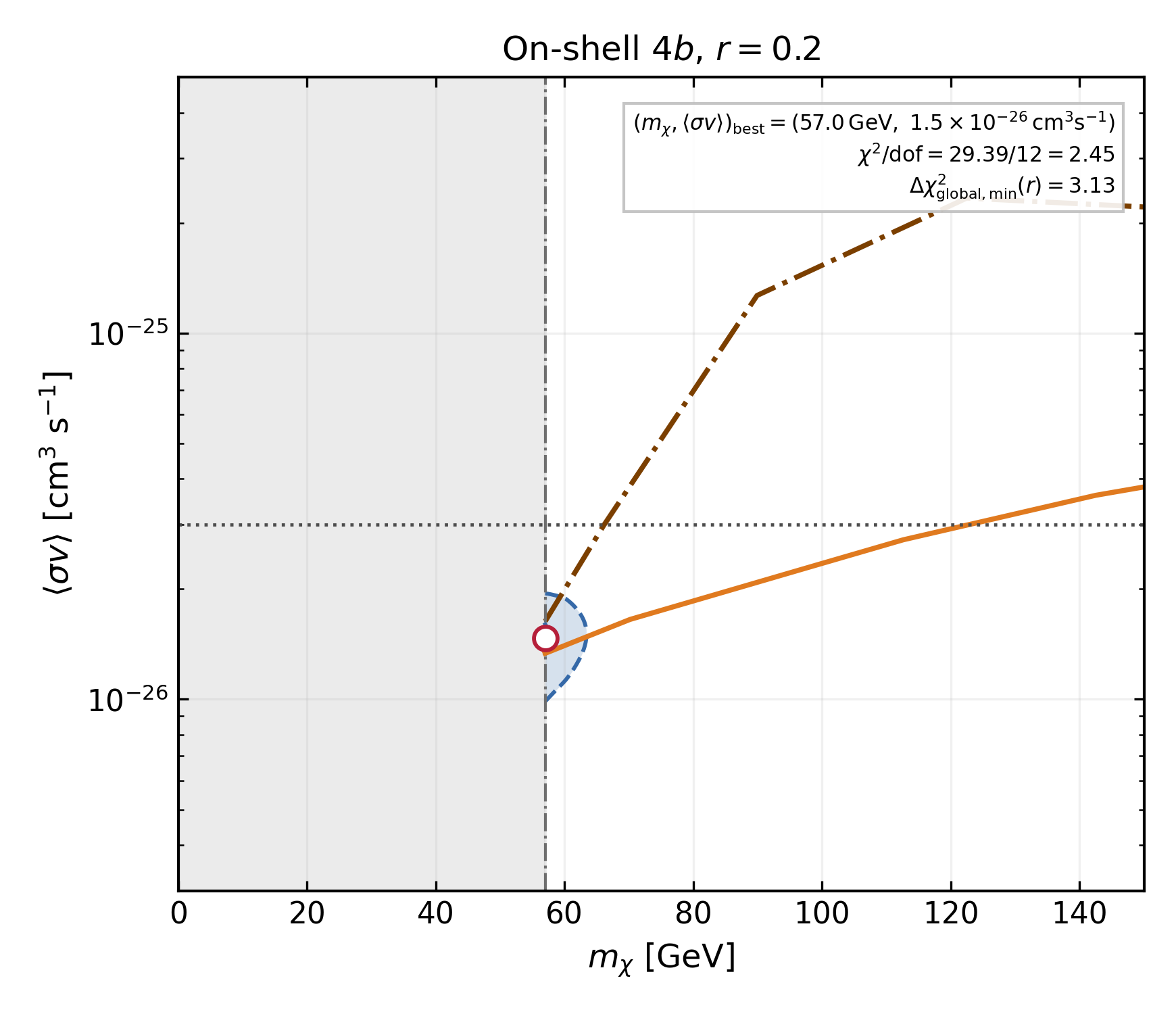}\hfill%
  \includegraphics[width=0.32\textwidth]{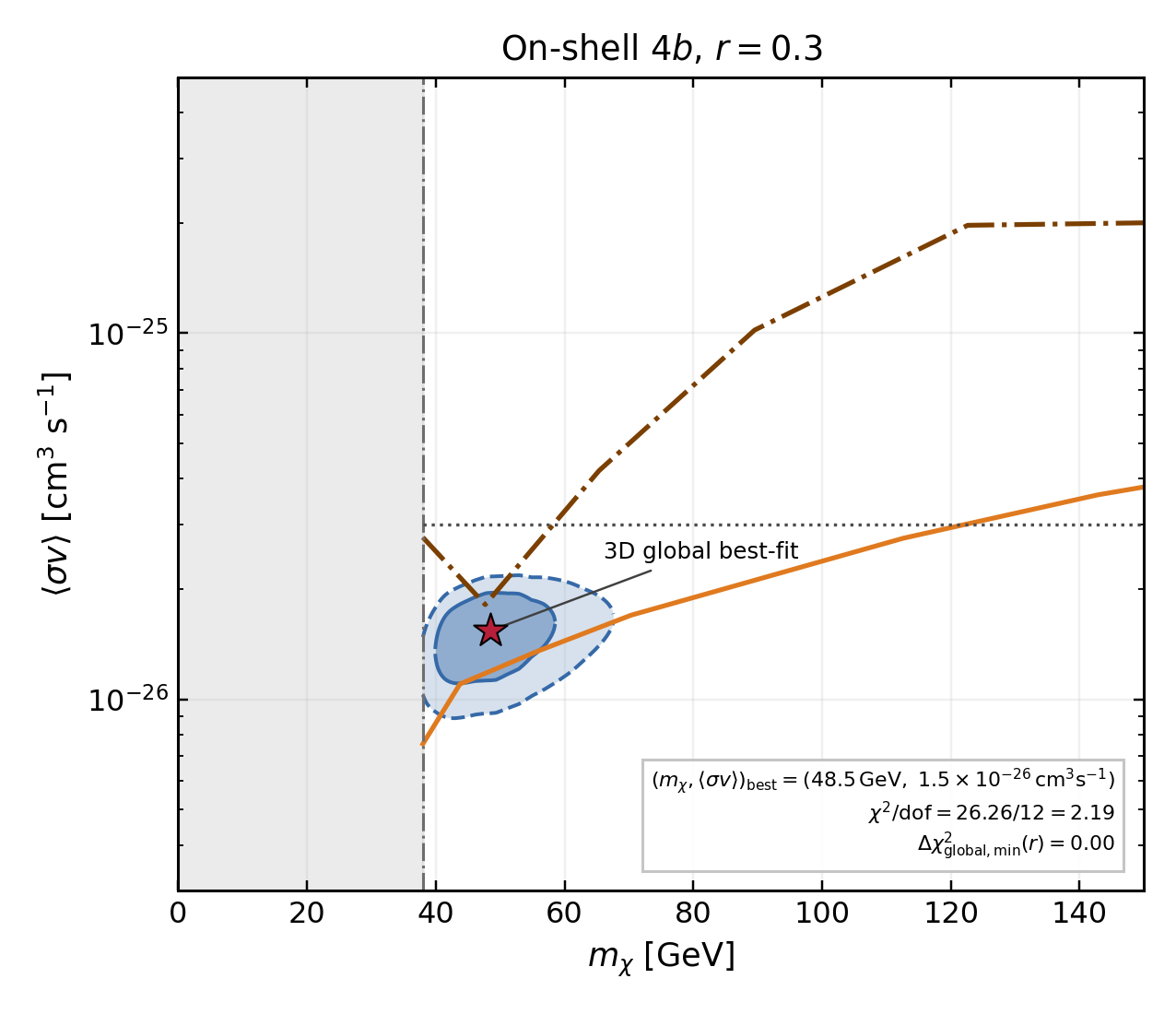}\hfill%
  \includegraphics[width=0.32\textwidth]{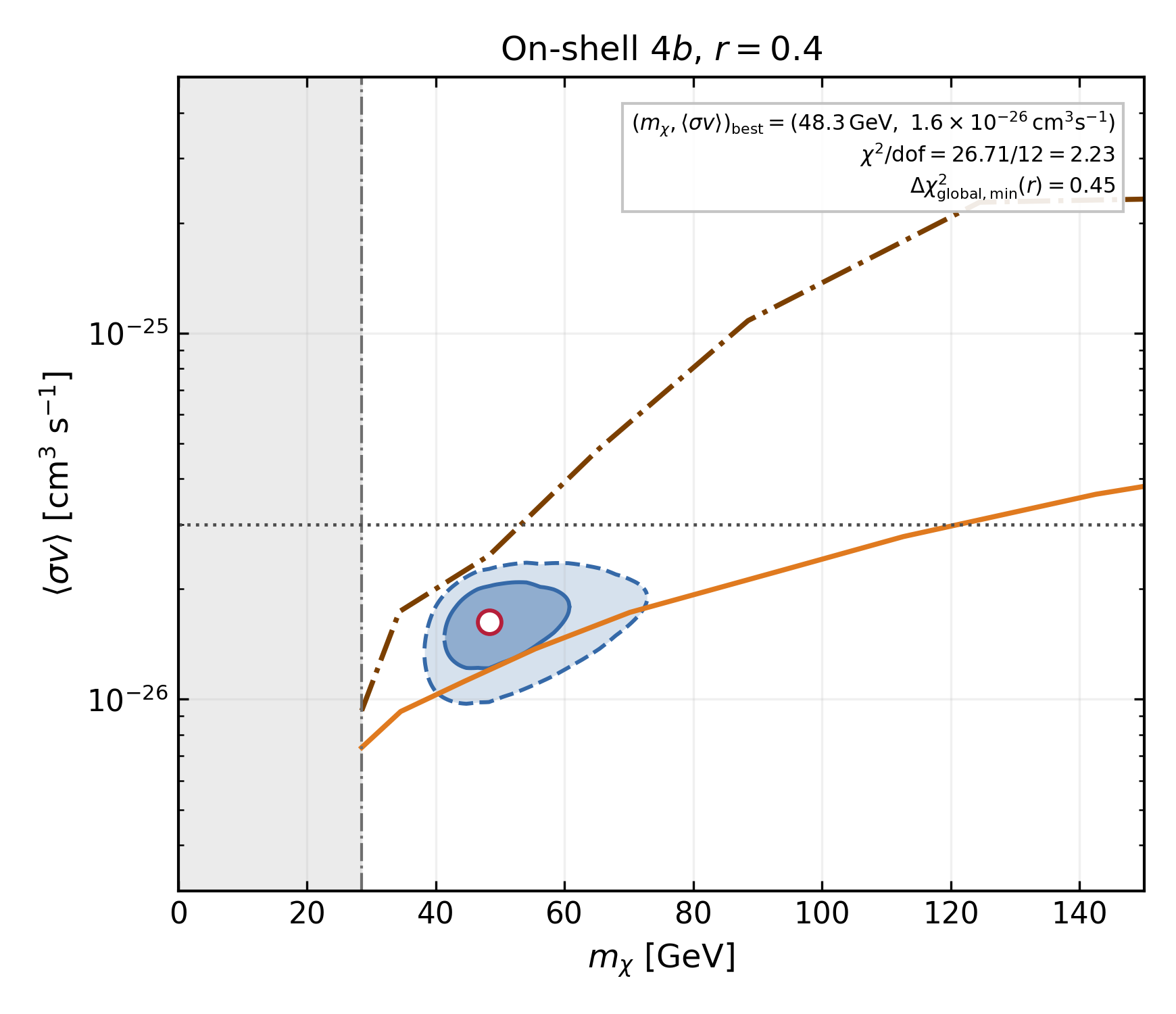}
  \\[2pt]
  \includegraphics[width=0.32\textwidth]{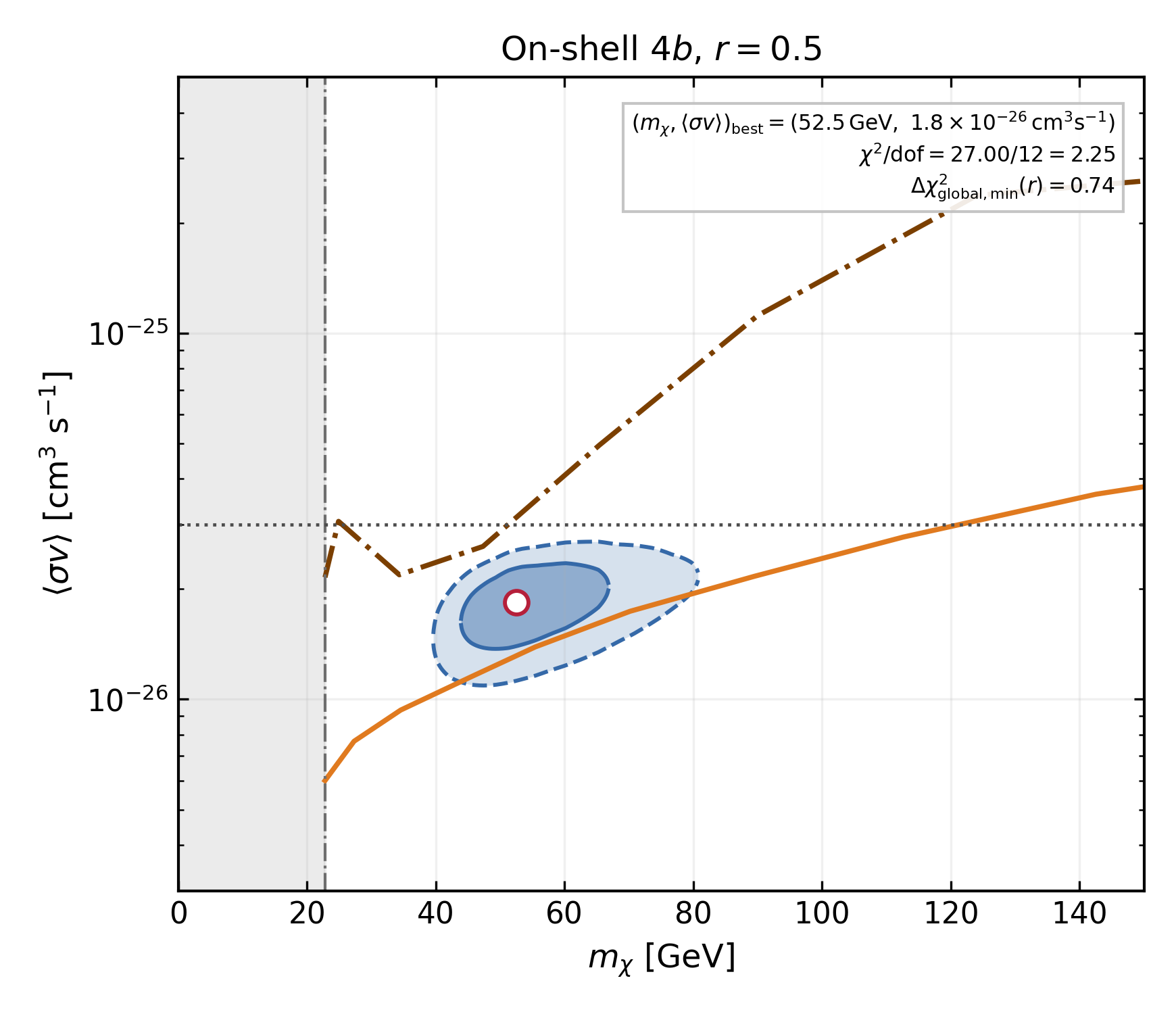}\hfill%
  \includegraphics[width=0.32\textwidth]{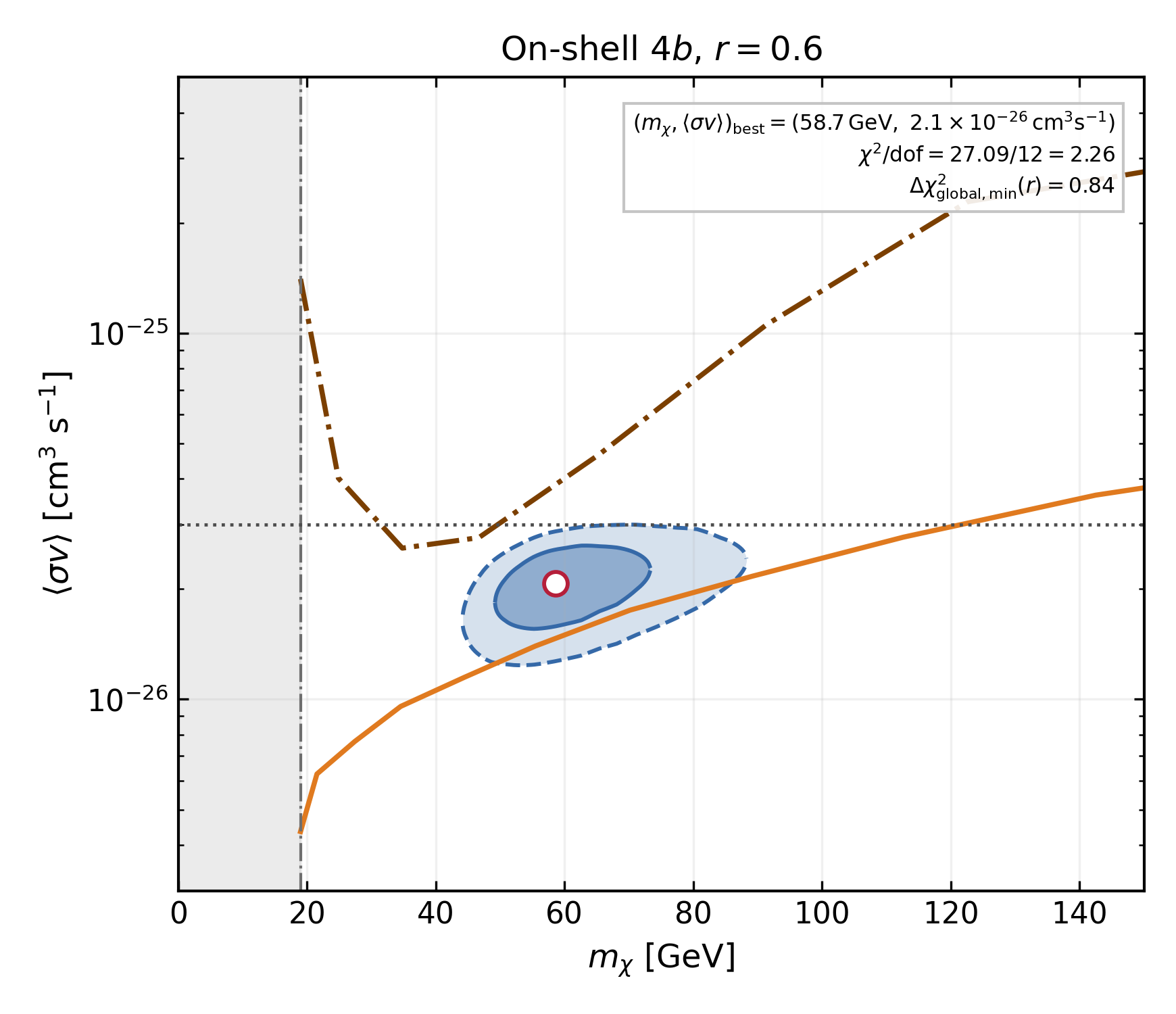}\hfill%
  \includegraphics[width=0.32\textwidth]{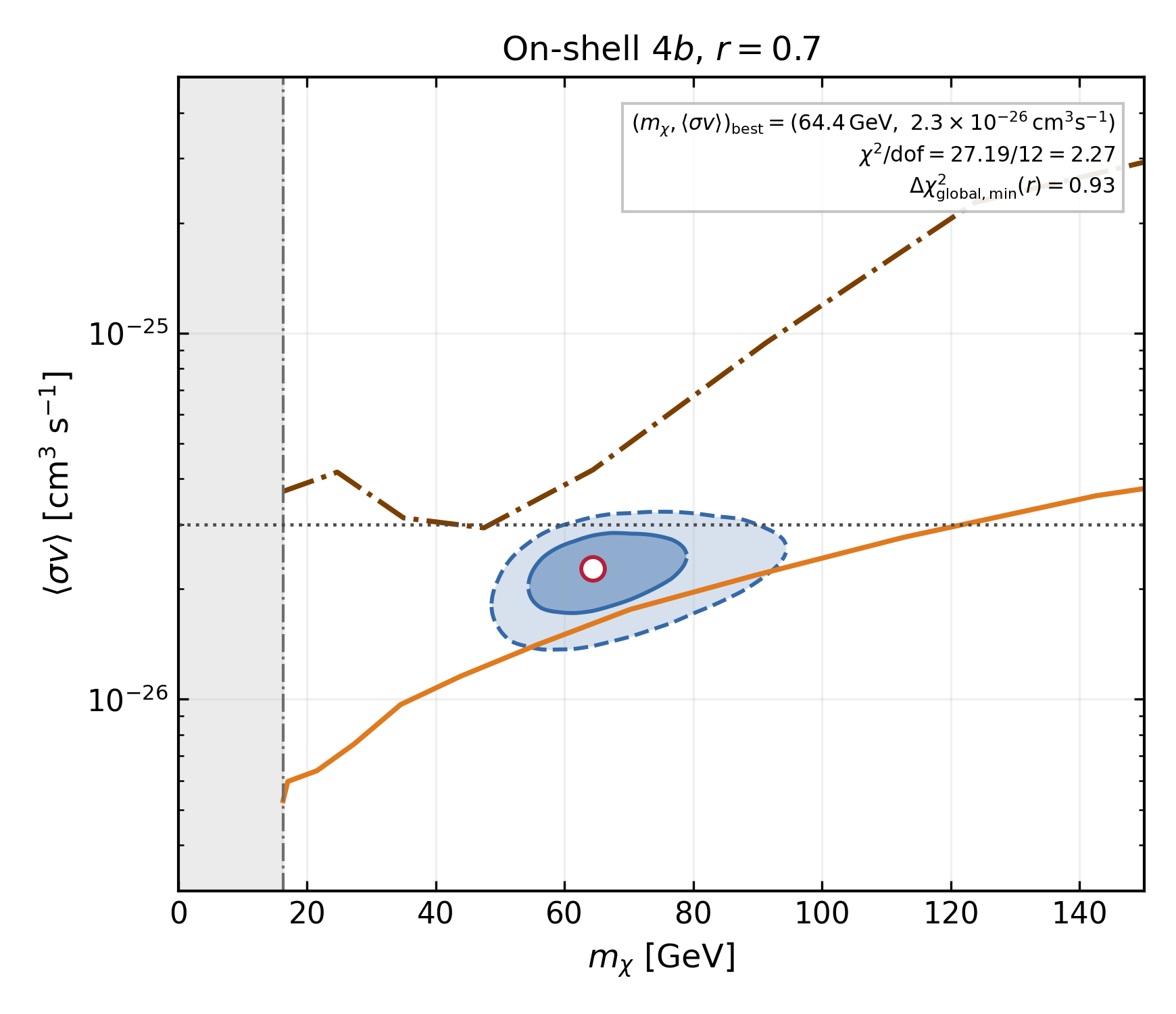}
  \\[2pt]
  \includegraphics[width=0.32\textwidth]{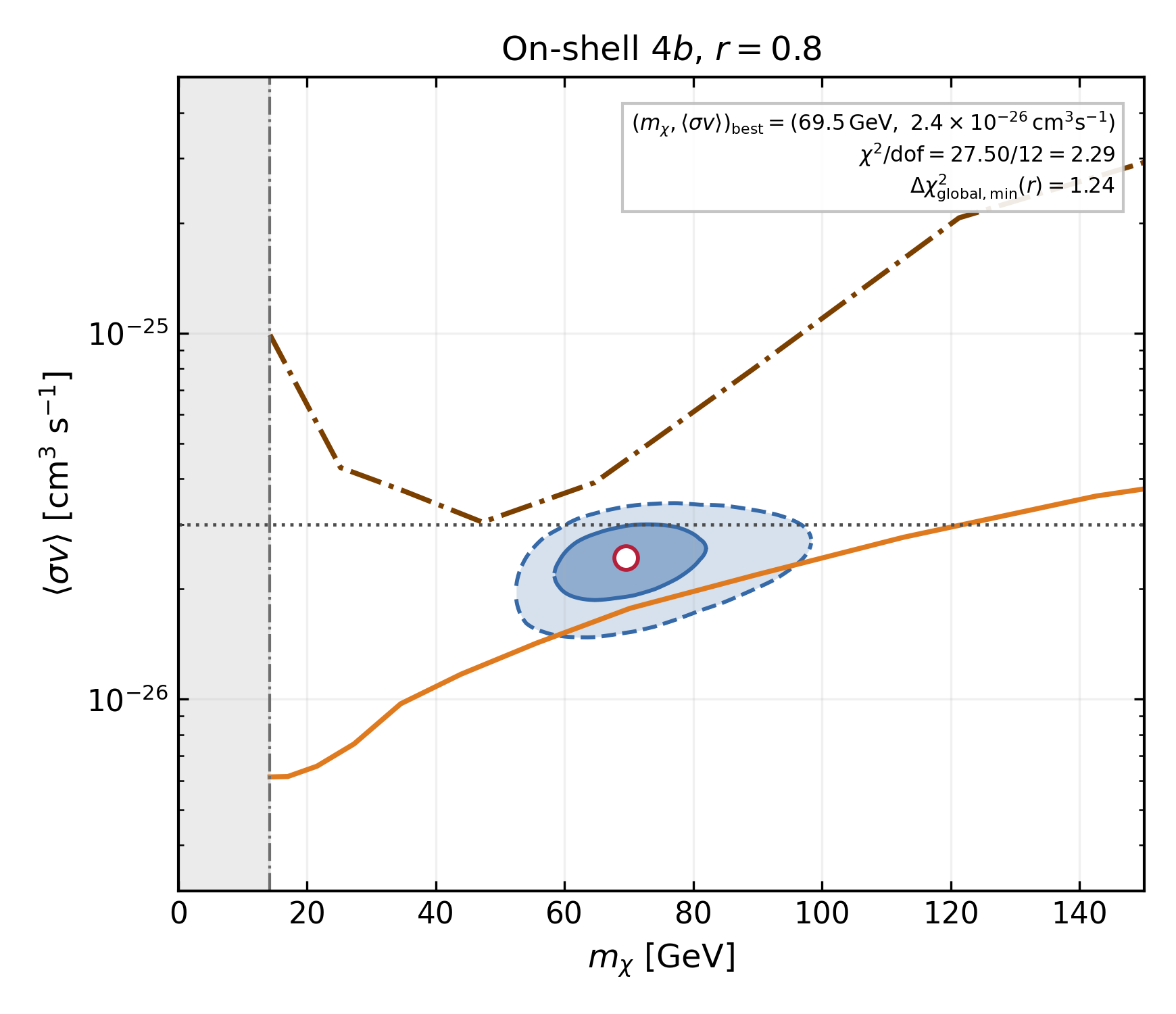}\hfill%
  \includegraphics[width=0.32\textwidth]{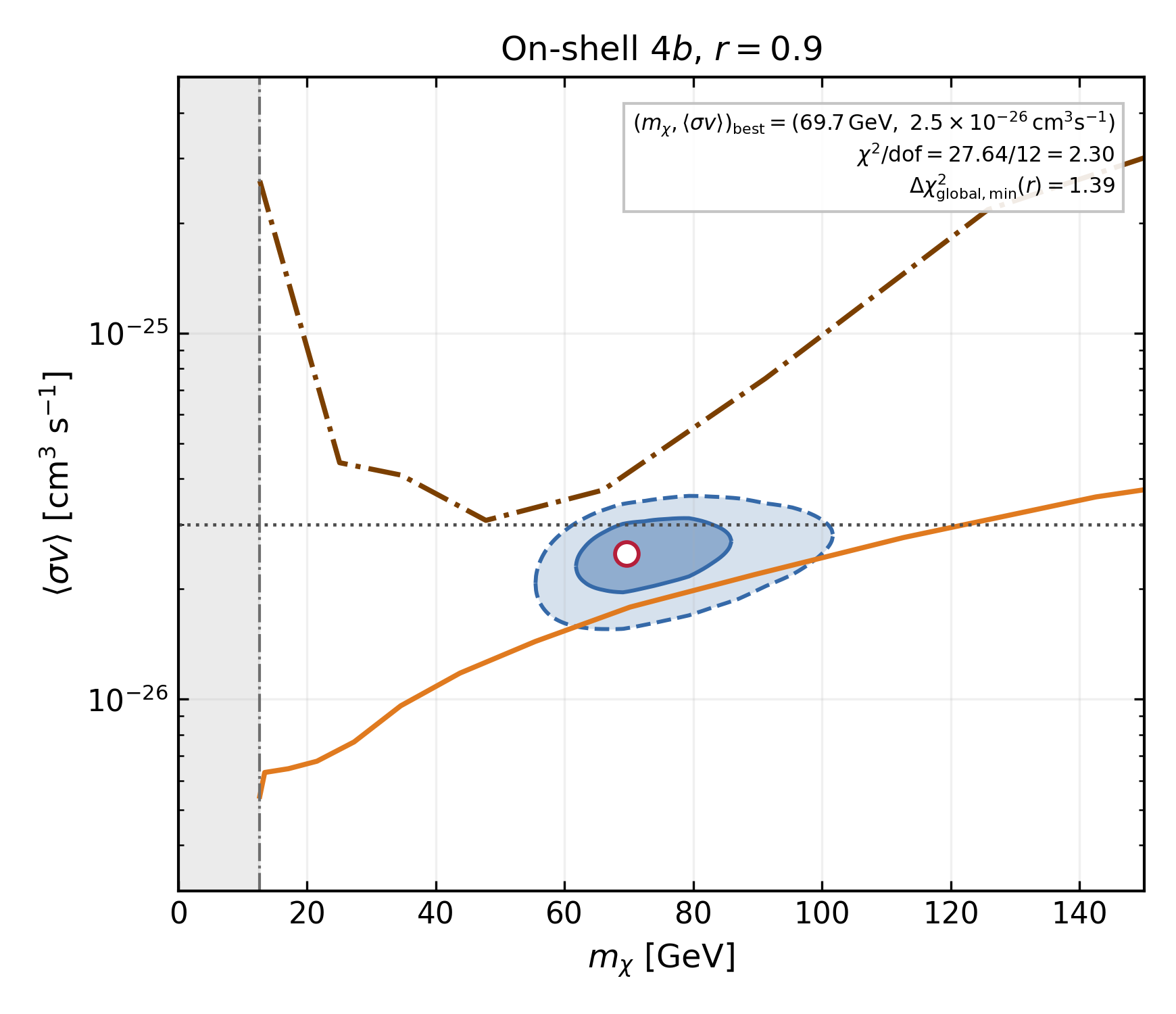}\hfill%
  \includegraphics[width=0.32\textwidth]{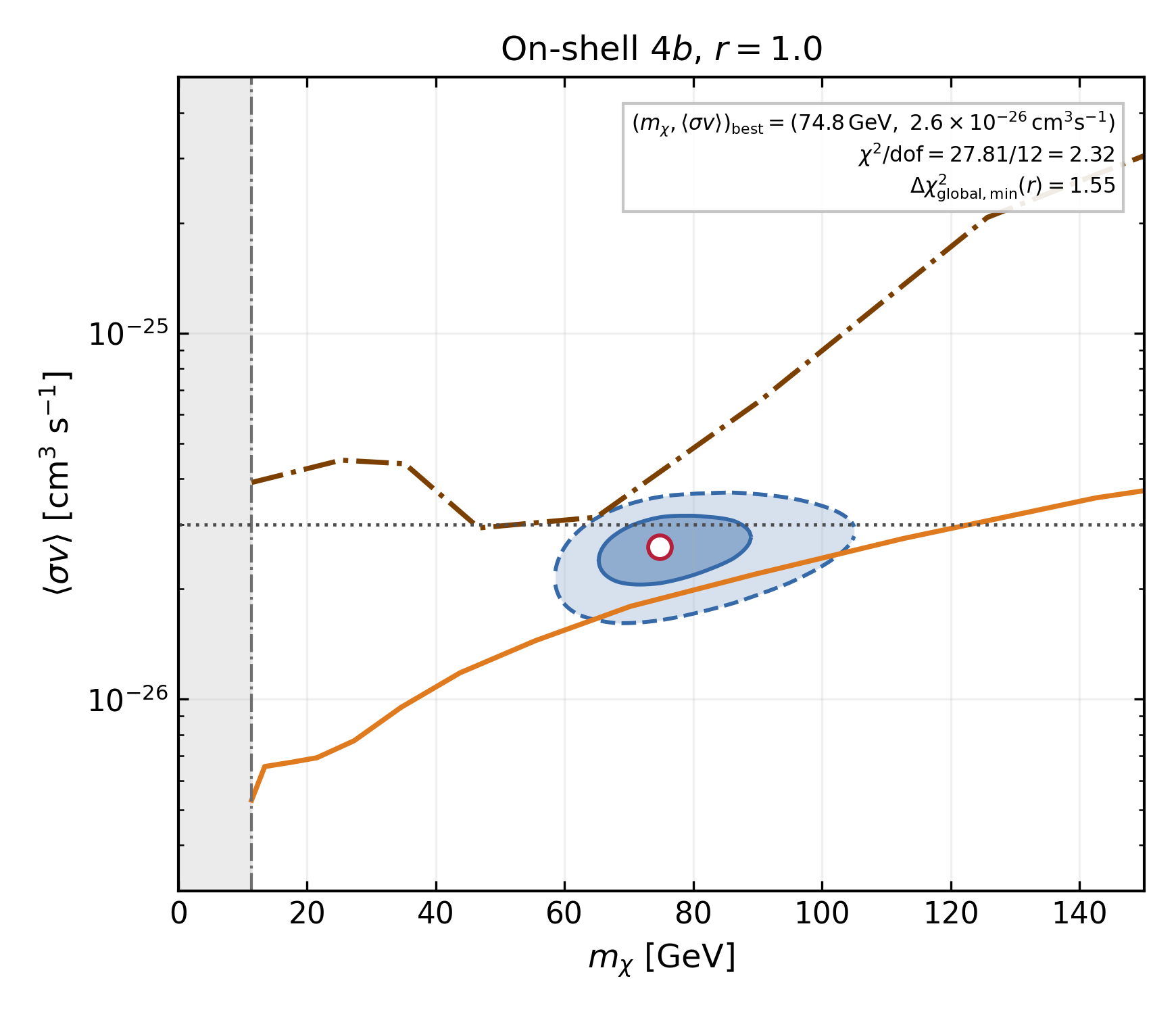}
  \\[4pt]
  \includegraphics[width=0.66\textwidth]{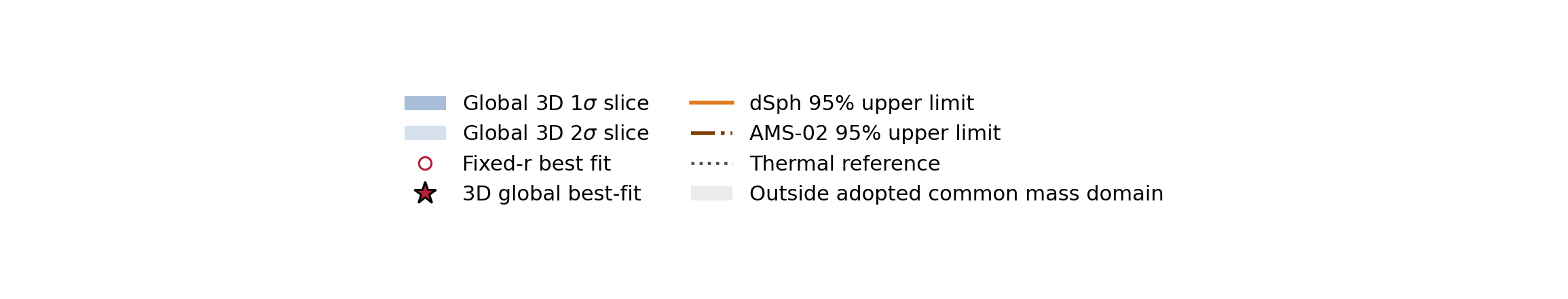}
  \caption{\justifying
Fixed-$r$ slices of the joint three-parameter GCE regions
for $\chi\chi\to\phi\phi\to4b$, with
$r=m_\phi/m_\chi=0.2,0.3,\ldots,1.0$ from left to right
and top to bottom.
Dark and light blue show $1\sigma$ and $2\sigma$ slices,
using three-parameter thresholds relative to the global
$\chi^2$ minimum.
Fits use Model~X and the full covariance.
Open red circles mark fixed-$r$ best fits; insets give
their parameters, $\chi^2_{\min}(r)/\mathrm{dof}$, and
$\Delta\chi^2$ from the global minimum.
Orange solid and brown dot-dashed curves show the combined
dSph and AMS-02 antiproton 95\% CL upper limits at the same $r$.
The horizontal dotted line marks the thermal relic reference,
$\langle\sigma v\rangle=3\times10^{-26}\,
\mathrm{cm^3\,s^{-1}}$.
Gray shading indicates masses outside the adopted
mediator-mass validity range.
}
  \label{fig:msv_per_r_4b}
\end{figure}

The comparisons at fixed $r$ clarify which GCE solutions are constrained.
For the $4b$ channel, Fig.~\ref{fig:msv_per_r_4b} shows
that the antiproton limits approach or intersect the GCE-preferred regions for the smaller ratios,
$r=0.2$--$0.4$, while remaining above them for the larger ratios shown.
The relative strength of the antiproton bounds depends on
the cascade spectra and the adopted cosmic-ray propagation
model~\cite{Elor:2015bho, Calore:2022stf}.
We discuss the fixed-$r$ results for each channel below.

\begin{figure}[htbp]
  \centering
  \includegraphics[width=0.32\textwidth]{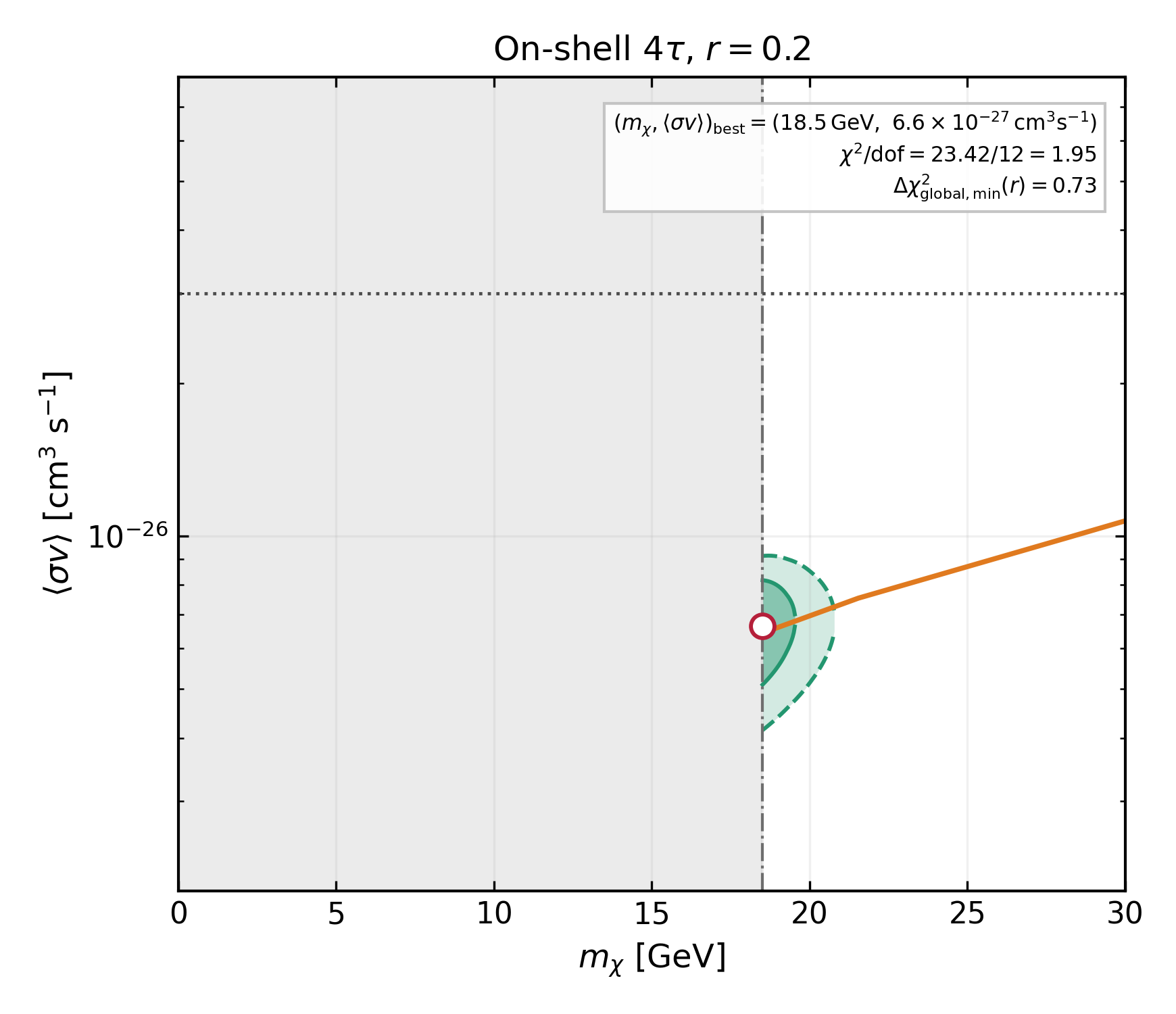}\hfill%
  \includegraphics[width=0.32\textwidth]{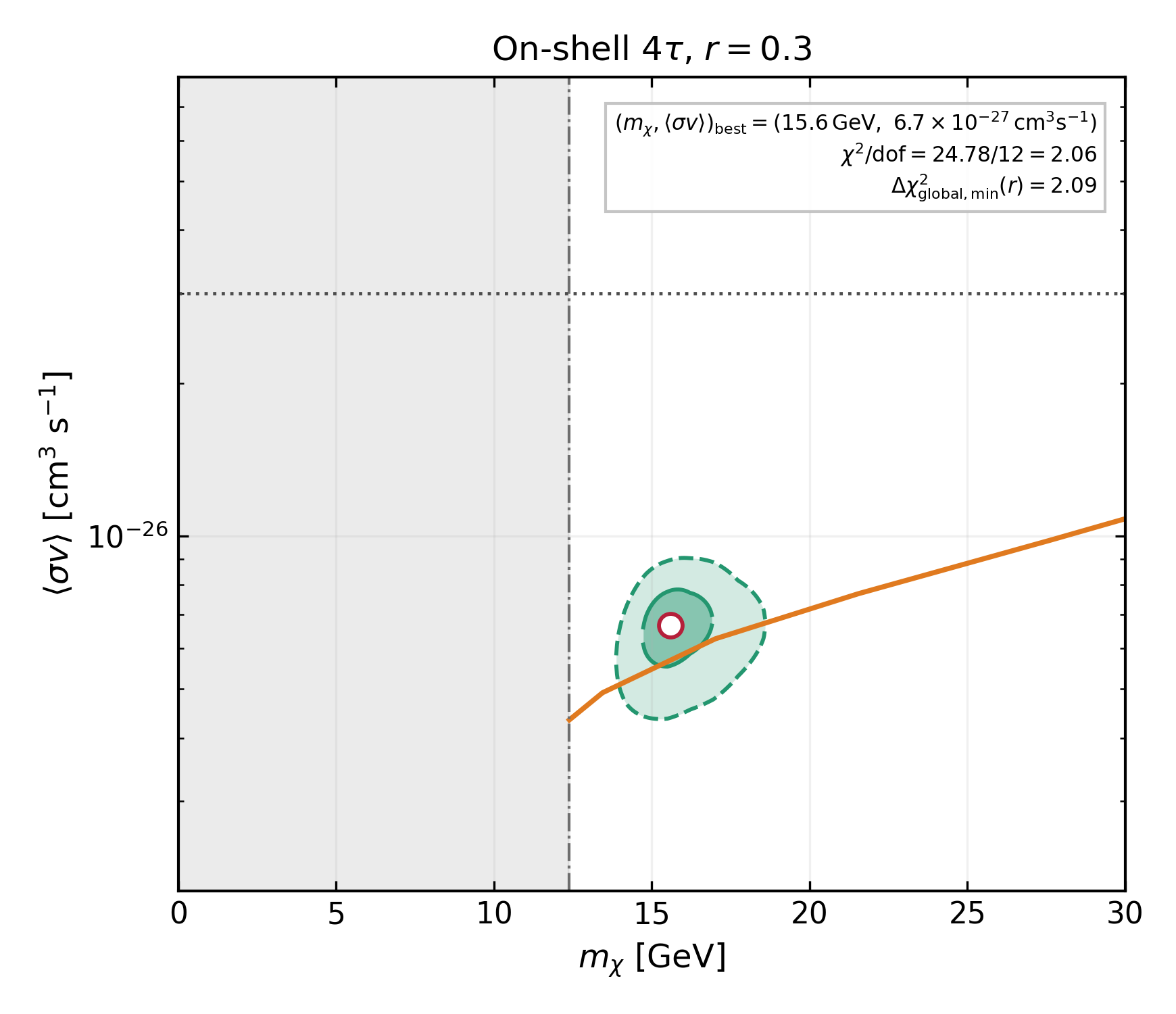}\hfill%
  \includegraphics[width=0.32\textwidth]{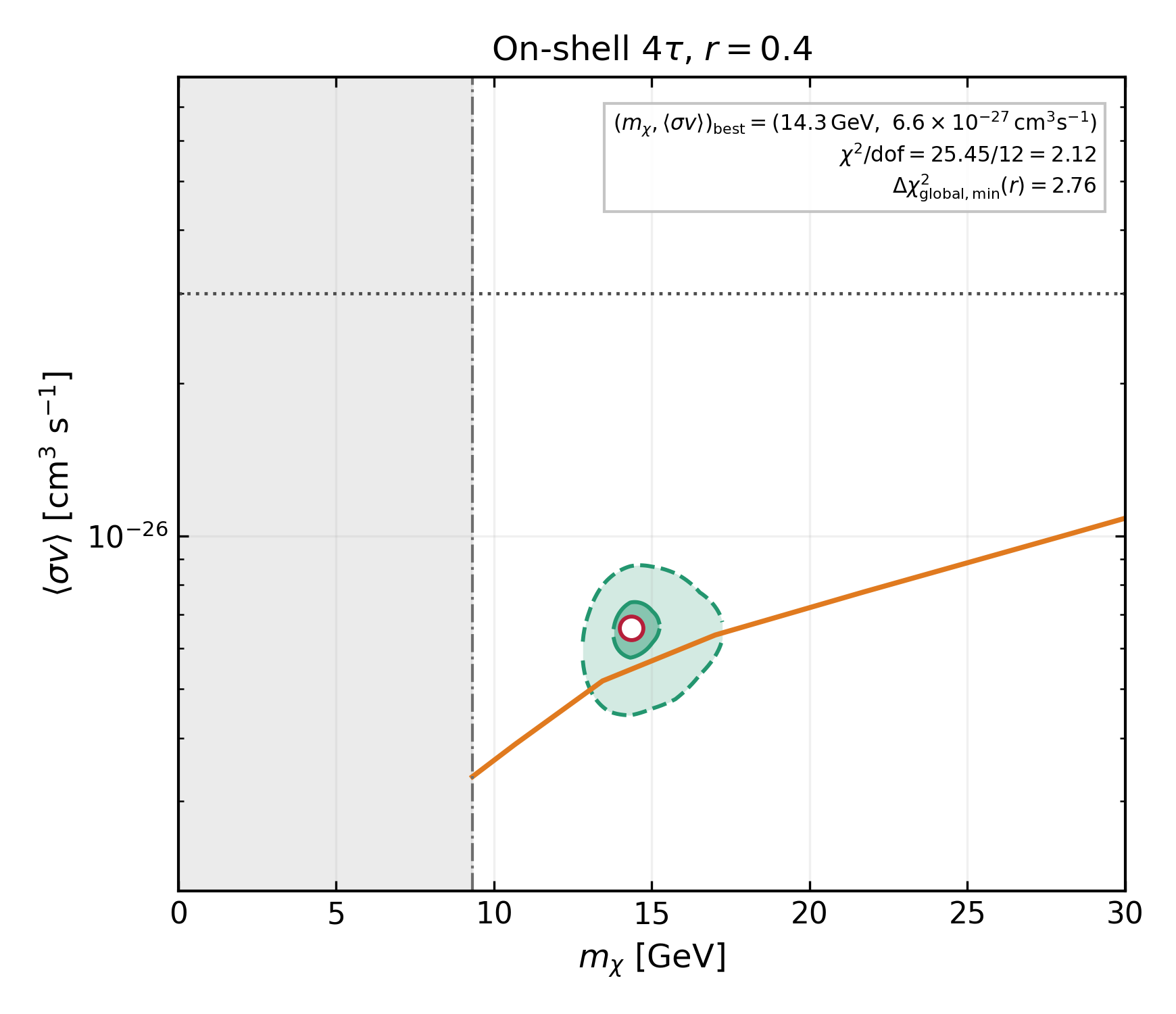}
  \\[2pt]
  \includegraphics[width=0.32\textwidth]{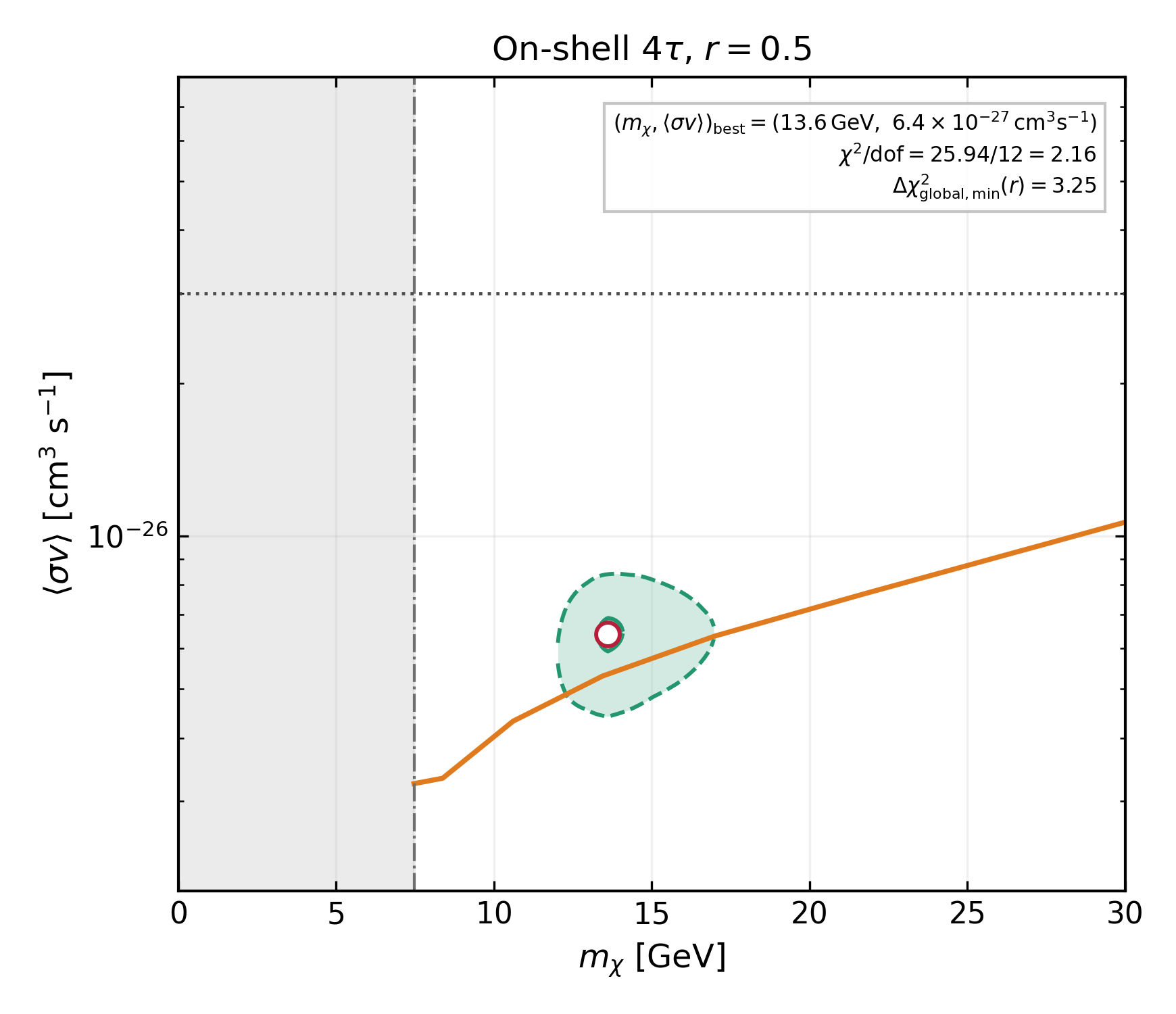}\hfill%
  \includegraphics[width=0.32\textwidth]{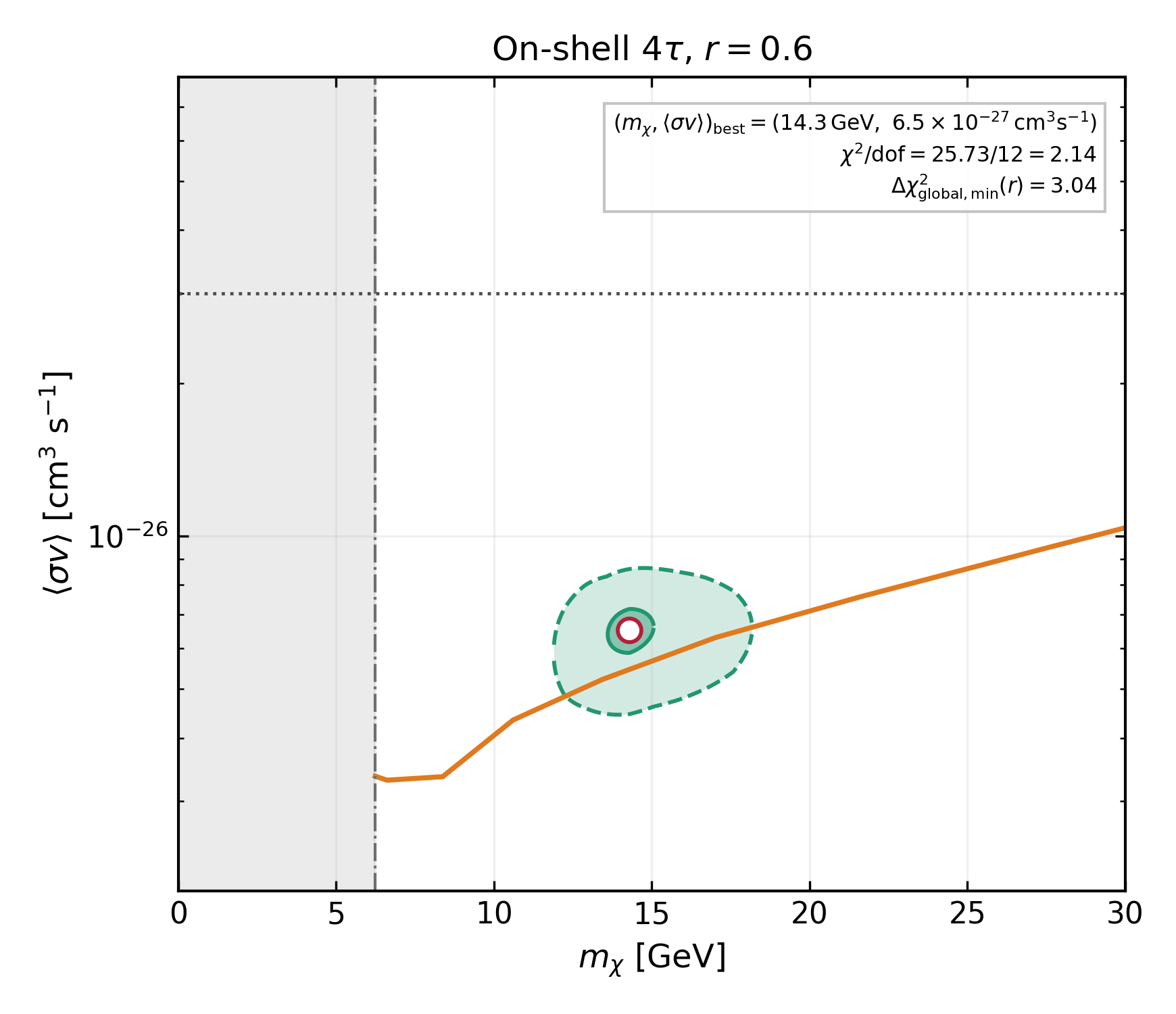}\hfill%
  \includegraphics[width=0.32\textwidth]{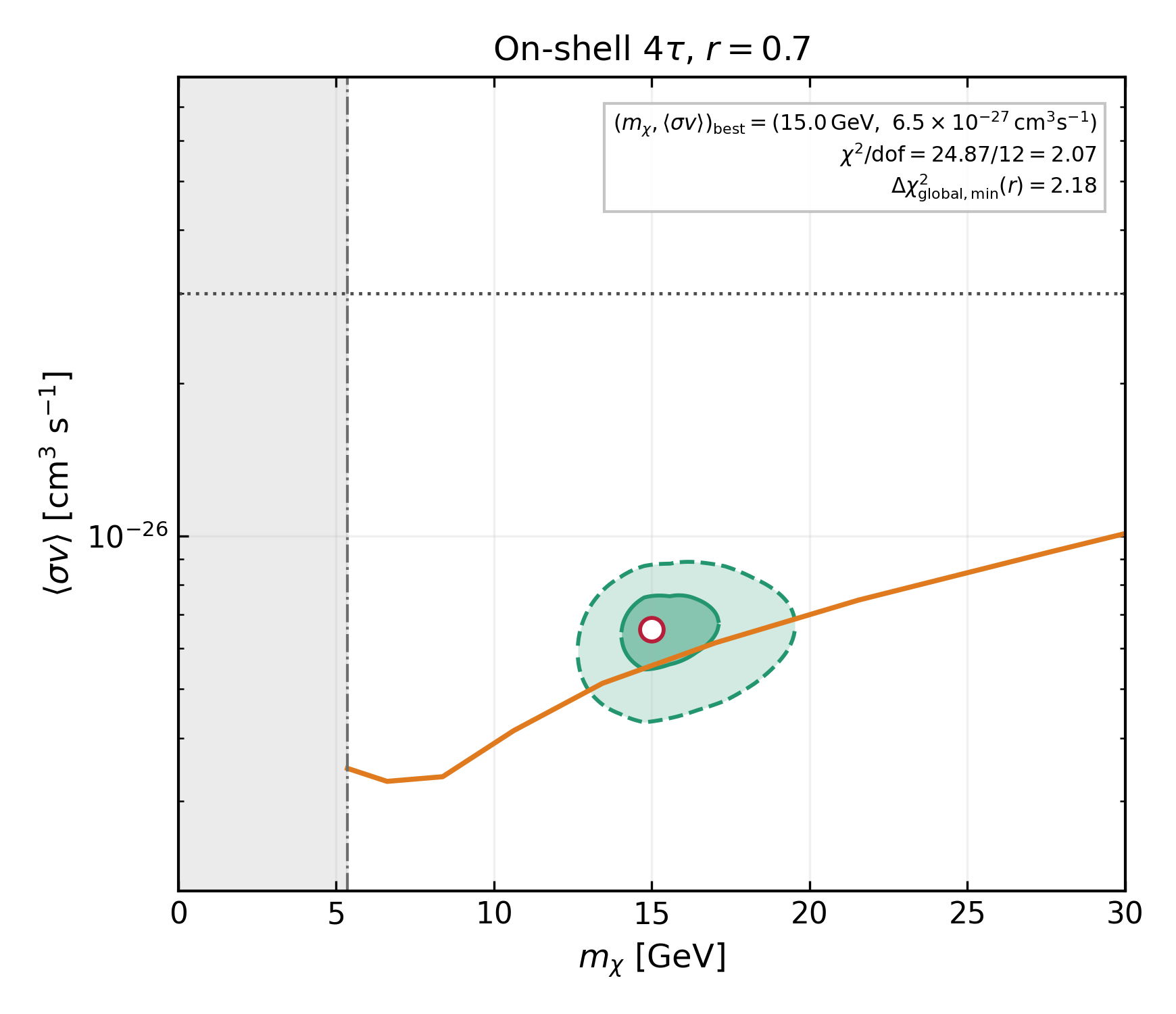}
  \\[2pt]
  \includegraphics[width=0.32\textwidth]{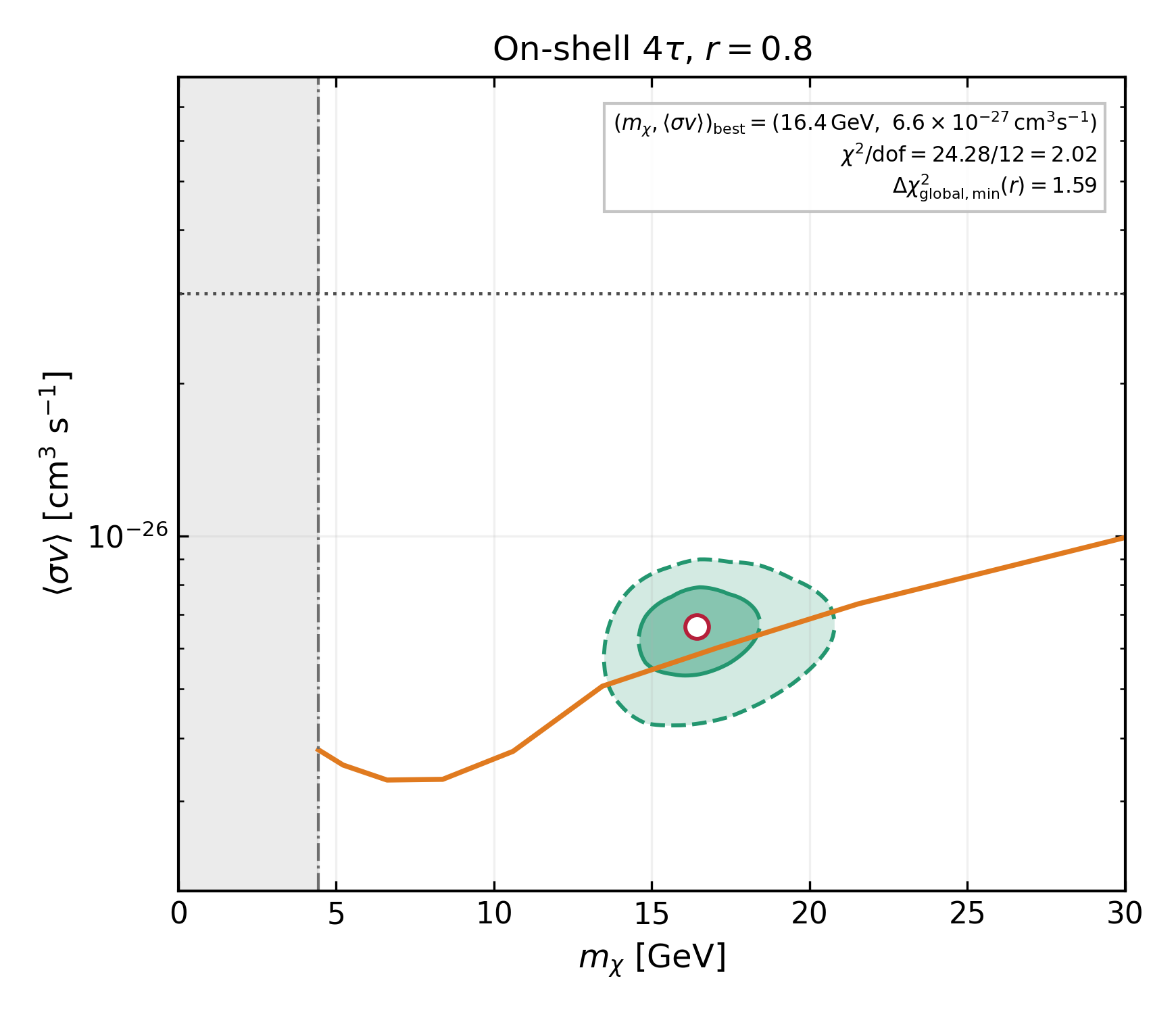}\hfill%
  \includegraphics[width=0.32\textwidth]{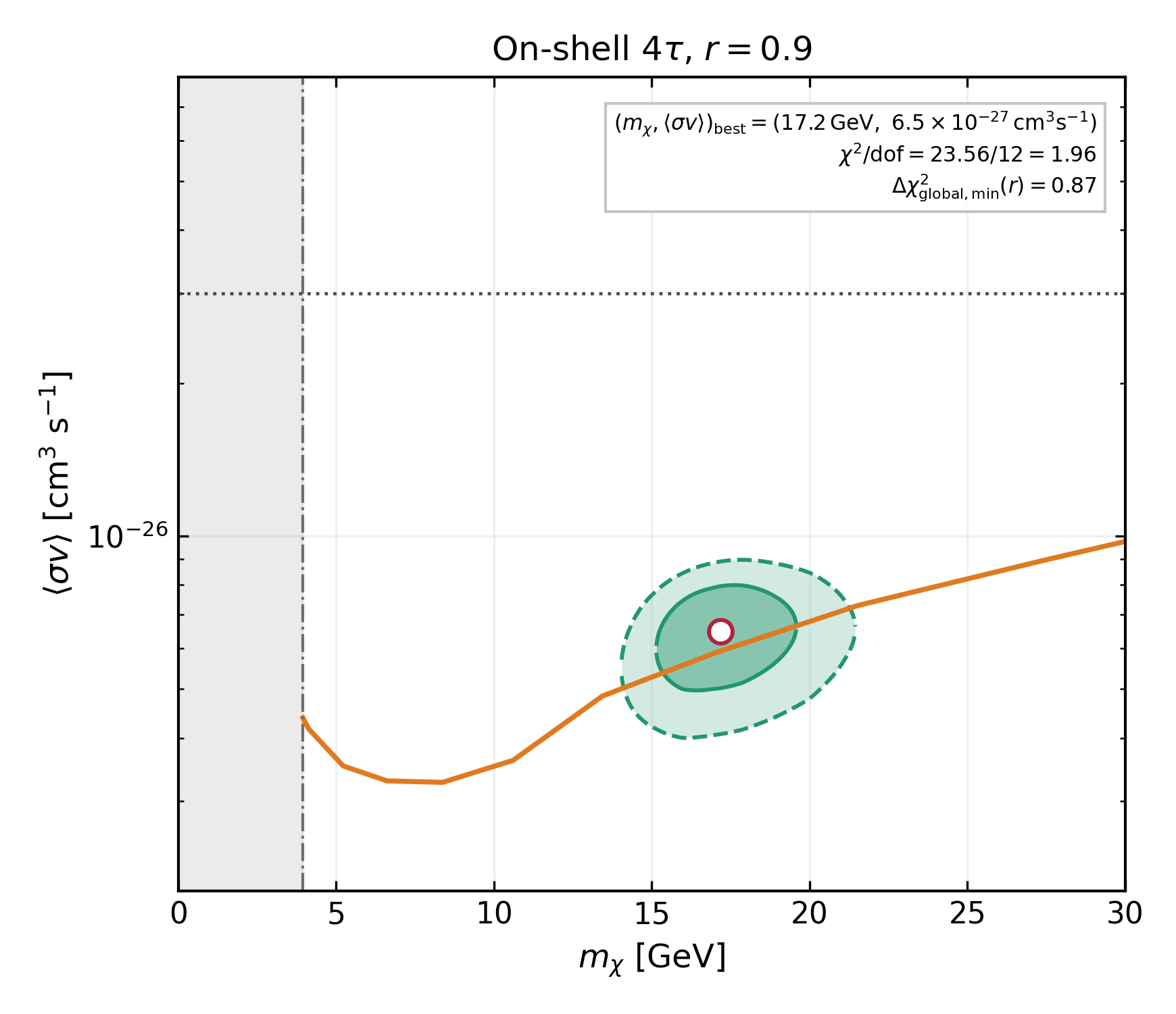}\hfill%
  \includegraphics[width=0.32\textwidth]{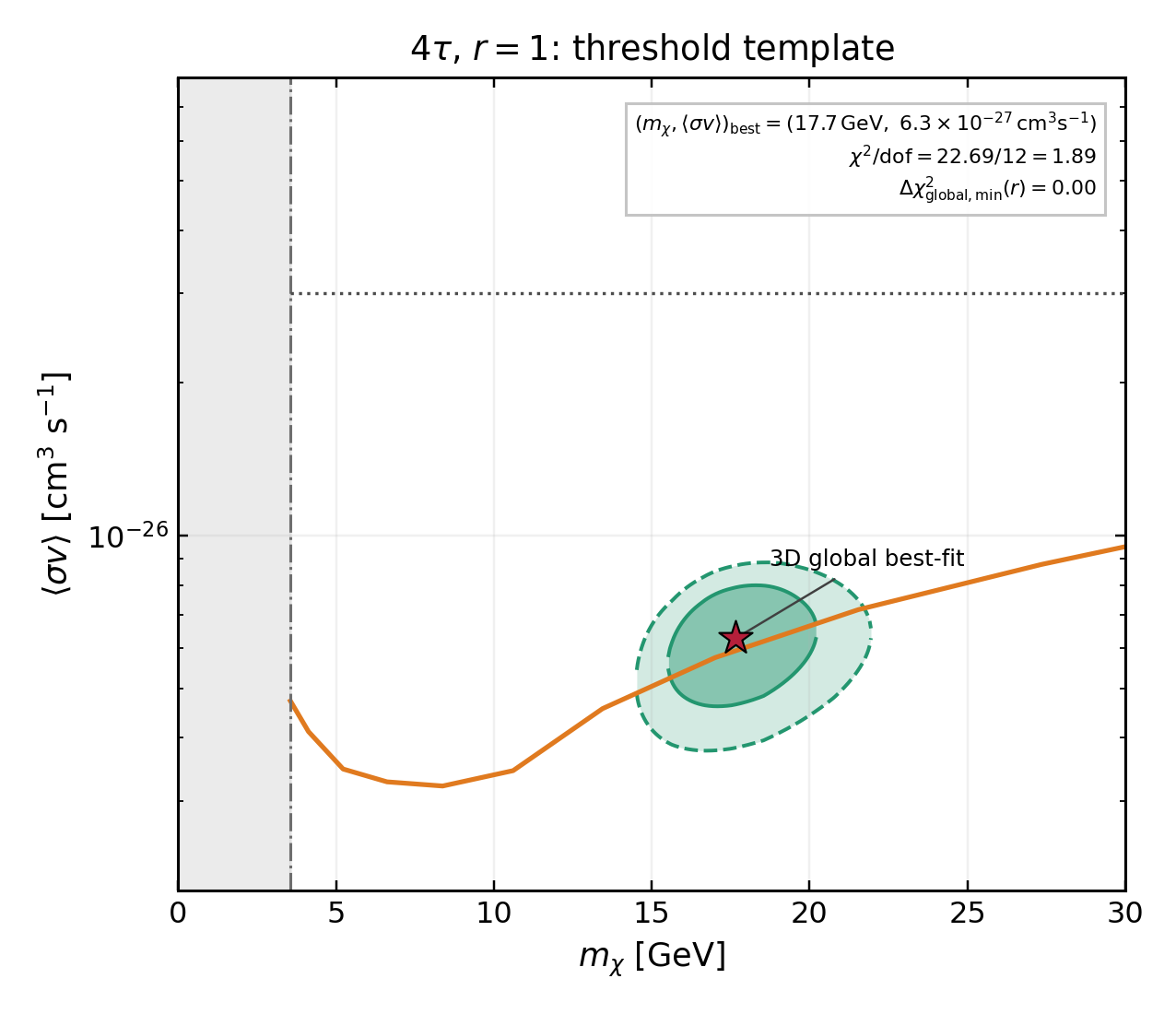}
  \\[4pt]
  \includegraphics[width=0.66\textwidth]{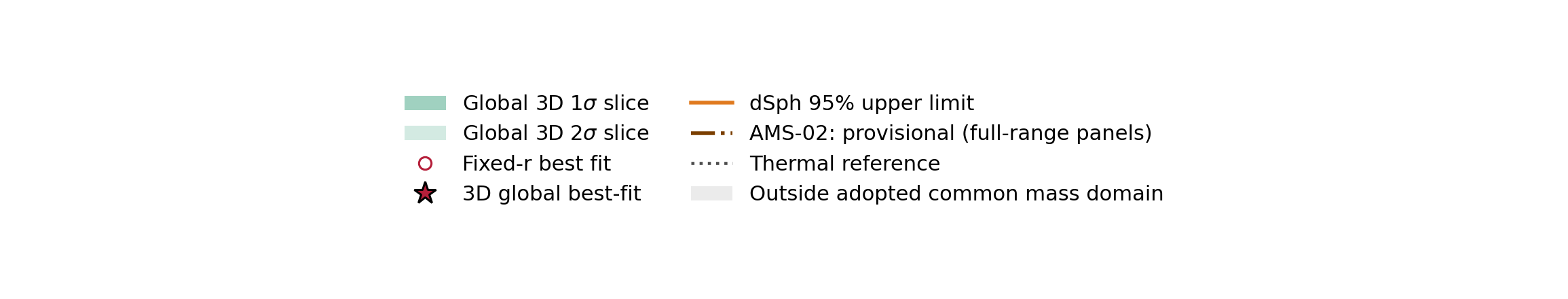}
  \caption{\justifying
  Same as Fig.~\ref{fig:msv_per_r_4b}, for the $4\tau$ on-shell cascade. The
antiproton curve is omitted because it does not constrain the displayed GCE-preferred
region in the adopted calculation. The $r=1$ panel uses the at-rest-mediator limiting
spectrum.
  }
  \label{fig:msv_per_r_4tau}
\end{figure}

\begin{figure}[htbp]
  \centering
  \includegraphics[width=0.32\linewidth]{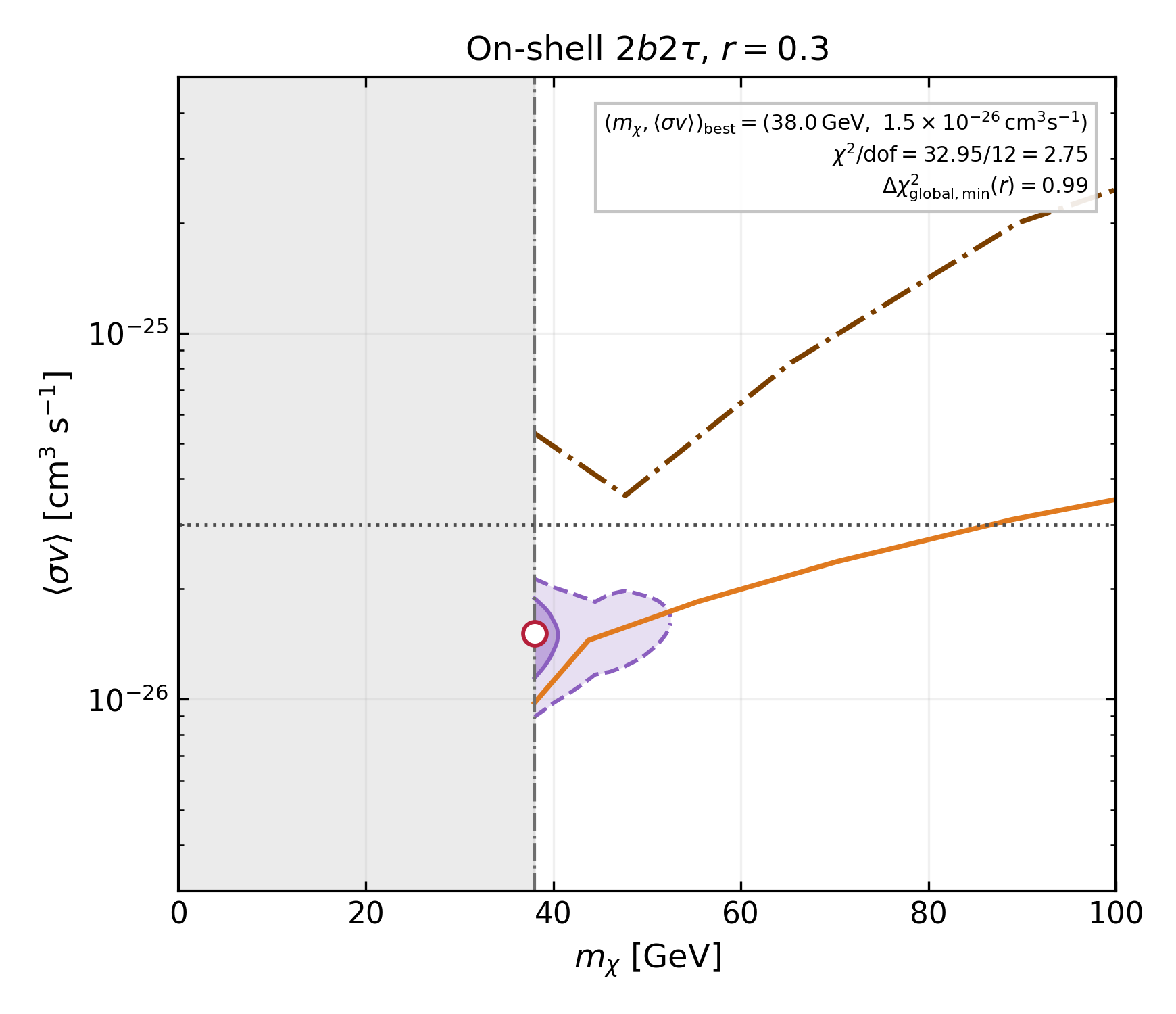}\hfill
  \includegraphics[width=0.32\linewidth]{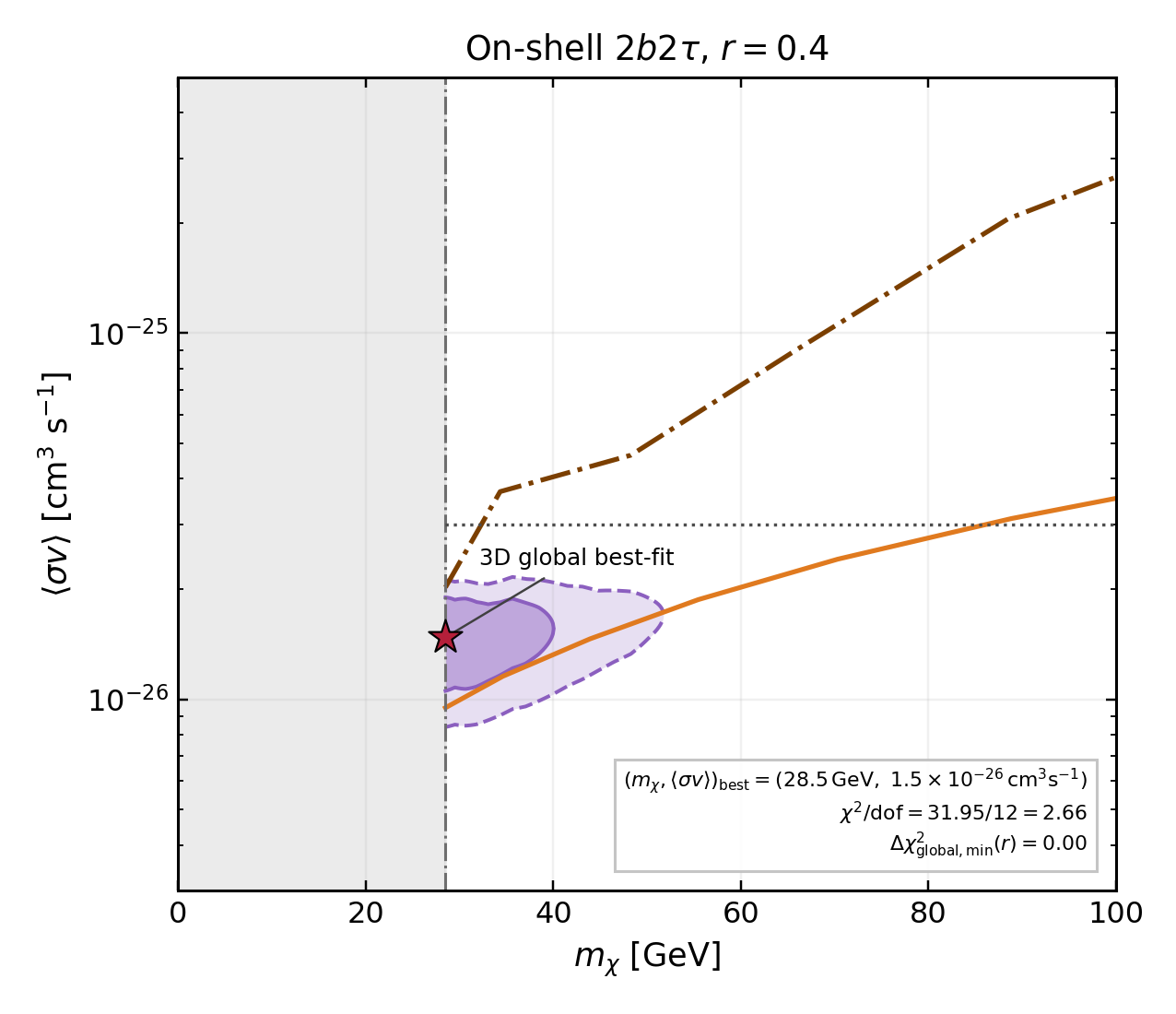}\hfill
  \includegraphics[width=0.32\linewidth]{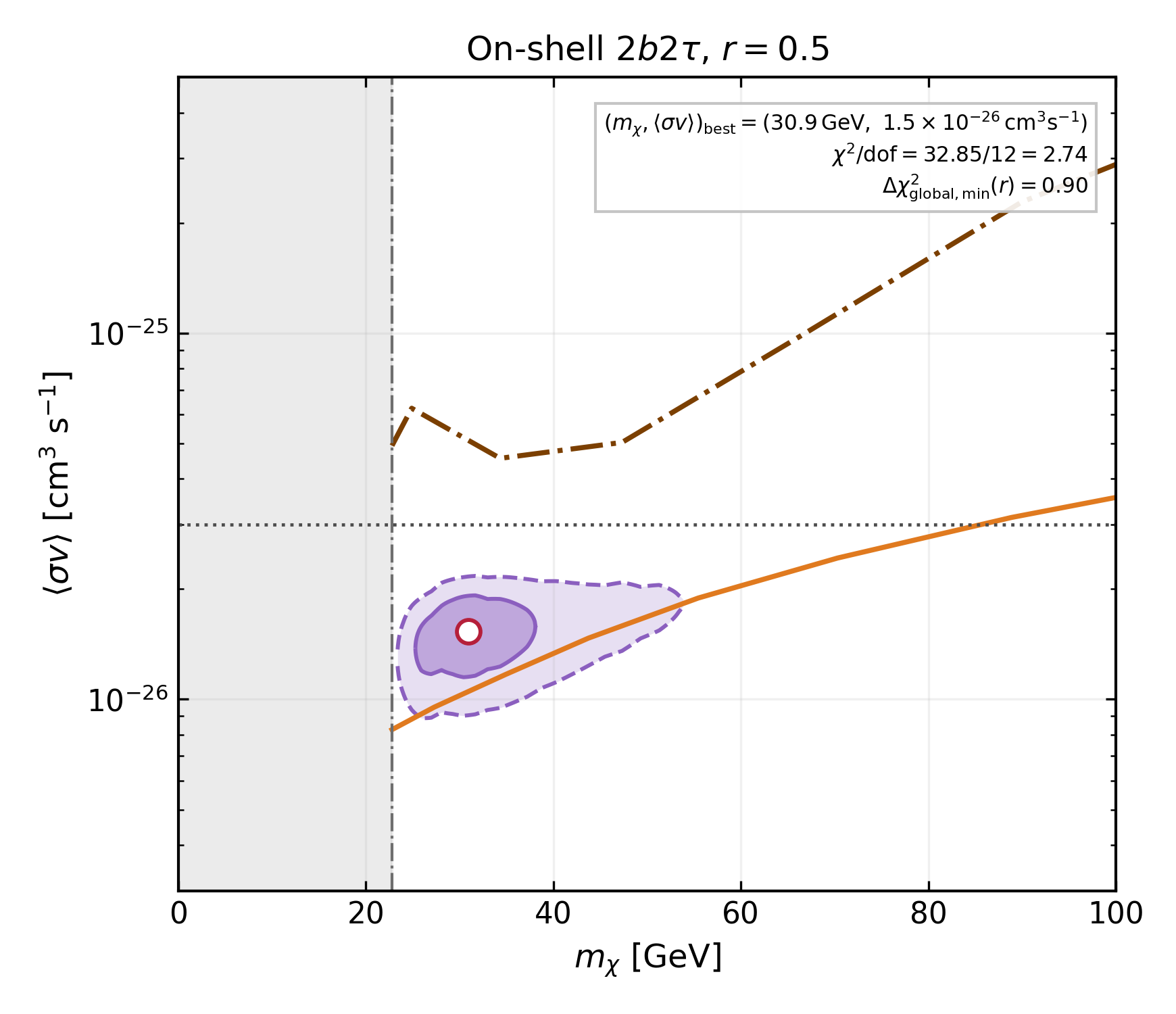}
  \par\vspace{2pt}
  \includegraphics[width=0.32\linewidth]{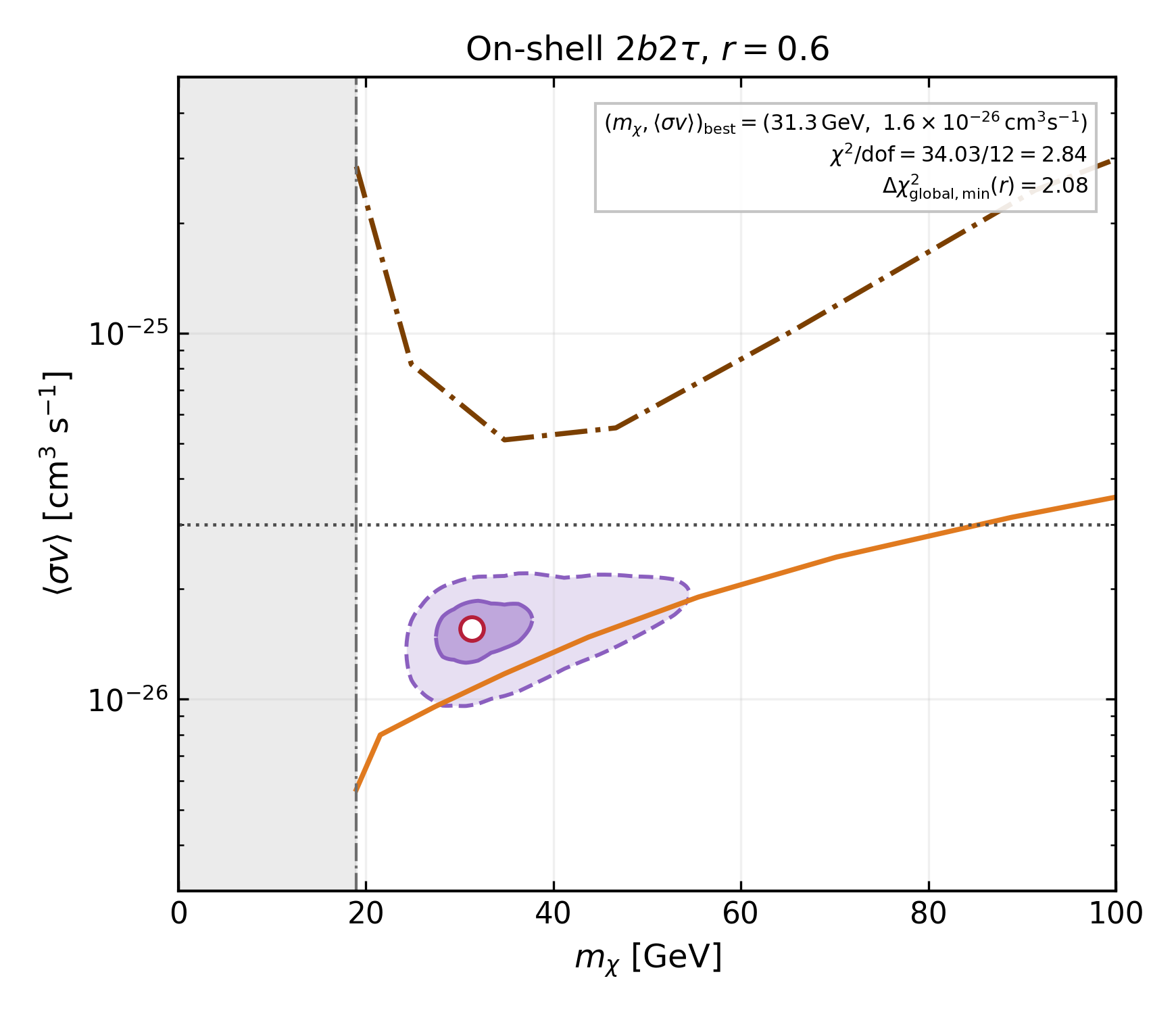}\hfill
  \includegraphics[width=0.32\linewidth]{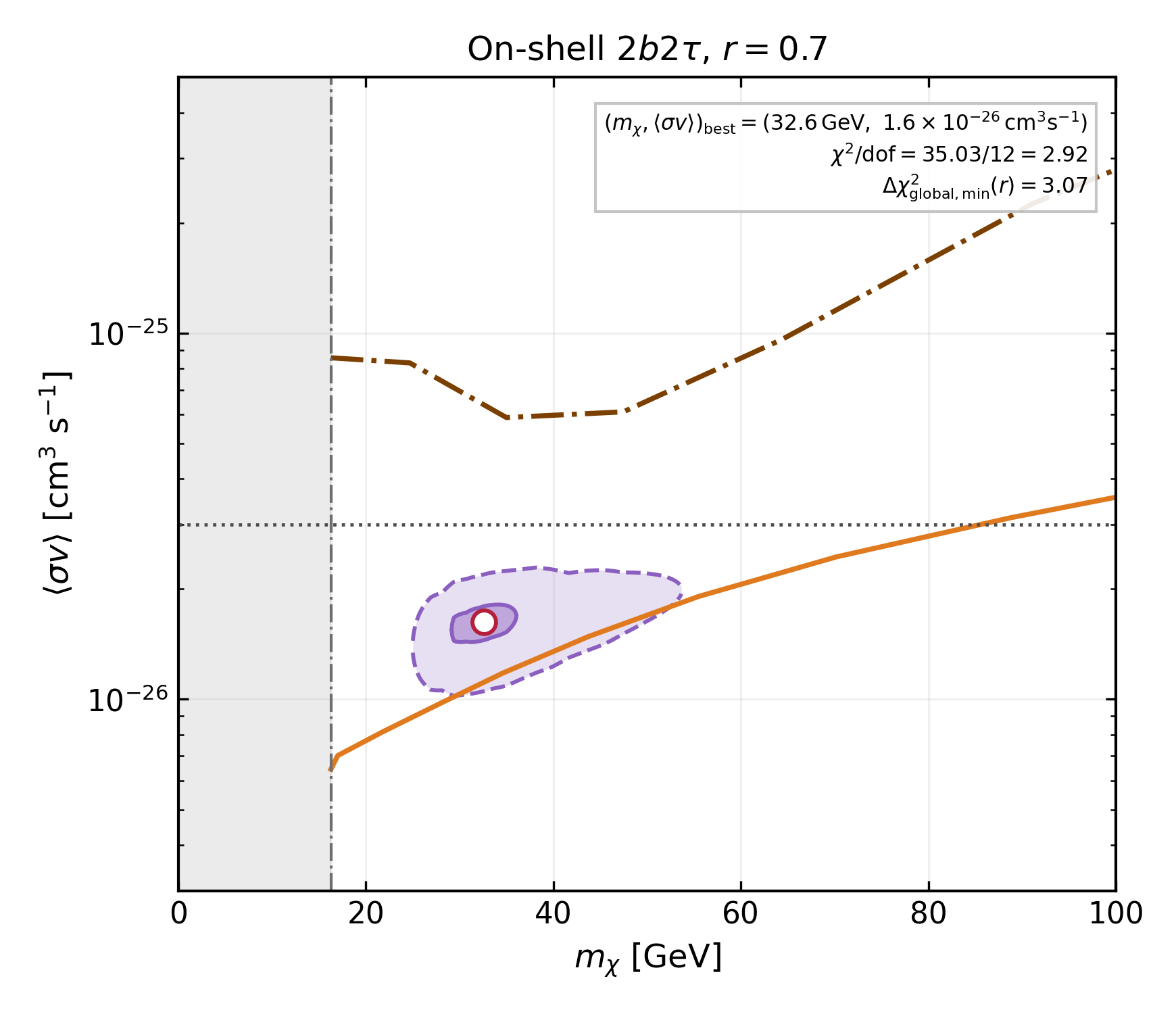}\hfill
  \includegraphics[width=0.32\linewidth]{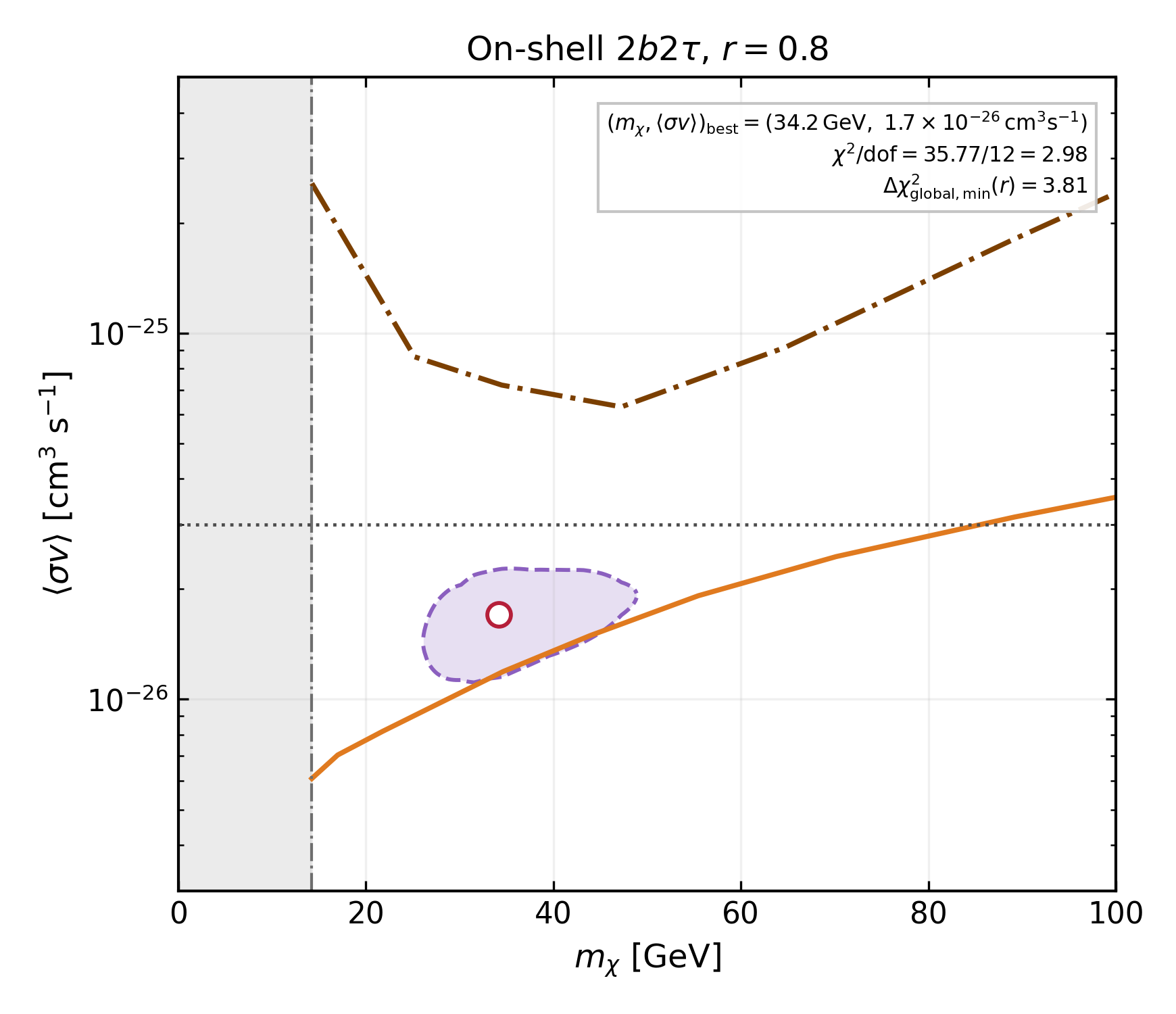}
  \par\vspace{2pt}
  \makebox[\linewidth][l]{%
  \includegraphics[width=0.32\linewidth]{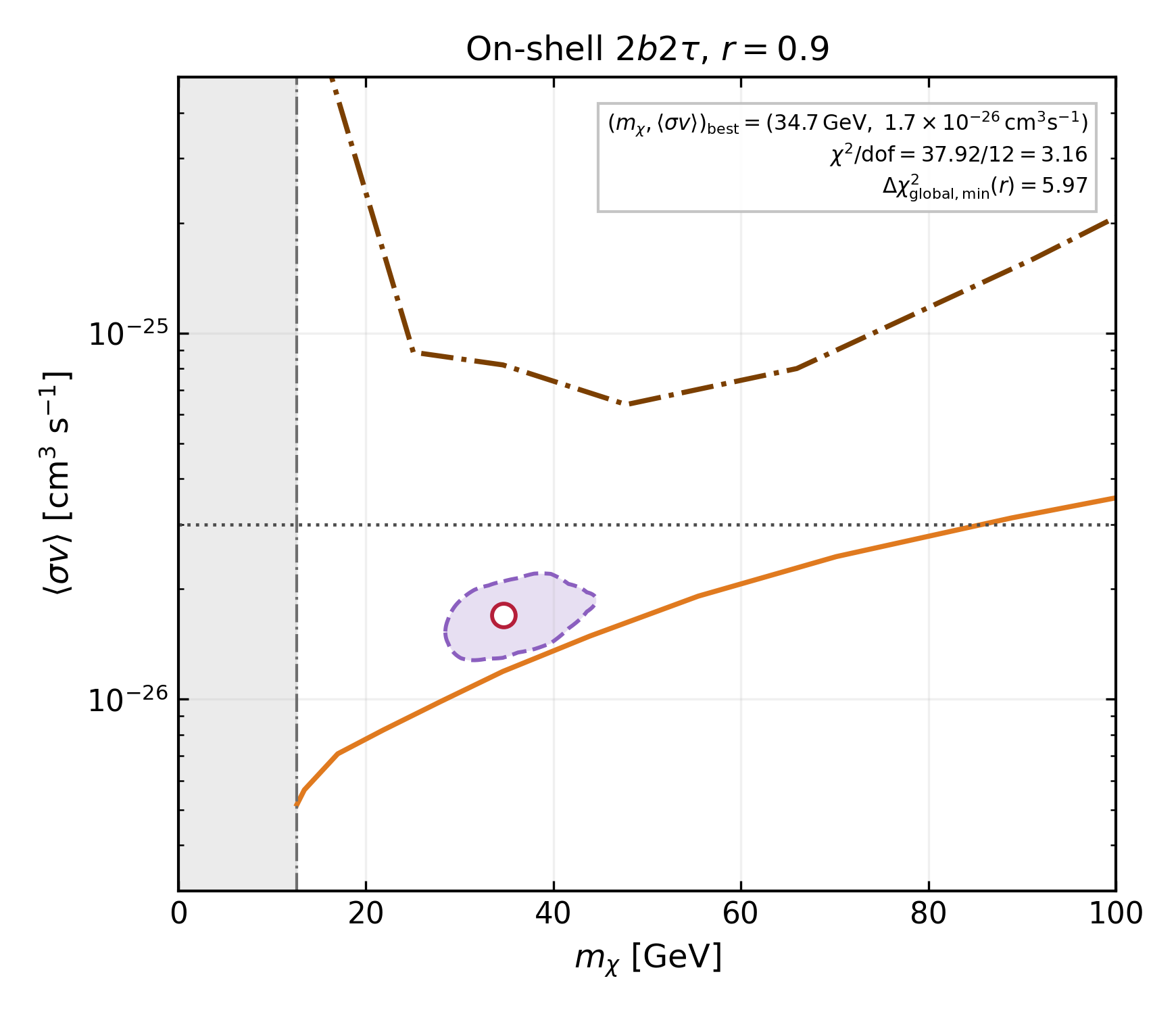}%
  \hfill
  \raisebox{15mm}{%
    \includegraphics[width=0.64\linewidth]{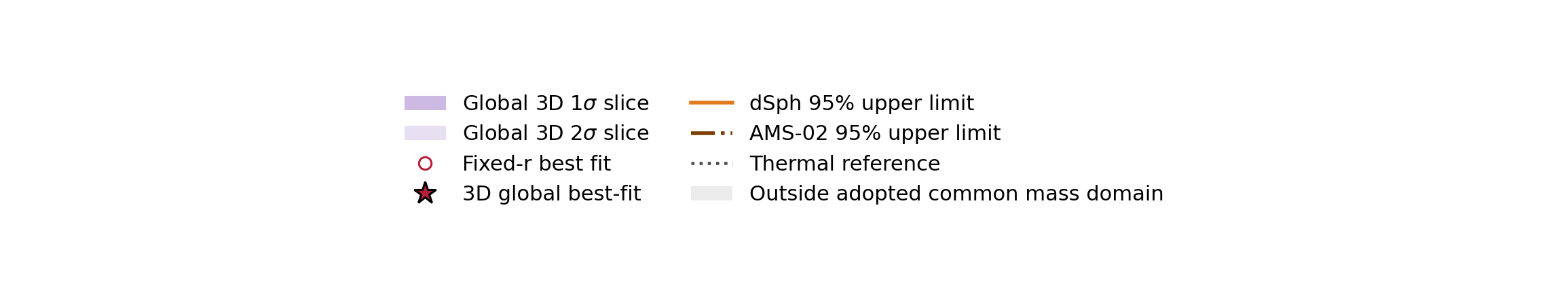}%
  }%
}
  \caption{\justifying
  Same as Fig.~\ref{fig:msv_per_r_4b}, for
  the $2b2\tau$ on-shell cascade, including the antiproton limit.}
  \label{fig:msv_per_r_2b2tau}
\end{figure}

Figures~\ref{fig:msv_per_r_4b}--\ref{fig:msv_per_r_2b2tau} show fixed-$r$ slices of the joint three-parameter GCE-preferred regions in the $(m_\chi,\langle\sigma v\rangle)$ plane.
Each panel gives the best-fit mass and annihilation cross section at the specified mediator-to-DM mass ratio $r$, the corresponding $\chi^2/\mathrm{dof}$, and the difference $\Delta\chi^2$ between the fixed-$r$ and global minimum
$\chi^2$ values.
The combined 95\% CL upper limit from dSphs is shown at the same $r$.
For the $4b$ and $2b2\tau$ channels, the corresponding
AMS-02 antiproton limit is also shown.
No antiproton limit is displayed for $4\tau$, whose antiproton yield is negligible in the mass range considered.

The fits at $r=0.1$ are substantially poorer than the
global best fits, with $\Delta\chi^2=27.3$, $28.6$, and
$15.7$ for $4b$, $4\tau$, and $2b2\tau$, respectively.
These cases are not shown.
All displayed panels within each figure use common axis ranges.


\section{A singlet fermionic dark matter benchmark for the GCE}
\label{sec:SFDM}

In this section, we apply our updated analysis to the singlet
fermionic dark matter model, a representative Higgs-portal
WIMP scenario~\cite{Kim:2008pp,Kim:2009ke}.
The DM candidate is a Dirac singlet fermion $\psi$, whose
interactions with SM particles arise through mixing between
the SM Higgs field and a real singlet scalar.
We revisit the secluded SFDM scenario with a pseudoscalar
DM coupling considered in Ref.~\cite{Kim:2016csm}.
This model realizes the cascade annihilation
$\psi \bar \psi \to h_2 h_2$, followed by decays of the singlet-like scalar $h_2$ into SM particles.
Therefore, it provides a specific framework for examining the predominantly $4b$ interpretation discussed in the previous section, including contributions from other
mediator decay modes.

\subsection{Model details}

The dark sector contains a real scalar field $S$ and a
Dirac fermion $\psi$, both of which are SM gauge singlets.
Following Ref.~\cite{Kim:2016csm}, we include both scalar
and pseudoscalar Yukawa interactions.
The dark sector Lagrangian is
\begin{align}
     \mathcal L^{\rm dark} = \bar\psi (i \slashed{\partial}
   - m_{\psi_0}) \psi + \frac{1}{2} \partial_\mu S \partial^\mu S  - g_S
  (\cos\theta \,\bar\psi \psi + \sin\theta \,\bar\psi i \gamma^5 \psi) S
  - V_S (S, \, H)\,
  \label{eq:SFDMlagrangian}
\end{align}
where the dark sector scalar potential is
\begin{align}
  V_S (S, \, H) = \frac12 m_0^2 S^2 + \lambda_1 H^\dagger H S +
  \lambda_2 H^\dagger H S^2 + \frac{\lambda_3}{3!} S^3 +
  \frac{\lambda_4}{4!} S^4 \,.
\end{align}

After electroweak symmetry breaking, we write $H^0=(v_h+h)/\sqrt{2}$ and $S=v_s+s$.
The scalar mass eigenstates are defined by
\begin{align}
h_1 &= h\cos\theta_s + s\sin\theta_s, \\
h_2 &= -h\sin\theta_s + s\cos\theta_s,
\end{align}
where $h_1$ is the SM-like Higgs boson and $h_2$ is
the singlet-like scalar.

In the resulting basis after chiral rotations making the DM mass $m_\psi$ real and positive, the DM interaction is
\begin{align}
  -\mathcal{L}^{\rm dark}_\mathrm{int} = g_S \cos\xi \,s \bar\psi \psi + g_S \sin\xi \,s
  \bar\psi i \gamma^5 \psi \,,
\end{align}
where the angle $\xi$ determines the dark scalar and pseudoscalar interaction ratio and is obtained from the Lagrangian parameters in Eq.~(\ref{eq:SFDMlagrangian}).

\subsection{An SFDM benchmark for the four-bottom cascade}

We consider a benchmark dominated by
$\psi\bar\psi\to h_2h_2$, with the mediator-to-DM mass ratio fixed to $r=m_{h_2}/m_\psi=0.3$.
The benchmark masses are $m_\psi=47.75$~GeV and $m_{h_2}=14.325$~GeV.
We choose a purely pseudoscalar DM coupling, $\xi=\pi/2$, with $g_S=0.458$.
The scalar trilinear interactions allow an $s$-wave annihilation contribution through $s$-channel scalar exchange~\cite{Kim:2016csm}.
For this benchmark, the present-day pair annihilation cross section is
\begin{align}
\langle\sigma v\rangle_{\psi\bar\psi}
=
2.194\times10^{-26}\,
\mathrm{cm^3\,s^{-1}}.
\end{align}
The relic density calculation gives
$\Omega_\psi h^2=0.12$, with the freeze-out parameter $x_f\equiv m_\psi/T_f=24.01$.
The relic density and annihilation spectra are calculated using \texttt{micrOMEGAs 7.1.4}~\cite{Belanger:2026asz}.

Since $h_2$ couples to SM particles through Higgs mixing, its partial widths into SM final states are suppressed by $\sin^2\theta_s$ relative to those of an SM Higgs boson with the same mass.
The fermionic partial widths also depend on the squared SM Yukawa couplings and the available phase space.
For the benchmark, the calculated branching fractions are
\begin{align}
\mathrm{Br}(h_2\to b\bar b) &= 0.7910, \\
\mathrm{Br}(h_2\to c\bar c) &= 0.1281, \\
\mathrm{Br}(h_2\to\tau^+\tau^-) &= 0.0797.
\end{align}
Consequently, the $4b$ final state accounts for approximately
$62.6\%$ of the $h_2h_2$ decays.
The remaining decay combinations, including $2b2c$ and
$2b2\tau$, also contribute to the photon and antiproton spectra.

The benchmark parameters satisfy the imposed tree-level conditions on the full scalar potential: boundedness from below, stationarity conditions, positive scalar squared
masses, a local minimum at the electroweak vacuum, scalar
unitarity, and perturbativity.
The electroweak vacuum is the unique physical global minimum of the tree-level potential.
These checks do not include renormalization-group evolution,
loop corrections, or finite-temperature effects.

\begin{figure}[h]
 \centering \includegraphics[width=0.345\textheight]{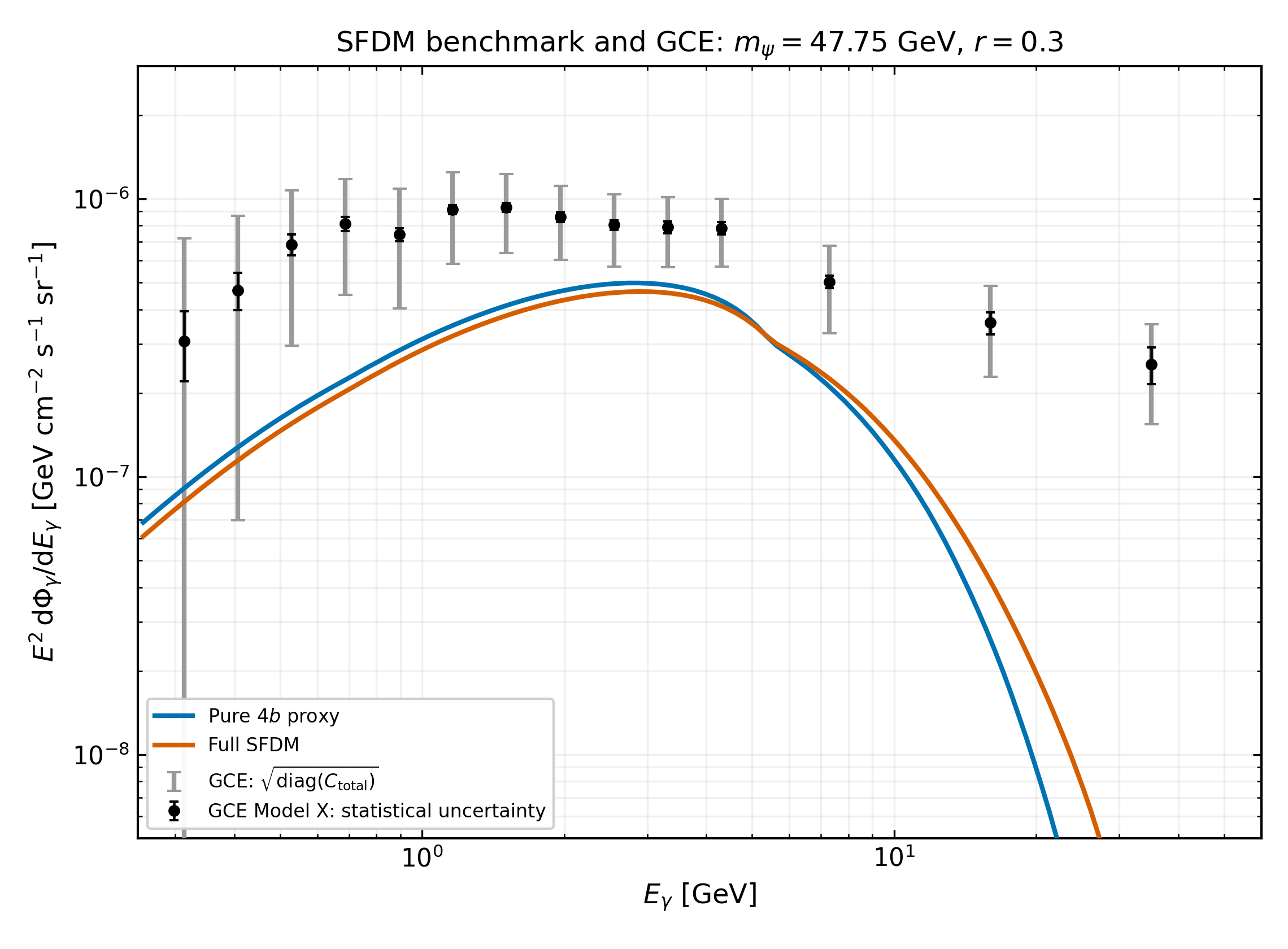} \includegraphics[width=0.3\textheight]{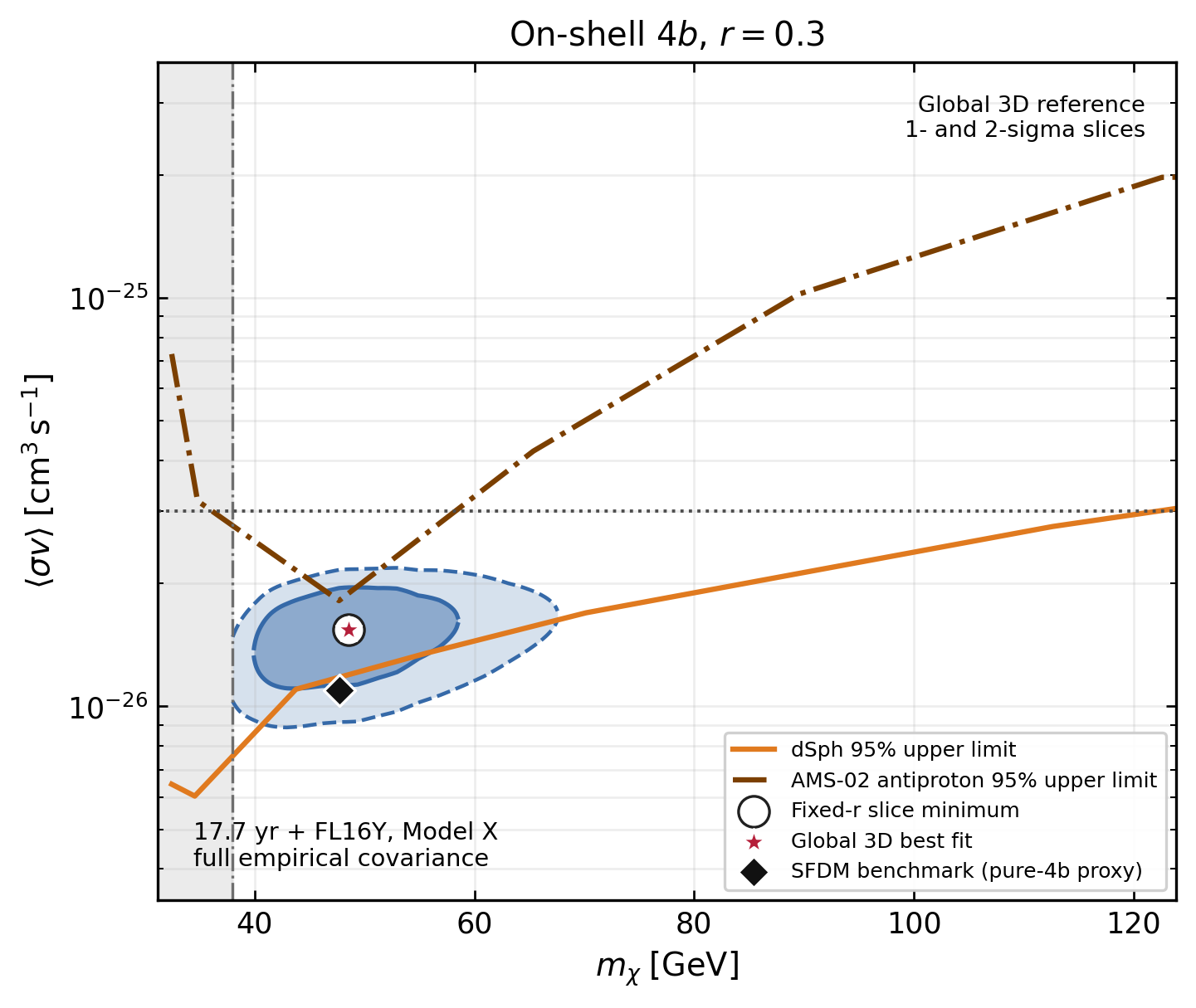}
 \caption{\justifying 
 SFDM benchmark and GCE fit.
\emph{Left}: SFDM photon spectrum (orange, rescaled by $1/2$)
and pure-$4b$ spectrum at $r=0.3$ (blue).
The fixed benchmark gives $\chi^2=28.05$ across 14 GCE energy bins.
\emph{Right}: Fixed-$r=0.3$ slice of the joint three-parameter
GCE confidence volume for pure $4b$ annihilation.
The black diamond marks the SFDM benchmark, and the dotted
line denotes the conventional thermal relic reference.}
 \label{fig:sfdm-benchmark}
\end{figure}

For the indirect detection calculation, we assume equal
abundances of $\psi$ and $\bar\psi$, which together account
for the total DM density.
Note that for a Dirac fermion and symmetric DM, we need to divide $\langle \sigma v \rangle $ by factor of 2 to match the convention of predicted flux of self-conjugate DM.
Figure~\ref{fig:sfdm-benchmark} compares the benchmark prediction with the GCE spectrum and the pure-$4b$ cascade results.
In the left panel, the full SFDM prediction and the pure $4b$ template have similar spectral shapes over $0.3$--$3$~GeV, while their differences become more apparent at higher energies. 
The fixed benchmark gives a total $\chi^2=28.05$ across the 14 GCE energy bins when evaluated with the full covariance matrix; no parameters are refitted in this evaluation. Although the photon spectra of the SFDM and the pure $4b$ do not perfectly match with the black dots and the bars of the GCE, the values of $\chi^2$ can be small enough taking into account the off-diagonal covariance matrix elements.

The right panel shows the fixed-$r=0.3$ slice of the joint three-parameter GCE confidence volume for the pure $4b$ cascade.
The SFDM benchmark is plotted at
$(m_\psi,\langle\sigma v\rangle_{\rm eff})$.
It lies near the lower boundary of the $1\sigma$ slice and below the displayed pure $4b$ limits from dSphs
and AMS-02 antiproton measurements.
This comparison is approximate because the SFDM spectrum contains several final states.
A quantitative assessment of the benchmark requires evaluating the dSph and antiproton likelihoods with
the full SFDM spectra.
For $\xi=\pi/2$, the tree-level DM--nucleon scattering amplitude through the $h_2$ exchange is suppressed by the momentum transfer and the mixing angle $\theta_s = 0.0076$, giving a negligible cross section in direct detection experiments.

The benchmark also predicts invisible and exotic decays
of the SM-like Higgs boson.
The calculated invisible branching fraction, $\mathrm{Br}(h_1 \to \psi \bar \psi)=0.006237$,
is below the quoted ATLAS 95\% CL limit $\mathrm{Br}_{\rm inv} < 0.107$~\cite{ATLAS:2023tkt}.
Similarly, the calculated branching fraction $\mathrm{Br}(h_1 \to h_2 h_2)=0.01936$ is below the limit
$\mathrm{Br}_{\rm und}<0.12$ from the ATLAS Higgs coupling fit, under the assumptions of that analysis~\cite{ATLAS:2022vkf}.
The latter is a bound on undetected Higgs decays, rather than a dedicated bound on the $4b$ final state.

Higgs coupling measurements, exotic Higgs decays, and direct production of $h_2$ provide further collider
tests of this scenario~\cite{Kim:2018uov}.
We leave a dedicated study of the sensitivity of future lepton colliders to this benchmark for future work.
The benchmark parameters and calculated observables are summarized in Table~\ref{tab:sfdm-r03-benchmark-paper}.

\begin{table}[h]
 \centering
 \small
 \begin{tabular}{lll}
 \hline
 Quantity & Value & Unit \\
 \hline
 $m_\psi$ & $47.75$ & GeV \\
 $m_{h_1}$ & $125.0$ & GeV \\
 $m_{h_2}$ & $14.325$ & GeV \\
 $r$ & $0.3000$ & --- \\
 $\lambda_0$ & $0.1289$ & --- \\
 $\lambda_1$ & $-0.1781$ & GeV \\
 $\lambda_2$ & $0.002389$ & --- \\
 $\lambda_3$ & $-10.09$ & GeV \\
 $\lambda_4$ & $0.1375$ & --- \\
 $g_S$ & $0.4580$ & --- \\
 $v_s$ & $136.7$ & GeV \\
 $\sin\theta_s$ & $0.007584$ & --- \\
 $c_{222}$ & $8.675$ & GeV \\
 $\xi$ & $\pi/2$ & rad \\
 $\Omega_\psi h^2$ & $0.1200$ & --- \\
 $x_f$ & $24.01$ & --- \\
 $\langle\sigma v\rangle$ & $2.194\times10^{-26}$ & $\mathrm{cm^3\,s^{-1}}$ \\
 $\mathrm{BR}(h_1\to\psi\bar\psi)$ & $0.006237$ & --- \\
 $\mathrm{BR}(h_1\to h_2h_2)$ & $0.01936$ & --- \\
 $\mathrm{BR}(h_2\to b\bar b)$ & $0.7910$ & --- \\
 Invisible $95\%$ CL bound & $\mathrm{BR}_{\mathrm{inv}}<0.107$ & Ref.~\cite{ATLAS:2023tkt} \\
 Undetected $95\%$ CL bound & $\mathrm{BR}_{\mathrm{und}}<0.12$ & Ref.~\cite{ATLAS:2022vkf} \\
 \hline
 \end{tabular}
 \caption{\justifying
Input parameters and derived observables for the SFDM benchmark.
The scalar trilinear coupling is defined by $V\supset c_{222}h_2^3/3!$, and
$x_f=m_\psi/T_f$.
The quoted experimental limits apply under the assumptions of the cited analyses.
 }
 \label{tab:sfdm-r03-benchmark-paper}
\end{table}

\section{Conclusions}
\label{sec:conclusions}

In this work, we investigated DM interpretations of the GCE using 17.7 years of Fermi-LAT data and the FL16Y source catalog.
We compared the GCE-preferred parameter regions with constraints from dSph gamma-ray observations and the AMS-02 antiproton flux.
Using the same observation interval, we derived updated combined 95\% CL upper limits from 30 dSphs with stellar-kinematic $J$-factor estimates, including their uncertainties.

For direct annihilation into two-body final states, the 68\% CL GCE-preferred regions for $b\bar b$, $c\bar c$, $q\bar q$, and $gg$ lie above the combined 95\% CL dSph limits.
Portions of the preferred regions for $\tau^+\tau^-$, $W^+W^-$, $ZZ$, $t\bar t$, and $hh$ remain below these limits.
However, the best-fit point for $\tau^+\tau^-$ is excluded, while the heavy final states provide substantially poorer fits to the GCE spectrum.
Compatibility with the dSph limits therefore does not by itself establish a successful GCE interpretation.
In our adopted antiproton analysis, the AMS-02 bounds do not exclude the GCE-preferred regions for the two-body channels considered here.
Their strength depends on the cosmic-ray propagation parameters and the treatment of uncertainties.

We also examined the effective four-body annihilation channels $4b$, $4\tau$, and $2b2\tau$.
The combined dSph limits exclude the 68\% CL GCE-preferred regions for the effective $4b$ and $2b2\tau$ channels.
Part of the corresponding $4\tau$ region remains compatible with the dSph limits, although its best-fit point is excluded.
With the BIG propagation model and the uncertainty treatment adopted here, the AMS-02 antiproton bounds do not exclude the GCE-preferred regions of these effective four-body channels.

We then investigated cascade annihilation through on-shell
scalar mediators, varying the mediator-to-DM mass ratio $r=m_\phi/m_\chi$.
We used the benchmark decay assignments defined above to
produce the $4b$, $4\tau$, and $2b2\tau$ final states.
The mediator mass changes the photon and antiproton spectra and hence both the GCE-preferred regions and the complementary constraints.
We presented the joint GCE-preferred regions in $(m_\chi,\langle\sigma v\rangle,r)$, their projections, and fixed-$r$ slices compared with limits evaluated at the same mediator mass ratio.

\begin{table}[tb]
\centering
\small
\begin{tabular}{lrrrrl}
\hline
Channel & $r_{\rm best}$ & $m_{\chi,\rm best}$ & $\langle\sigma v\rangle_{\rm best}$ & $\chi^2/{\rm dof}$ & dSphs \\
 & & [GeV] & [cm$^3$ s$^{-1}$] & & \\
\hline
$\tau^+\tau^-$ & -- & 9.3 & $3.2\times10^{-27}$ & $22.82/12=1.90$ & Fail \\
$4\tau$ (effective) & -- & 16.6 & $6.7\times10^{-27}$ & $24.29/12=2.02$ & Fail \\
$4\tau$ (on-shell) & 1.0 & 17.7 & $6.3\times10^{-27}$ & $22.69/11=2.06$ & Fail \\
$b\bar b$ & -- & 39.5 & $1.2\times10^{-26}$ & $27.50/12=2.29$ & Fail \\
$4b$ (effective) & -- & 67.4 & $2.5\times10^{-26}$ & $27.92/12=2.33$ & Fail \\
$c\bar c$ & -- & 28.4 & $7.9\times10^{-27}$ & $28.11/12=2.34$ & Fail \\
$4b$ (on-shell) & 0.3 & 48.5 & $1.5\times10^{-26}$ & $26.26/11=2.39$ & Fail \\
$gg$ & -- & 41.9 & $1.3\times10^{-26}$ & $31.02/12=2.59$ & Fail \\
$q\bar q$ & -- & 17.6 & $5.0\times10^{-27}$ & $31.36/12=2.61$ & Fail \\
$2b2\tau$ (on-shell)$^{\dagger}$ & 0.4 & 28.5 & $1.5\times10^{-26}$ & $31.95/11=2.90$ & Fail \\
$2b2\tau$ (effective) & -- & 35.1 & $1.8\times10^{-26}$ & $35.13/12=2.93$ & Fail \\
$hh$ & -- & 130.0 & $2.9\times10^{-26}$ & $38.64/12=3.22$ & Pass \\
$ZZ$ & -- & 91.2 & $2.6\times10^{-26}$ & $42.32/12=3.53$ & Pass \\
$t\bar t$ & -- & 172.4 & $3.9\times10^{-26}$ & $44.29/12=3.69$ & Pass \\
$W^+W^-$ & -- & 80.4 & $2.1\times10^{-26}$ & $44.34/12=3.69$ & Pass \\
\hline
\end{tabular}
\caption{\justifying
GCE best-fit parameters and dSph 95\% upper-limit tests. Pass means that the unrounded best-fit annihilation rate does not exceed the interpolated dSph upper limit at the same mass and, where applicable, the same mediator ratio. Two-body spectra use CosmiXs; effective and on-shell cascade spectra retain the custom templates using Pythia. The GCE-only degrees of freedom are 12 for two-body/effective fits and 11 for on-shell fits; the latter select $r$ on the computed grid. $\dagger$ denotes a mass/support boundary-limited best fit.}
\label{tab:gce-bestfits-dsph}
\end{table}

On the adopted $r$ grid, the lowest GCE $\chi^2$ values occur at $r=0.3$ for $4b$, $r=1.0$ for $4\tau$, and
$r=0.4$ for $2b2\tau$.
Here $r=1.0$ denotes the zero-boost limit of the cascade spectral templates.
All three best-fit points lie above the corresponding combined dSph limits.
Nevertheless, portions of the GCE-preferred regions for the $4b$ and $4\tau$ cascades remain compatible with these limits.
The antiproton bounds additionally constrain part of the $4b$ cascade region at some mediator mass ratios.
Among the four-body benchmarks considered here, the $4\tau$ channel retains a comparatively large GCE-preferred region consistent with both constraints under our analysis assumptions.
The best-fit parameters, fit statistics, and dSph comparisons are summarized in Table~\ref{tab:gce-bestfits-dsph}.

Finally, we examined a benchmark of the secluded SFDM model with $m_\psi=47.75$~GeV and $m_{h_2}=14.325$~GeV.
The dominant annihilation process,
$\psi \bar \psi \to h_2 h_2$, produces a predominantly four-bottom state through the decays of the
singlet-like scalar.
This benchmark connects the cascade interpretation to a renormalizable Higgs portal model in which the
relic abundance, mediator branching fractions, direct detection signals, and Higgs decays are related.
A global assessment using the full model predictions and complementary constraints is left for future work.

\section*{Acknowledgements}

The authors thank the participants of the 7th and 8th workshops on Dark Matter as a Portal to New Physics (DMPNP 2025 and 2026), the Focus Program [APCTP-2025-F01] and [APCTP-2026-F01], held at APCTP, Pohang, Korea for fruitful discussions.
The authors also thank Mattia Di Mauro and Patrick J. Fox for useful discussion.
This work is supported by the National Research Foundation of Korea (NRF) with Grant No. RS-2025-00562917.
SS is supported by the IBS fund IBS-R018-D1.

\appendix

\section{Validation of the dSphs analysis}
\label{app:validation}

\begin{figure}[t]
\centering
\includegraphics[width=0.78\linewidth]{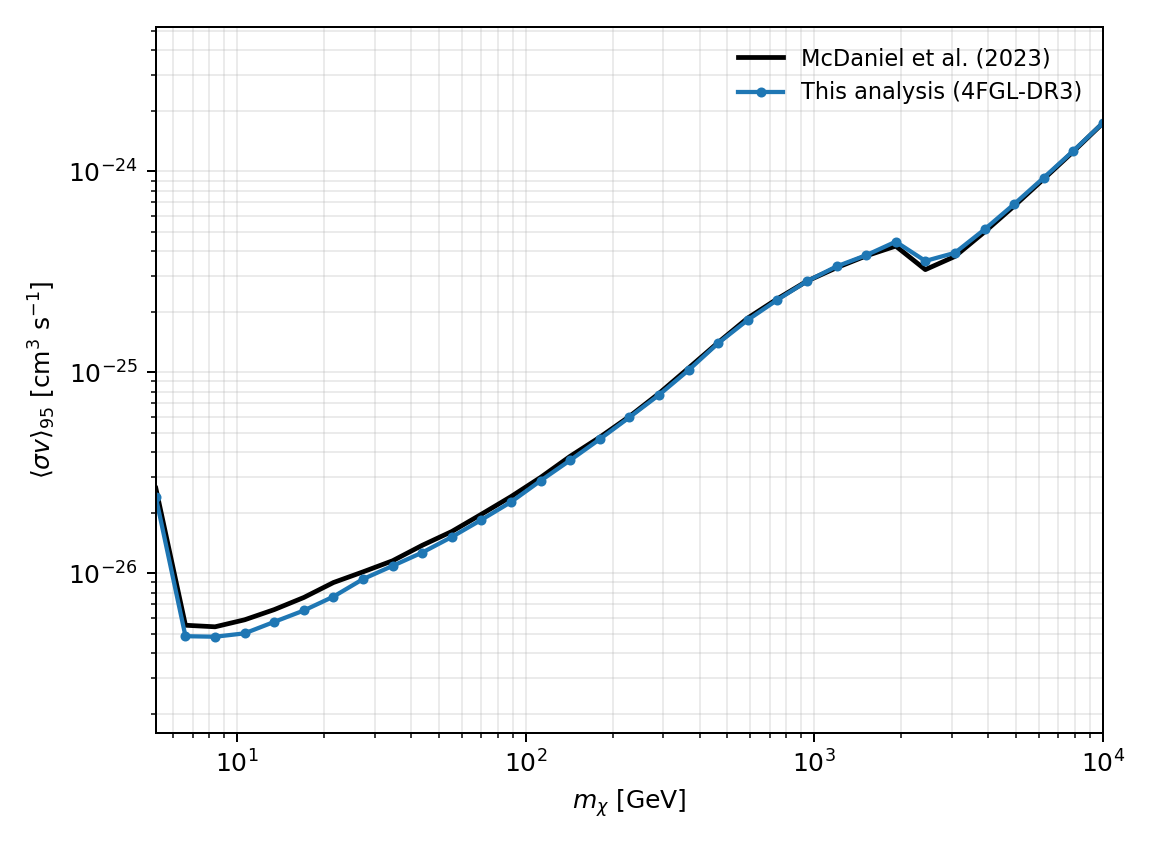}
\caption{\justifying
Combined measured-30 J-prior \(b\bar b\) limit from the 14-year
4FGL-DR3 analysis compared with Ref.~\cite{McDaniel:2023bju}.  Same data duration, dSphs samples, final state model, and background model inputs are used.}
\label{fig:bb-14yr-mauro-comparison}
\end{figure}

\begin{figure}
    \centering
    \includegraphics[width=0.5\linewidth]{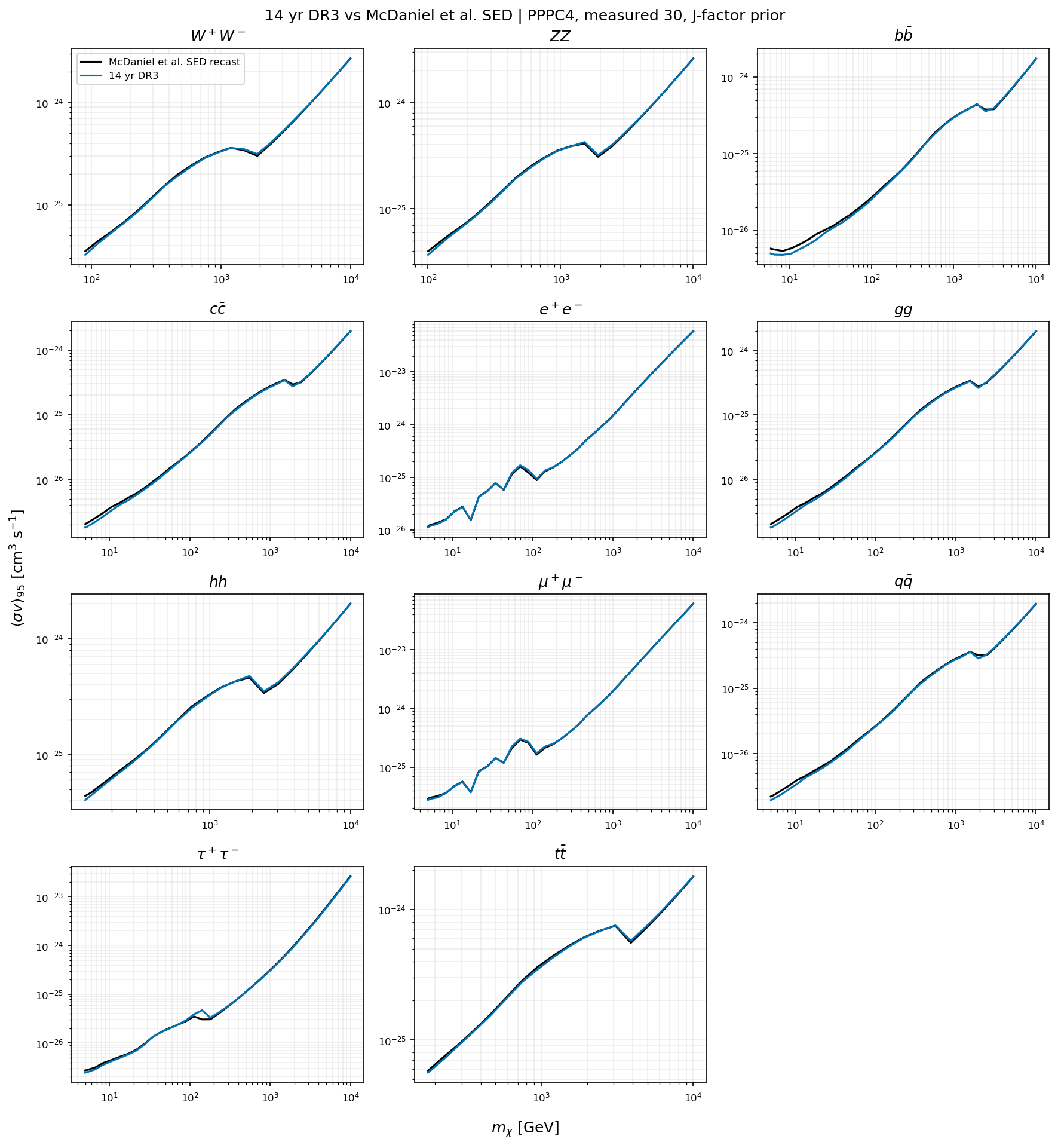}
    \caption{\justifying
    The comparison with Ref. \cite{McDaniel:2023bju} results for possible two-body annihilation channel. In this case, we used PPPC4 table for the prompt annihilation spectra, same as used in Ref. \cite{McDaniel:2023bju}. For most annihilation channels and across a wide range of DM masses, the results show good agreement with those of Ref.~\cite{McDaniel:2023bju}. }
    \label{fig:4FGL_DR3_comparision}
\end{figure}

We checked our 14.3-year $b\bar b$ analysis using 4FGL-DR3
against the reference TS profiles released in Ref.~\cite{McDaniel:2023bju}.
For both our analysis and the reference products, we summed
the individual TS profiles of the same 30 dwarfs at each
DM mass and annihilation cross section, then derived the
combined 95\% confidence-level upper limits.
This comparison used the $J$-factor constraints described above.
The generated and reference TS arrays had matching dimensions,
and all TS entries were finite. Figure \ref{fig:bb-14yr-mauro-comparison} shows the comparison of $b\bar{b}$ limit using same method as Ref. \cite{McDaniel:2023bju}. By comparing them, we verify that our analysis method is consistent with others. Figure \ref{fig:4FGL_DR3_comparision} compares our result with Ref. \cite{McDaniel:2023bju} for other two-body final states.

\section{Comparison of the FL16Y and 4FGL-DR2 catalogs}
\label{app:catalog_comparison}

We compare the FL16Y source list adopted in this work~\cite{Ballet:2026kpq} with the 4FGL-DR2 catalog used in Ref.~\cite{Cholis:2021rpp}.
The comparison covers the $60^\circ\times60^\circ$ region centered on the Galactic center, which contains the $40^\circ\times40^\circ$ fitting region.
In this outer region, FL16Y contains 1130 catalog entries, comprising 1103 entries modeled as pointlike sources and 27 as extended sources.
The corresponding numbers for 4FGL-DR2 are 918 and 26, respectively.

\begin{figure*}[!tb]
    \centering
    \includegraphics[width=0.39\linewidth]{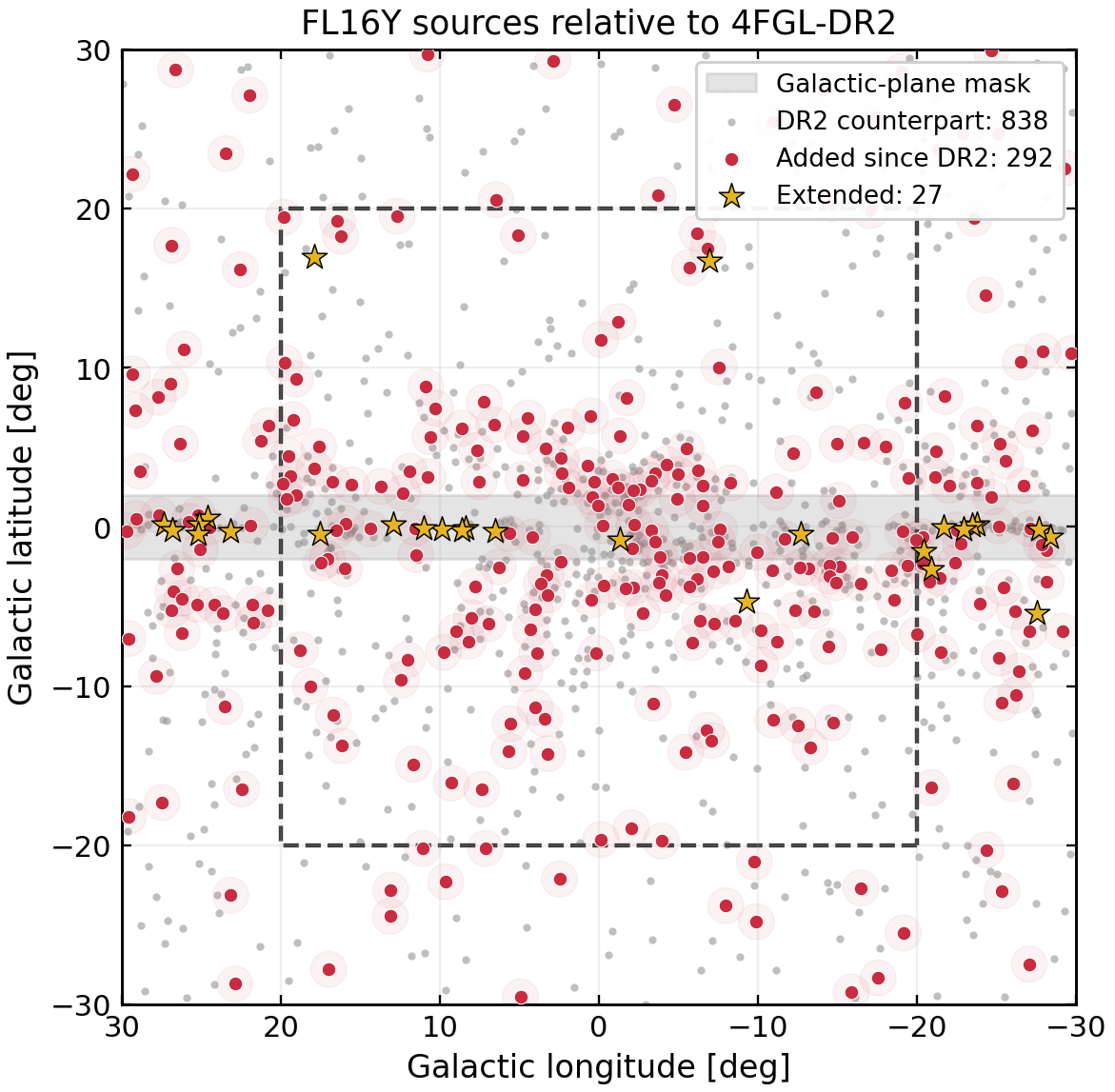}\hfill%
    \includegraphics[width=0.48\linewidth]{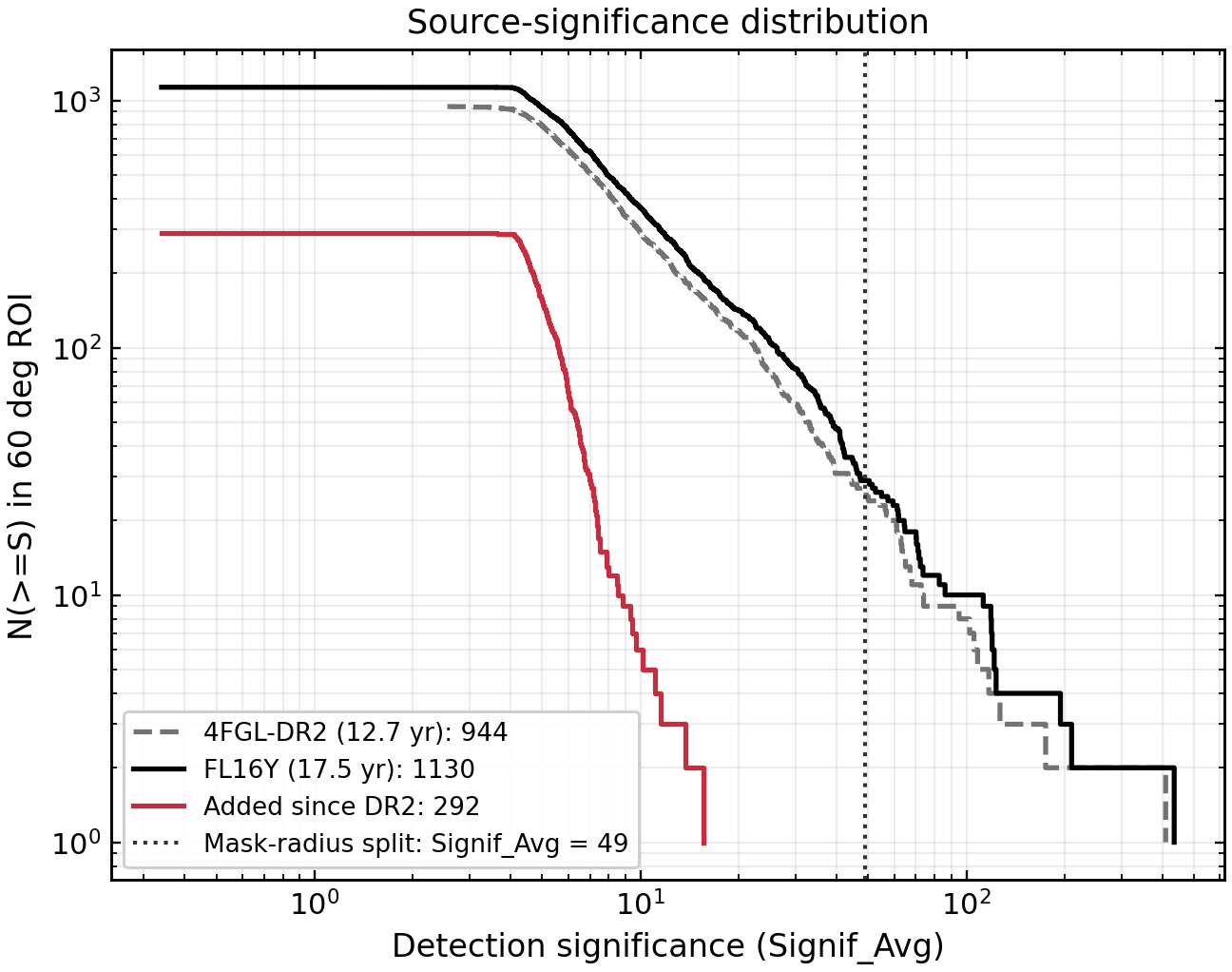}
    \caption{\justifying
    Comparison of the FL16Y and 4FGL-DR2 catalogs within the $60^\circ\times60^\circ$ region centered on the Galactic center.
    \emph{Left:} Positions of the 1130 FL16Y catalog entries. Gray markers indicate entries with a nearest-neighbor DR2 match within $0.2^\circ$, while crimson markers indicate entries without a match under this criterion. Yellow stars identify the 27 entries modeled as extended sources. The translucent crimson disks correspond to the lowest-energy-bin masking radius, $\theta_s=1.125^\circ$, around the unmatched entries. The shaded horizontal band marks the Galactic-plane mask, $|b|<2^\circ$, and the dashed square outlines the $40^\circ\times40^\circ$ fitting region.
    \emph{Right:} Cumulative detection-significance distributions, $N(\geq S)$, where $S$ denotes the catalog quantity $\mathrm{Signif\_Avg}$, for FL16Y (black solid), 4FGL-DR2 (gray dashed), and the unmatched FL16Y entries (crimson). The vertical dotted line marks $S=49$; entries above this threshold receive the larger masking radius.}
\label{fig:fl16y_catalog}
\end{figure*}

To compare the source positions, we use nearest-neighbor matching with a maximum angular separation of $0.2^\circ$.
The left panel of Fig.~\ref{fig:fl16y_catalog} shows the FL16Y entries with and without a DR2 match under this criterion.
Here, an unmatched entry does not necessarily represent a newly detected source, since changes in localization or source modeling can also affect the positional correspondence between the catalogs~\cite{Ballet:2026kpq}.
The shaded band and dashed square indicate the Galactic-plane mask and the fitting region, respectively.

The right panel of Fig.~\ref{fig:fl16y_catalog} compares the cumulative distributions of the catalog detection significance, \texttt{Signif\_Avg}, for the two catalogs and for the unmatched FL16Y entries.
The unmatched entries are concentrated at low detection significance.
The vertical line marks the threshold above which we apply the larger masking radius, following the prescription described in the main text.
The resulting changes in the masked photon-count maps and retained pixel fractions are illustrated in Fig.~\ref{fig:mask_comparison}.

\section{Front and Back foil data}
\label{app:frontback}

\begin{figure}
    \centering
    \includegraphics[width=0.5\linewidth]{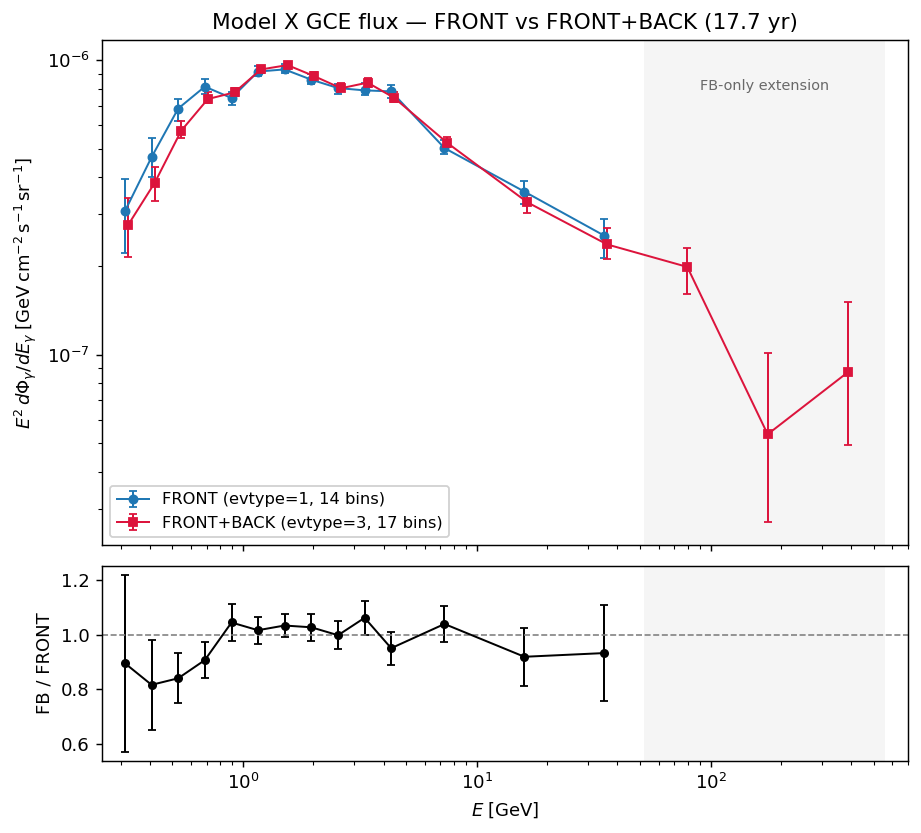}
    \caption{\justifying
    Model~X GCE spectrum from \texttt{FRONT}-converting events (\texttt{evtype=1}, 14 energy bins; blue circles), as in the main analysis, and from \texttt{FRONT+BACK} events (\texttt{evtype=3}, 17 bins; crimson squares). Points show the best-fit values and error bars the 16th--84th percentiles of the posterior. The first 14 bins have identical edges ($0.275$--$51.9$~GeV) for both event selections; the shaded region marks the three bins available only for \texttt{FRONT+BACK}, which extend to $556$~GeV. \emph{Bottom:} ratio of the two spectra, with uncertainties added in quadrature; since the \texttt{FRONT} events are a subset of the \texttt{FRONT+BACK} events, this procedure overestimates the uncertainties. For $E_\gamma>51.9~\mathrm{GeV}$, we extrapolate the masking radii beyond the energy range tabulated in Ref.~\cite{Cholis:2021rpp}.}
    \label{fig:gce_front_vs_frontback}
\end{figure}

The choice of LAT event selection involves a tradeoff between angular resolution and photon statistics. 
\texttt{FRONT}-converting events generally provide a narrower point-spread function, which helps reduce point-source contamination in the Galactic center analysis. 
Including \texttt{BACK}-converting events increases the photon sample and improves the statistics at high energies, where fewer photons are available. 
In this appendix, we compare the GCE spectra obtained with the \texttt{FRONT} and \texttt{FRONT+BACK} selections and extend the latter to energies above the range used in the main analysis.

To test the dependence on the event selection, we repeat the GC fit for GDE Model~X with \texttt{FRONT+BACK} events (\texttt{evtype=3}), keeping all other analysis choices fixed, and extend the binning by three bins with edges at $51.9$, $114.9$, $252.8$, and $556.1$~GeV\footnote{We extended to the higher photon energy as \texttt{BACK} has better sensitivity in that region compared to $\texttt{FRONT}$}. The same FL16Y-based masks are used for both selections, although the radii of Table~III of Ref.~\cite{Cholis:2021rpp} were chosen for \texttt{FRONT} events and the PSF of \texttt{BACK} events is broader. Above $51.9$~GeV the radii are extrapolated linearly in energy from the two highest-energy bins of that table and set to zero where the extrapolation becomes negative: in the $51.9$--$114.9$~GeV bin this gives small- and large-mask radii of about $0.04^\circ$ and $0.12^\circ$, comparable to or smaller than the $0.1^\circ$ pixel size, and at higher energies at most the pixel containing each source is masked. Figure~\ref{fig:gce_front_vs_frontback} compares the two spectra. Above $0.78$~GeV, the ratio of the \texttt{FRONT+BACK} to the \texttt{FRONT} intensity lies between $0.92$ and $1.06$ in all bins, within about $1\sigma$, and the weighted mean of the ratio over the 14 common bins is $1.00$. In the four lowest-energy bins ($0.275$--$0.784$~GeV), the \texttt{FRONT+BACK} intensity is lower by $9$--$18\%$; each bin deviates by less than $2\sigma$ (at most $1.7\sigma$), but all four deviate in the same direction. Because the \texttt{FRONT} events are a subset of the \texttt{FRONT+BACK} sample, the two spectra are positively correlated and the uncertainties of the ratio are overestimated, so this low-energy offset is more significant than these numbers suggest. The \texttt{FRONT+BACK} selection extends the spectrum to $556$~GeV, with intensities of $2.0\times10^{-7}$, $5.4\times10^{-8}$, and $8.7\times10^{-8}~\mathrm{GeV\,cm^{-2}\,s^{-1}\,sr^{-1}}$ at $77$, $170$, and $375$~GeV; these three bins are not used in the DM fits.

\section{Explanation of weak limit of 17.7 yr}\label{app:dSphs_weak}

\begin{figure}
    \centering
    \includegraphics[width=1.0\linewidth]{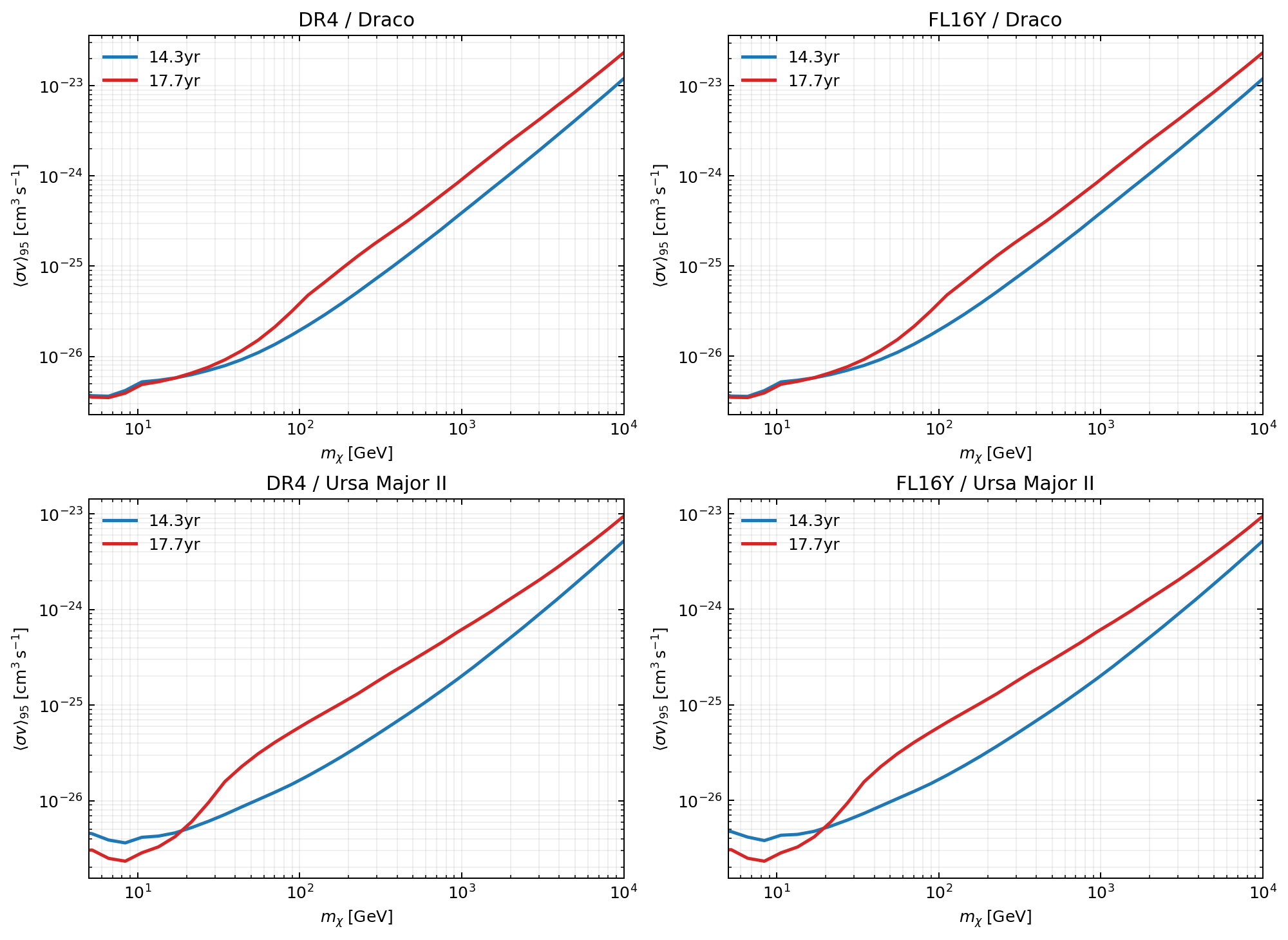}
    \caption{\justifying
    Comparison of the limits from Draco and Ursa Major II galaxies, individually not combined. Regardless to the catalog, those galaxies show weaker limit with 17.7yr compared to the 14.3yr. This can be possibly explained by the observation of time-varying point source in the vicinity of the target center.}   \label{fig:Draco_UrsaMajorII_ind_limit}
\end{figure}

\begin{figure}
    \centering
    \includegraphics[width=0.6\linewidth]{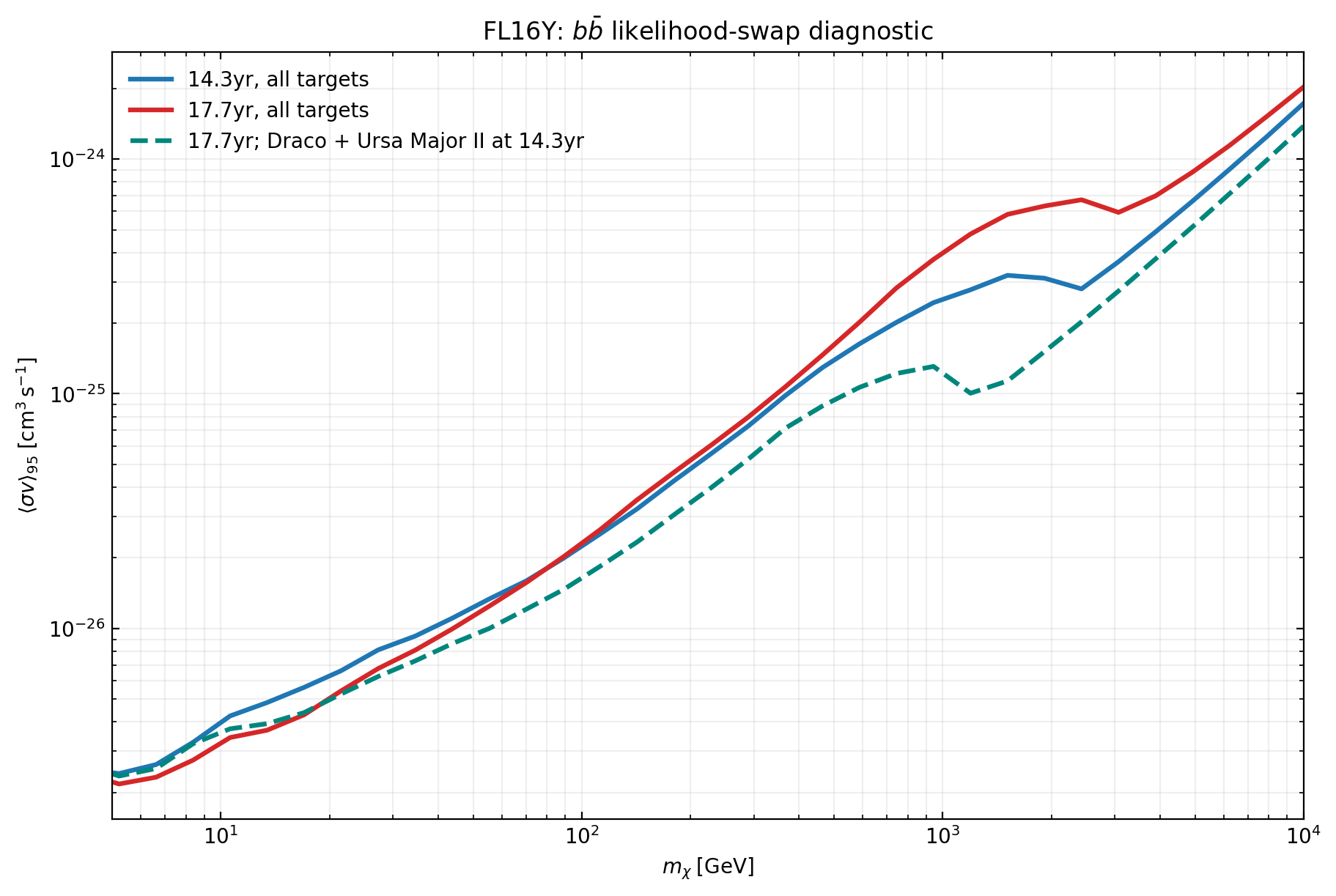}
    \caption{\justifying
    Comparison of the cases where only Draco and Ursa Major II are kept as 14.3yr, while other samples are based on 17.7yr. The dashed green line is the dSphs limit for 28 dSphs with 17.7yr, and Draco and Ursa Major II with 14.3yr, using same FL16Y catalog. After fixing the two galaxy samples, the 17.7yr limit follows the expected behavior of having stronger limit even on the higher DM mass.}
    \label{fig:17yr_Draco_UrsaMajorII_revert}
\end{figure}

Figure \ref{fig:matrix_comparison_bb} shows general weakening of the 17.7yr dSphs limits compared to the 14.3yr even with same point source catalogs. The main contribution to this result is from Draco and Ursa Major II galaxies. Figure \ref{fig:Draco_UrsaMajorII_ind_limit} compares the $b\bar{b}$ limits under same observation period and point source catalog. As Draco and Ursa Major II generally shows weak limits under 17.7yr compared to 14.3yr and their limits are relatively stronger compared to the other targets, such as Pisces II and Sagittarius II, their contributions are dominant on the combined likelihood profile. Figure \ref{fig:17yr_Draco_UrsaMajorII_revert} demonstrates the direct effect of the two galaxies, by fixing their likelihood sample as 14.3yr while updating all the other 28 galaxy samples. 
Retaining the 14.3-year likelihood profiles for Draco and Ursa Major II while updating the other 28 targets strengthens the combined high-mass limits.

The weakening of the limits for these two targets could potentially be explained by a time-varying source near their respective centers. For example, we identified a photon event near the center of Draco at MET 792266927.532209 (UTC 2026-02-08 18:08:42), with a reconstructed energy of approximately 14.8 GeV and an angular separation of $0.019^\circ$ from the target center.

\begin{figure}
    \centering
    \includegraphics[width=0.6\linewidth]{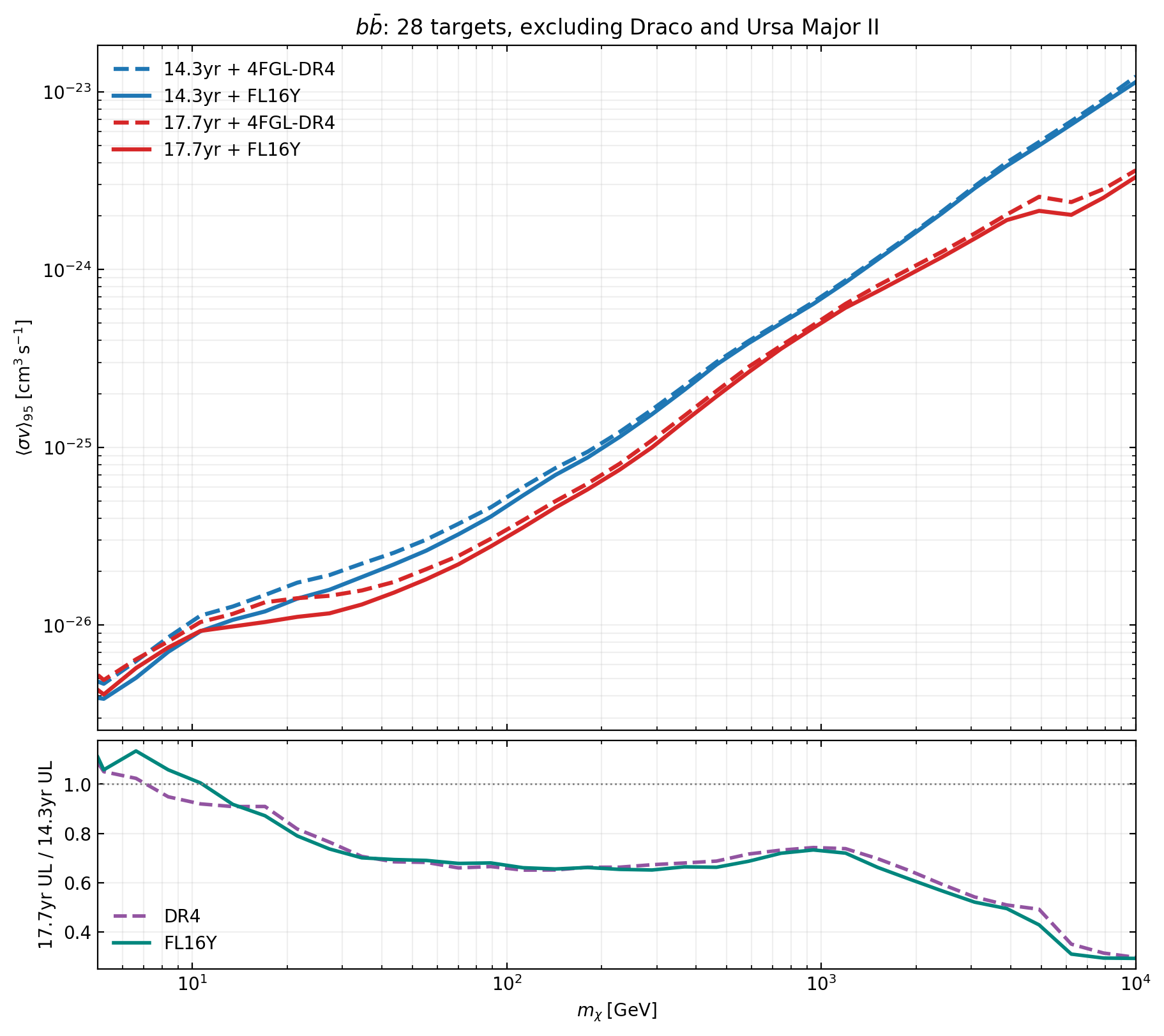}
    \caption{\justifying
    The dSphs limits for each different observation duration and catalog. Draco and Ursa Major II targets are excluded in this figure.}
    \label{fig:placeholder}
\end{figure}

\bibliography{ref}

\end{document}